\documentstyle[epsfig]{mn}
 at 10truept

\def\m@th{\mathsurround=0pt }
\def\eqalign#1{\null\,\vcenter{\openup1\jot \m@th
 \ialign{\strut\hfil$\displaystyle{##}$&$\displaystyle{{}##}$\hfil
 \crcr#1\crcr}}\,}

\title [Stable clustering, the halo model and nonlinear cosmological power spectra]
{\vglue-3.0truecm
\centerline{\it Accepted for publication in  Monthly Notices}
\vglue 2.5truecm
      Stable clustering, the halo model and nonlinear cosmological power
spectra}
\author 
     [{\it Smith et al.}]{
     \parbox[t]{\textwidth}{
     R. E. Smith$^{1,2\thanks{E-mail: robert.e.smith@nottingham.ac.uk}
}$, 
     J. A. Peacock$^{1}$, 
     A. Jenkins$^{3}$, 
     S. D. M. White$^{4}$, 
     C. S. Frenk$^{3}$,\\
     F. R. Pearce$^{2}$, 
     P. A. Thomas$^{5}$, 
     G. Efstathiou$^{6}$ 
     and H. M. P. Couchman$^{7}$ \\
     {\Large(The Virgo Consortium)}}\\\\
     $^1$ 
     Institute for Astronomy, 
     University of Edinburgh,
     Royal Observatory,
     Blackford Hill, 
     Edinburgh, 
     U.K.\\   	
     $^2$
     School of Physics and Astronomy,         
     University of Nottingham,                
     University Park,                         
     Nottingham, 
     NG7 2RD\\   
     $^3$ 
     Department of Physics, 
     University of Durham, 
     South Road, 
     Durham, 
     DH1 3LE\\
     $^4$ 
     Max-Planck-Institut f\"{u}r Astrophysik, 
     Garching, 
     D-85740 M\"{u}nchen,
     Germany\\
     $^{5}$
     Astronomy Centre,
     CPES, 
     University of Sussex, 
     Falmer, 
     Brighton, 
     BN1 9QH\\
     $^{6}$	
     Institute of Astronomy, 
     Madingley Road,
     Cambridge\\
     $^{7}$
     Department of Physics and Astronomy, 
     McMaster University, 
     Hamilton, 
     Ontario 
     L8S 4M1, 
     Canada}

\def\be{\begin{equation}}
\def\ee{\end{equation}}
\def\pppm{P$^3$M}
\def\rhob{\bar{\rho}}

\def\simless{\mathbin{\lower 3pt\hbox
  {$\rlap{\raise 5pt\hbox{$\char'074$}}\mathchar"7218$}}}   
\def\simgreat{\mathbin{\lower 3pt\hbox  
   {$\rlap{\raise 5pt\hbox{$\char'076$}}\mathchar"7218$}}}  
\def \scs{\scriptscriptstyle}
\def \thinthin{\kern 0.05em}
\def \L{\thinthin{\scs{\rm L}}}
\def \NL{\thinthin{\scs{\rm NL}}}
\def \b{\thinthin{\rm b}}
\def \eff{\thinthin{\rm eff}}

\def \initial{\thinthin{\rm initial}}
\def \final{\thinthin{\rm final}}
\def \err{\thinthin{\rm err}}
\def \Nq{\thinthin{\rm Ny}}
\def \shot{\thinthin{\rm shot}}
\def \glass{\thinthin{\rm glass}}
\def \true{\thinthin{\rm true}}
\def \delk{\delta_{\bf k}}
\def \kdotx{{\bf k}\cdot{\bf x}}
\def \H{\thinthin{\scs{\rm H}}}
\def \Q{\thinthin{\scs{\rm Q}}}

\def \obs{\thinthin{\rm obs}}
\def \disc{\thinthin{\rm disc}}
\def \JMW{\thinthin{\scs{\rm JMW}}}
\def \PD{\thinthin{\scs{\rm PD}}}
\def \hompc{{\,h\,\rm Mpc^{-1}}}
\def \mpcoh{{\,h^{-1}\,\rm Mpc}}
\def\ba{\begin{eqnarray}}
\def\ea{\end{eqnarray}}

\newbox\voidbox
\def \LL{\unhbox\voidbox\setbox0=\hbox{L}\hbox to \wd0{\hss\char'40L}}

\begin{document}

\maketitle


\begin{abstract}\vspace{-0.3cm}\\
We present the results of a large library of cosmological $N$-body
simulations, using power-law initial spectra. The nonlinear evolution
of the matter power spectra is compared with the predictions of
existing analytic scaling formulae based on the work of Hamilton et
al.  The scaling approach has assumed that highly nonlinear structures
obey `stable clustering' and are frozen in proper coordinates.  Our
results show that, when transformed under the self-similarity scaling,
the scale-free spectra define a nonlinear locus that is clearly
shallower than would be required under stable clustering.
Furthermore, the small-scale nonlinear power increases as both the
power spectrum index $n$ and the density parameter $\Omega$ decrease,
and this evolution is not well accounted for by the previous scaling
formulae.  This breakdown of stable clustering can be understood as
resulting from the modification of dark-matter haloes by continuing
mergers. These effects are naturally included in the analytic `halo
model' for nonlinear structure; we use this approach to fit both
our scale-free results and also our previous CDM data.  
This method is more accurate than the
commonly-used Peacock--Dodds formula and should be applicable to more
general power spectra.  Code to evaluate nonlinear power spectra using
this method is available from {\tt
http://as1.chem.nottingham.ac.uk/$\sim$res/software.html}.  Following
publication, we will make the power-law simulation data publically
available through the Virgo website {\tt
http://www.mpa-garching.mpg.de/Virgo/}.

\end{abstract}

\begin{keywords}
Cosmology: theory -- large scale structure of Universe --
Galaxies: gravitational clustering 
\end{keywords}


\section{Introduction}

In the current cosmological paradigm, structures grow through the
gravitational instability of collisionless dark matter
fluctuations. This occurs in a hierarchical way, with small-scale
perturbations collapsing first and large-scale perturbations
later. One of the most direct manifestations of this nonlinear process
is the evolution of the power spectrum of the mass, $P(k)$, where $k$
is the wavenumber of a given Fourier mode.  Understanding this
evolution of the power spectrum is one of the key problems in
structure formation, being directly related to the abundance and
clustering of galaxy systems as a function of mass and redshift. If
the processes that contribute to the evolution can be captured in an
accurate analytic model, this opens the way to using observations of
the nonlinear mass distribution (from large-scale galaxy clustering or
weak gravitational lensing) in order to recover the primordial
spectrum of fluctuations.

One of the most influential attempts at such an analytic description
of clustering evolution was the `scaling ansatz' of Hamilton et
al. (1991; HKLM), which is described in Section \ref{scfmodels}.  This
scaling procedure was generalized to models with $\Omega\ne1$ and
given a more accurate $N$-body calibration by Peacock \& Dodds (1996;
PD96).  HKLM assumed that a nonlinear collapsed object would decouple
from the global expansion of the Universe to form an isolated system
in virial equilibrium -- the `stable clustering' hypothesis of Davis
\& Peebles (1977). This assumption has been widely adopted, and yet it
appears somewhat inconsistent with hierarchical models -- in which
objects are continuously accreting mass and growing through mergers.
Indeed, the validity of stable clustering has been increasingly
questioned in recent years (e.g. Yano \& Gouda 2000; Caldwell et
al. 2001).  One of our aims in this paper is thus to establish whether
stable clustering is relevant for understanding the small-scale
evolution of the power spectrum.
 
We therefore explore the gravitational instability of dark matter
fluctuations through a series of large $N$-body simulations of
clustering from power-law initial conditions, with
\begin{equation}
P(k) \propto k^n.
\end{equation}
We consider both $\Omega=1$ models, in which the evolution can obey a
similarity solution, and also low-density models with and without a
cosmological constant.  We demonstrate that the resolution of the
simulations is sufficient to measure the power well into the regime at
which the HKLM procedure predicts a well-defined slope for the power
spectrum determined by stable clustering.  In practice, we find that
the power spectra are generally shallower than would be required for
clustering to be stable on small scales.  Furthermore, as both $n$ and
$\Omega$ decrease, the amplitude of the small-scale spectrum increases
in a manner that is not well described by any of the previous fitting
formulae.  In light of these results, a new method for predicting
nonlinear spectra is proposed.  This method is based on the `halo
model' (e.g. Seljak 2000; Peacock \& Smith 2000), which does not
assume stable clustering. This allows us to fit our data and also the
cold dark matter (CDM) data of Jenkins et al. (1998; J98) with a high
degree of accuracy.

The paper is structured as follows. In Section \ref{scfmodels} we
provide a brief overview of the theoretical understanding of nonlinear
evolution. In particular, a description of the stable clustering
hypothesis, the nonlinear HKLM scaling relations and the halo model
are given, as these ideas are central to this paper. We also discuss
the scale-free models and their self-similarity properties.  In
Section \ref{scfsimulations} we describe the numerical simulations and
we provide a visual comparison of the growth of structure in the
different scale-free models. In Section \ref{scfmeasuring} we describe
an improved method for measuring power spectra and in Section
\ref{scfresults} we present the power spectra data and contrast them
with the current nonlinear fitting formulae. In Section \ref{scfhalo}
we describe a new approach to fitting power spectra and its
generalization to CDM, and then compare our new globally optimized
formula with the results from Section \ref{scfresults} and also the
CDM data. Finally, in Section \ref{scfdiscussion} we draw our
conclusions and discuss our findings in a wider context.

\vfill


\section{Description of Nonlinear evolution}\label{scfmodels}


\subsection{From linear theory to stable clustering}

The mass density field, at comoving position ${\bf x}$ and time $t$, is defined as
\be \rho({\bf x},t) = \rhob(t)\left[1+\delta({\bf x},t)\right]\  ,\ee
where $\delta$ is the density fluctuation about the homogeneous
background $\rhob$. The 2-point auto-correlation function of the
density field is
\be \xi({\bf r})=\left<\delta({\bf x})\delta({\bf x}+{\bf r})\right> \
,\ee
which in three dimensions is related to the dimensionless power
spectrum $\Delta^2(k)$ through the integral relation
\be \xi(r)=\int\Delta^2(k) \, \frac{\sin kr}{kr} \, \frac{dk}{k}\ ,\ee
where we have assumed that the field is isotropic and homogeneous.
$\Delta^2$ is the contribution to the fractional density variance per
unit ln $k$. In the convention of Peebles (1980), this
is
\be \Delta^2(k)\equiv \frac{d\sigma^2}{d \ln
k}=\frac{V}{(2\pi)^3}4\pi k^3 P(k)\ ,\ee 
$V$ being a normalization volume.

When $\delta({\bf x},t)\ll1$ the temporal evolution of the fluctuation is
separable and the field scales as
\be \delta({\bf x},t) = \frac{D(t)}{D(t_0)}\: \delta({\bf x},t_0)\  ,\ee
where $D(t)$ is a growth factor whose exact form can be determined
from linear theory. As $\delta({\bf x},t)\rightarrow 1$, increasingly
higher orders of perturbation theory are required (see Bernardeau et
al. 2001 for a thorough review). Eventually, perturbation theory fails
and numerical methods must be applied. Even so, it was proposed
\cite{Peebles1974a,DavisPeebles1977,Peebles1980} that clustering in
the very nonlinear regime might be understood by assuming that regions
of high density contrast undergo virialization and subsequently
maintain a fixed proper density.  The correlation function for a
population of such systems would then simply evolve according to
$\xi(r,t)\propto 1/\rhob \propto a^3$, where $r$ is a proper distance.
This evolution was termed `stable clustering'. Peebles went on to show
that if the initial power spectrum was a pure power-law in $k$ with
spectral index $n$, $P(k)\propto k^n$, and if $\Omega=1$, then under
the stable clustering hypothesis, the slope of the nonlinear
correlation function would be directly related to the spectral index
through the relation
\be \xi(r,t)\propto r^{-\gamma} \ ; \hspace{0.5cm}
\gamma=\frac{3(3+n)}{5+n}\label{stab-gam} .\ee
Hence, if stable clustering applies, then the nonlinear density field
retains some memory of its initial configuration, and in principle can
be used to measure the primordial spectrum of fluctuations.


\subsection{The HKLM scaling relations}

HKLM developed a method for interpolating between linear theory on
large scales and the nonlinear predictions of the stable clustering
hypothesis on small scales. They showed that the nonlinear volume
averaged two-point correlation function,
\be \bar{\xi}(x) \equiv \frac{3}{x^3}\int_0^x y^2\xi(y) 
\;dy \ ,\ee
measured from the scale-free simulations of Efstathiou et
al. \shortcite{Efstathiouetal1988}, could be parameterized by a simple
function of the linear correlation function, provided that nonlinear
evolution were to induce a change of scale.

The transformation of scales follows from an intuitive continuity
argument, based upon the `spherical top-hat' model.  Let the mass
enclosed within a spherical overdensity in the initial stages of
evolution be $m_0\left(<\ell\right)$ and its mass at some later time
be $m\left(<x\right)$. As each shell evolves, it will reach a maximum
expansion point, turn around and collapse. If there is no shell
crossing, then mass is conserved and
\be m_0\left(<\ell\right)=\frac{4}{3}\pi\rho\left(<\ell\right) \;
\ell^3=\frac{4}{3}\pi \rho\left(<x\right)\; x^3 = m\left(<x)\right)\
.\ee
The argument now identifies $1+\bar{\xi}$ as the factor by which the
density is enhanced relative to the mean \cite{Peebles1980}.  Provided
$\bar{\xi}_{\L} \ll 1$, this implies the scaling
\be x^3\left[1+\bar{\xi}_{\NL}(x,t)\right]=\ell^3\ ,
\label{HKLMscaling}\ee
where $x$ represents a nonlinear scale and $\ell$ a Lagrangian scale.

Finally, after this rescaling, the nonlinear correlations are
taken to be a universal function of the linear ones:
\be \bar{\xi}_{\NL}(x,t)=f\left[\bar{\xi}_{\L}(\ell,t)\right].\ee
HKLM then assumed that the functional form of $f(y)$ could be
determined analytically in two regimes: in the linear regime, where
$\bar{\xi}_{\L} \ll 1$, $f(y)=y\ ;$ when $\bar{\xi}_{\L}\gg1$, galaxy
groups would exhibit `stable clustering', for which
$\Delta_{\NL}^2\propto a^3$ and since $\Delta_{\L}^2\propto a^2$, this
implied that $f(y)\propto y^{3/2}$. The interpolation between these
two regimes, where $y \sim 1$, was determined empirically by HKLM, by
comparison with numerical simulation.  However, Padmanabhan
\shortcite{Padmanabhan1996} proposed that the quasilinear regime could
also be understood analytically.  He considered the point at which a
spherical perturbation would reach its maximum radius, which is
$x_{\max}=l/\delta_{\L}\propto l/\bar{\xi}_{\L}$, according to the
spherical model. Padmanabhan thus conjectured that
\be \bar{\xi}_{\Q} \propto \rho(<x_{\max}) \propto \frac{m}{x_{\max}^3}
\propto \frac{m_0}{x_{\max}^3} \propto
\frac{l^3}{l^3/\bar{\xi}_{\L}^3}\propto \bar{\xi}_{\L}^3 \label{paddy}\ee
(in effect rediscovering the argument of Gott \& Rees 1975).  Although
useful heuristically in explaining why the quasilinear regime of
$f_{\NL}$ should be steeper than either the linear or nonlinear
regime, it is not clear that this expression matches the observed
quasilinear slope very well \cite{Padmanabhanetal1996,Jain1997}.  We
investigate this further in Section \ref{scfresults}.

HKLM's nonlinear scaling argument was further developed by Peacock \&
Dodds (1994; PD94), who proposed that the scaling ansatz could be used
for predicting power spectra by simply replacing
$\bar{\xi}\rightarrow\Delta^2$ and letting the linear and nonlinear
scales represent linear and nonlinear wavenumbers: $\ell=k_{\L}^{-1}$
and $x=k_{\NL}^{-1}$. This suggested the formalism
\[ \Delta^2_{\NL}(k_{\NL})=f_{\NL} \left[ \Delta^2_{\L}
(k_{\L})\right]\ ; \]
\be k_{\NL}=\left[1+\Delta_{\NL}^{2}(k_{\NL})\right]^{1/3}k_{\L} \label{NSR} .\ee

The accuracy of the HKLM and PD94 scaling formulae was tested by Jain,
Mo \& White (1995; JMW95). They performed a series of simulations with
$100^3$ particles as opposed to the previous $32^3$, and discovered
that the nonlinear locus described by the data exhibited a strong
$n$-dependence. The HKLM and PD94 functions underestimated the
measured correlation functions and power spectra, the fits being worse
for more negative $n$. JMW95 then showed that this $n$-dependence
could be removed by a simple scaling of the variables in the
$\log\bar{\xi}_{\NL}(x,t)-\log\bar{\xi}_{\L}(\ell,t)$ plane.  In order
for the model to be applied to curved spectra, such as the CDM model,
an effective spectral index $n_{\eff}$ was required. JMW95 proposed
that the appropriate $n$ should be given by
\be n_{\eff}=\left.\frac{d\ln P(k)}{d \ln k}\right|_{k=1/R_c} 
\label{JMW95neff}\ ,\ee
where $R_c$ is the scale on which the variance of the density field is
unity. This showed the right response with scale, and described their
data to a precision of $15-20\%$, which was adequate given the scatter
within the simulations.

Further refinements were again made by Peacock \& Dodds (1996; PD96),
who used a large ensemble of $80^3$ particle simulations to
investigate the $n$-dependence and the response of the clustering to
low density universes: $\Omega<1$ and $\Omega+\Lambda=1$, where
$\Omega$ and $\Lambda$ are the densities associated with matter and
the cosmological constant, relative to the Einstein--de Sitter
universe. PD96 concluded that nonlinear effects tend to increase the
power on small scales for spectra with more negative spectral indices
and for lower densities. PD96 also produced a fitting formula which
modelled their data, and also CDM-like spectra through defining an
effective spectral index that changed with each wavenumber
\be n_{\eff}(k_{\L})=\frac{d\ln P}{d \ln k}(k=k_{\L}/2)\ . 
\label{PD96neff}\ee

Subsequently, high resolution numerical simulations of CDM-like
universes have shown that the PD96 formulae match the observed
nonlinear power spectra closely (Mo, Jing \& B\"orner 1997; J98; Smith
et al. 1998), but with some significant deviations.  Jain \&
Bertschinger \shortcite{JainBertschinger1998} found a larger
discrepancy in their $256^3$ \pppm \ simulation of clustering from an
$n=-2$ power spectrum, with both the formula of JMW95 and PD96
underestimating the quasi-linear power. They also claimed that their
results for highly nonlinear clustering were in accordance with stable
clustering, although finite volume effects have drawn their results
into question \cite{MaFry2000a,Scoccimarroetal2001}. We discuss this
issue in further detail in Section \ref{finitevolume}. Recent attempts
to constrain cosmological parameters from weak gravitational lensing
studies, that require as input the nonlinear matter power spectrum,
have also uncovered deficiencies in the PD96 formula, with the poorest
performance for the $\Omega=1$ $\tau$CDM model
\cite{VanWaerbekeetal2001}.


\subsection{A dark matter halo approach}

More recently an entirely different analytical model for nonlinear
gravitational clustering has emerged: the `halo model'.  In this
model, the density field is decomposed into a distribution of clumps
of matter with some density profile.  This basic idea goes back to
Neyman \& Scott (1952), and recurs in more modern form in
\cite{ScherrerBertschinger1991}.  Following the realization that
galaxy bias was strongly influenced by the number of galaxies in a
halo \cite{JingMoBorner1998,Bensonetal2000}, a number of authors
\cite{Seljak2000,PeacockSmith2000,MaFry2000a,Scoccimarroetal2001}
resurrected the Neyman--Scott model with a modern mass function for
dark haloes \cite{PressSchechter1974,ShethTormen1999,Jenkinsetal2001},
plus realistic density profiles
\cite{NavarroFrenkWhite1996,NavarroFrenkWhite1997,Mooreetal1999}, and
a mass-dependent galaxy `occupation number'.  The inclusion of bias is
an attractive aspect of the halo model, but we will not be concerned
with this here.

In the halo model, the large-scale clustering of the mass arises
through the correlations between different haloes.  Prescriptions for
this clustering were given by Mo \& White \shortcite{MoWhite1996}; Mo,
Jing \& White \shortcite{MoJingWhite1997}; Sheth \& Lemson
\shortcite{ShethLemson1999}; Sheth \& Tormen
\shortcite{ShethTormen1999}; Sheth, Mo \& Tormen
\shortcite{ShethMoTormen2000}, and a recent example of their
effectiveness is shown clearly in Colberg et
al. \shortcite{Colbergetal2000}. On small scales, the correlations are
derived purely from the convolution of the density profile of the halo
with itself \cite{Peebles1974b,McClellandSilk1977,ShethJain1997}.
This model thus makes strong predictions about the clustering on small
scales. Unless the density profile and mass function obey a specific
relationship, the merger-driven evolution of the mass function means
that stable clustering approximation does not hold true
\cite{YanoGouda2000,MaFry2000b}.  For a more detailed review of the
halo model and its applications we refer the reader to
Sheth \& Cooray \shortcite{ShethCooray2002}.


\subsection{Scale-free models}\label{scfnonlinearscale}

An elegant way to study nonlinear evolution is to simulate
`scale free' universes that have no inbuilt
characteristic physical length scales.
We follow Efstathiou et al. \shortcite{Efstathiouetal1988} 
and require two conditions to be satisfied:
\begin{enumerate}
\item The initial power spectrum of fluctuations is a power law:
\be P(k)= A k^n ;\hspace{1cm} 1 < n <-3 .\ee 
\item The evolution of the scale factor for the cosmological model
power law in time:
\be a(t)\propto t^{\alpha}\ .\ee 
\end{enumerate}
The most interesting cosmological model that satisfies these
constraints is the Einstein--de Sitter model: $\alpha=2/3$, $\Omega=1$
and $\Lambda=0$, so that the linear-theory growth of the
power spectrum is $P(k)\propto a^2$.

In this case, the only natural way to define a characteristic length is
through the scale at which the fluctuations become nonlinear.
The variance of the linear density field, smoothed on some comoving length
scale $x$, is
\be \sigma^2(x,a)=\int \Delta^2_{\L}(k,a)\; |W(kx)|^2 \, \frac{dk}{k}\ , \ee
where $W$ is the filter function. If we assume
$\Delta^2(a,k)\propto a^2 k^{3+n}$, and that the
filter causes a cut-off at
some high spatial frequency $k_c \sim 1/x$, we find
\be \sigma^2(k_c,a)\propto \int^{k_c}_{0} a^2 k^{n+2}dk \propto a^2
x_c^{-(3+n)} \ .\ee
We now define a nonlinear wavenumber, $k_{\NL}$ such that
$\sigma^2(k_{\NL},a) = 1$, so that
\be k_{\NL}(a) \propto a^{-2/(3+n)}.  \label{self-sim} \ee 
Under this transformation, it is plausible that 
the statistics of gravitational clustering will be expressible
as a similarity solution:
\be P(k,a) = \tilde P(k/k_{\NL}) \ee
\cite{DavisPeebles1977,Peebles1980,Efstathiouetal1988,JainBertschinger1998}.
No formal proof of the similarity solution exists, and this conjecture
is something that must be tested empirically via simulation. We refer
the reader to the work of Colombi, Bouchet \& Hernquist
\shortcite{Colombietal1996} for further discussion of the range of
spectral indices for which self-similarity should be valid.

In practice, we present good evidence in this paper that the power
spectrum does scale in this way for $0 \ge n \ge -2$. Spectra outside
this range are harder to simulate and so not yet tested.  We may
however anticipate that only certain initial spectra will evolve in a
self-similar fashion.  For $n\ge 1$, the amplitude of gravitational
potential fluctuations diverges on small scales, so one might question
the idea of a hierarchy that grows via the merger and disruption of
small systems.  However, this argument is not definitive, since the
similarity solutions generally depart from $P\propto k^n$ for $k >
k_{\NL}$. We seek a function which is of this power-law form for $k <
k_{\NL}$ and some unknown form at larger $k$, and which evolves in a
self similar fashion. In practice, this function is found by starting
with exact power-law initial conditions, and hoping that the
simulation will relax into the desired self-similar form as it
evolves. The existence of a self-similar solution with $n>1$ on large
scales therefore remains an open question.  On large scales, the
peculiar velocity field diverges if $n \le -1$, so more negative
indices may seem problematic. This does not seem to be a problem in
practice, probably for the reasons discussed by Bernardeau et
al. \shortcite{Bernardeauetal2001}: the divergent modes of very long
wavelength really just cause a translation, and Galilean invariance
means that the statistics of smaller-scale clustering are unaffected.
Certainly, well-defined results can be obtained from perturbation
theory for $n$ more negative than $-1$, so the only clear limit is
$n\le -3$, for which the whole idea of asymptotic homogeneity breaks
down.

If we can find initial spectra for which self-similarity applies, this
is an extremely useful means of assessing the reliability of $N$-body
results.  Also, over limited ranges of mass, the scale-free models
correspond directly to more physically motivated models such as CDM,
whose spectral index is a slow function of scale.  As we shall show,
an analytic description of nonlinear evolution in the scale-free case
leads quite directly to a method that can also give an accurate
description of nonlinear evolution in CDM models.


\begin{table*}
\caption{\small{Parameters of the $256^3$ particle, scale-free
simulations. The r1 simulations represent glass initial
conditions and r2 simulations are grid starts.}}
\label{table-scf}
\raggedright
\begin{tabular}{lccccccl}
\hline
\hline
\noalign{\vglue 0.2em}
simulation & $ \epsilon/L$ & $\Delta^2(k_{\b},a=1)$ & 
$a_{\initial}$ & $a_{\final}$ & timesteps & energy error & output values of $a$ \\
\noalign{\vglue 0.2em}
\hline
\noalign{\vglue 0.2em}
$n=-2$\ r1 & 0.00025 & 0.133 & 0.025 & 0.62 & 831 & 0.04 \% &
$0.025, 0.1, 0.2, 0.3, 0.4, 0.5, 0.6, 0.62$\\
$n=-2$\ r2 & 0.00025 & 0.133 & 0.025 & 0.55 & 904 & 0.04 \% & 
$0.025, 0.1, 0.2, 0.3, 0.4, 0.5, 0.55$\\ 
$n=-1.5$\ r1 & 0.00023 & 0.046 & 0.010 & 0.96 & 991 & 0.16 \% &
$0.01, 0.25, 0.315, 0.4, 0.5, 0.63, 0.794, 0.96$\\
$n=-1.5$\ r2 &  0.00023 & 0.046 & 0.010 & 1.00 & 915 & 0.16 \% &
$0.01, 0.25, 0.315, 0.4, 0.5, 0.63, 0.794, 1.0$\\
$n=-1$\ r1 &  0.00023 & 0.017 & 0.010 & 0.83 & 991 & 0.31 \% &
$0.01, 0.25, 0.315, 0.4, 0.5, 0.63, 0.794, 0.83$\\
$n=-1$\ r2 &  0.00023 & 0.017 & 0.010 & 1.00 & 815 & 0.31 \% &
$0.01, 0.25, 0.315, 0.4, 0.5, 0.63, 0.794, 1.0$\\
$n= 0$\ r1 &  0.00025 & 0.003 & 0.025 & 0.66 & \llap{1}443 & 0.50 \% &
$0.025, 0.1, 0.2, 0.3, 0.4, 0.5, 0.6, 0.66$\\
$n= 0$\ r2 & 0.00025 & 0.003 & 0.025 & 0.50 & \llap{1}239 & 0.50 \% &
$0.025, 0.1, 0.2, 0.3, 0.4, 0.5$\\
\noalign{\vglue 0.2em}
\hline
\hline
\end{tabular}
\end{table*}

\section{The numerical simulations}\label{scfsimulations}

We have produced a large library of $N$-body simulations with
$N=256^3$ particles. We considered Einstein--de Sitter ($\Omega=1$)
models, and also low-density open and flat $\Lambda$ geometries. The
spectral indices that have been simulated are $n=-2,\; -1.5,\; -1$ and
$0$ and two realizations of each spectral index were carried out.  The
simulations were executed on either 128 or 64 processors of the
Edinburgh Cray T3E supercomputer, using the parallelized \pppm \
`shmem' version of HYDRA
\cite{Macfarlandetal1998,Couchmanetal1995,PearceCouchman1997}, in
purely collisionless dark matter mode.

The large-scale force calculation in HYDRA used a $512^3$ Fourier
mesh, supplemented by direct summation of close pairs to achieve the
desire total interparticle force. As usual, this is softened on small
scales in order to suppress two-body encounters.  In HYDRA, the
transition from pure Newtonian to constant force is achieved using a
`spline-kernel softening'; with this method, the interparticle forces
become precisely Newtonian after 2.34 times the softening length.  In
all cases, we adopted a comoving softening length that is simply a
fraction $f$ of the interparticle spacing
\be \epsilon= f L/N^{1/3}\ ,\ee
where $L$ is the side of the simulation box.  We used $f\simeq0.064$,
which is slightly smaller than the late-time value used by Efstathiou
et al. \shortcite{Efstathiouetal1988} and the small-box calculations
of J98 who used $f\simeq0.1$. However, it is slightly larger than the
values used by Jain \& Bertschinger \shortcite{JainBertschinger1998}
who used an effective value of $f=0.05$, and also the value chosen by
J98 for their big-box simulations, $f\simeq0.038$.  We ran a few test
simulations in which $f$ was varied, and we believe that the results
quoted here are not sensitive to the exact value adopted.

\begin{table*}
\caption{\small{Parameters of the $256^3$ particle, power-law $\Lambda$
and open simulations. Epochs include $a=$ 0.025, 0.05, 0.1, 0.2, 0.25,
0.3, 0.4, 0.5, 0.6, 0.7, 0.8, 0.9, 1.0}}
\begin{tabular}{lcccccccc} 
\hline
\hline
\noalign{\vglue 0.2em}
simulation & $\epsilon/L$ & $\Delta^2(k_{\b})$ & $\Omega$ &
 $\Lambda$ & $a_{\initial}$ & $a_{\final}$ & timesteps & energy error \\
\noalign{\vglue 0.2em}
\hline
\noalign{\vglue 0.2em}
 $n=-2$ & 0.00025 & 0.0479  & 0.26 & 0.74 & 0.025 & 
 1.0 & \llap{1}065 & 0.05 \% \\ 
 $n=-2$ & 0.00025 & 0.0479  & 0.2 & 0.0 & 0.025   &
 1.0 & 965  & 0.09 \% \\ 
 $n=-1.5$ & 0.00025 & 0.0240  & 0.26 & 0.74 & 0.025 &
 1.0 & 971  & 0.13 \%\\ 
 $n=-1.5$ & 0.00025 & 0.0240  & 0.2 & 0.0 & 0.025   &
 1.0 & 965  & 0.13 \%\\ 
 $n=-1$ & 0.00025 & 0.0101  & 0.26 & 0.74 & 0.025 &
 1.0 & \llap{1}342 & 0.28 \% \\ 
 $n=-1$ &  0.00025 & 0.0101  & 0.2  & 0.0  & 0.025 &
 1.0 & 965  & 0.87 \%\\ 
 $n= 0$ &  0.00025 & 0.0003 & 0.26 & 0.74 & 0.025 & 
 1.0 & \llap{1}020 & 0.86 \%\\ 
 $n= 0$ &  0.00025 & 0.0003 & 0.2 & 0.0 & 0.025 & 
 1.0 & 967  & 1.86 \%\\
\noalign{\vglue 0.2em}
\hline
\hline 
\end{tabular}\label{table-pow}
\end{table*}

For the initial particle load, a combination of `quiet' starts and
`glass' configurations was used. The quiet starts were produced
by simply placing particles onto a uniform grid with spacing
$L/N^{1/3}$. This method gives no contribution to the power spectrum
from particle placement except on scales of the order half a mesh
spacing (See Section \ref{scfmeasuring}). However, grid initial
conditions may lead to non-physical features on very small scales at
late times. An example of this occurs in the Warm Dark Matter
simulations of Bode, Ostriker \& Turok
\shortcite{BodeOstrikerTurok2001}, where the population of `secondary
objects' which they find to form by fragmentation of sheets and
filaments may actually be a numerical artefact induced by the grid.
An alternative approach is the glass-like distribution that is
obtained when a random distribution of particles is evolved with the
signs of the N-body accelerations reversed
\cite{White1993,Baughetal1995}.  The resultant particle distribution
displays no regular pattern, but is sub-random.  By construction, the
glass initial conditions are non-evolving in the absence of
perturbations.  The glass load was generated once, but can be used in
many different simulations by adding in the appropriate displacement
field.  This was generated from the initial density field using the
approximation of Zel'dovich \shortcite{Zel'dovich1970}.  The Fourier
modes of the density field were a Gaussian realization, with random
phases and amplitudes chosen from a Rayleigh distribution.

For both the grid and glass methods, particle discreteness on
the smallest scales leads to a spectrum that is comparable to that of
the shot-noise distribution on that scale. Numerical
evolution should proceed until the scales of interest are well above
this noise. For most spectra, memory of the initial small-scale 
discreteness is only truly lost after expansion
by roughly a factor of 10 (see Section \ref{scf-discrete}).


\subsection{Self-similar simulations} 

The normalization of the scale-free power spectra is 
most simply specified in
terms of the power on the box scale at the epoch when the expansion
factor $a$ is unity,
\be \Delta^2_{\L}(k)=\Delta^2(k_{\b})\left(\frac{k}{k_{\b}}\right)^{n+3}
\label{power}\ee
where $k_{\b}=2\pi/L$. The benefit of normalizing the spectrum in this
way is that the box-scale power is directly related to the error
induced through omitting modes with wavelength above $L$, and so
the effects can be monitored (see Section \ref{finitevolume}).

Table 1 displays all relevant simulation parameters for the scale-free
runs. A large degree of nonlinearity was achieved for all of the
simulations and the $n=-1$ and $-1.5$ calculations were completed to
the specified level of normalization. The $n=0$ calculations were
halted after the cube had expanded by roughly a factor of 25, due to
the intense demands on the cpu time from performing the PP part of the
calculation. Also, the $n=-2$ calculations were halted after a similar
factor of growth; this was due to the problems of finite volume
effects, which we discuss in detail in Section \ref{finitevolume}.


\begin{figure*}
\centering{
\epsfig{file=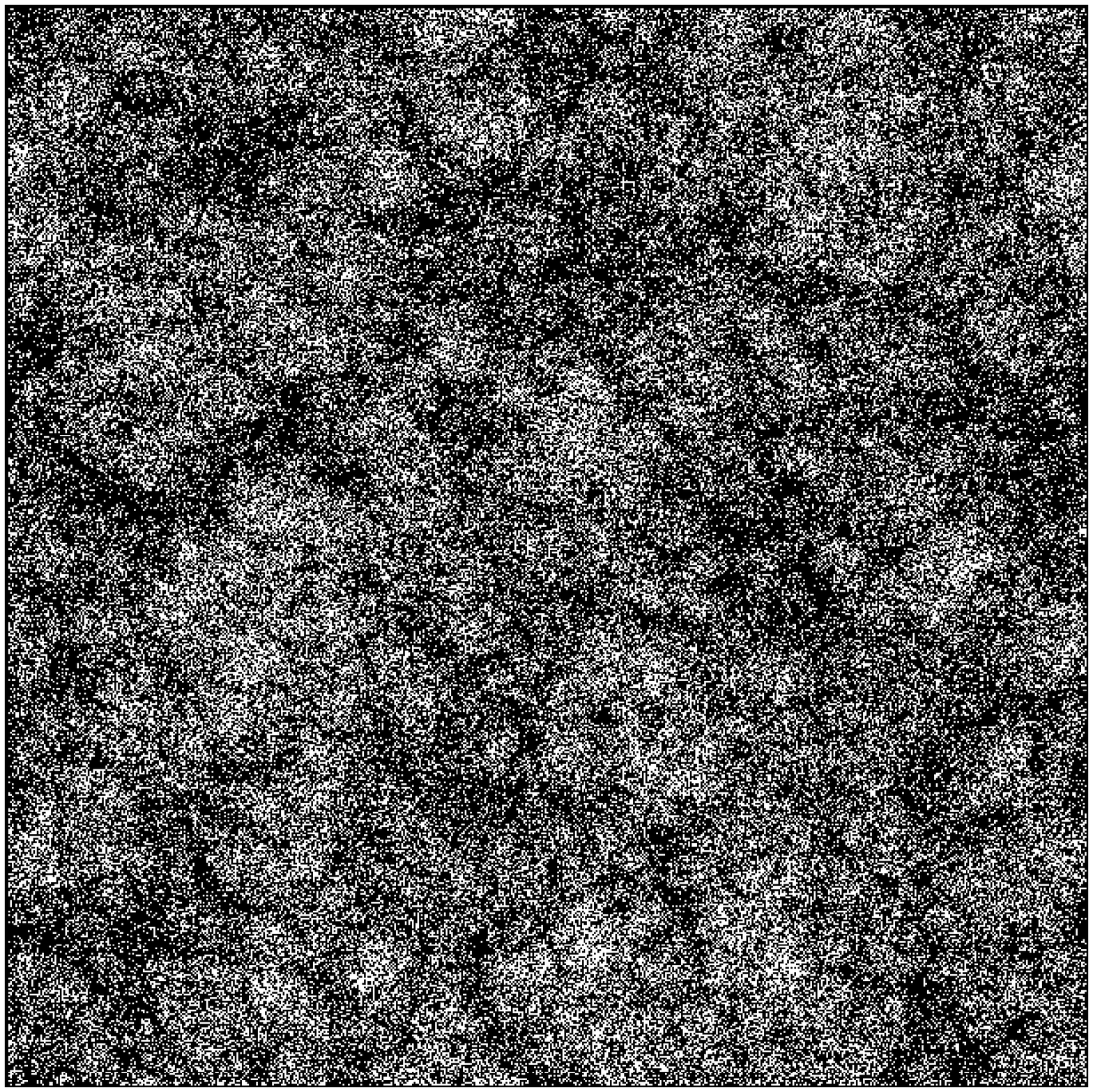,width=7.2cm,angle=0,clip=}
\epsfig{file=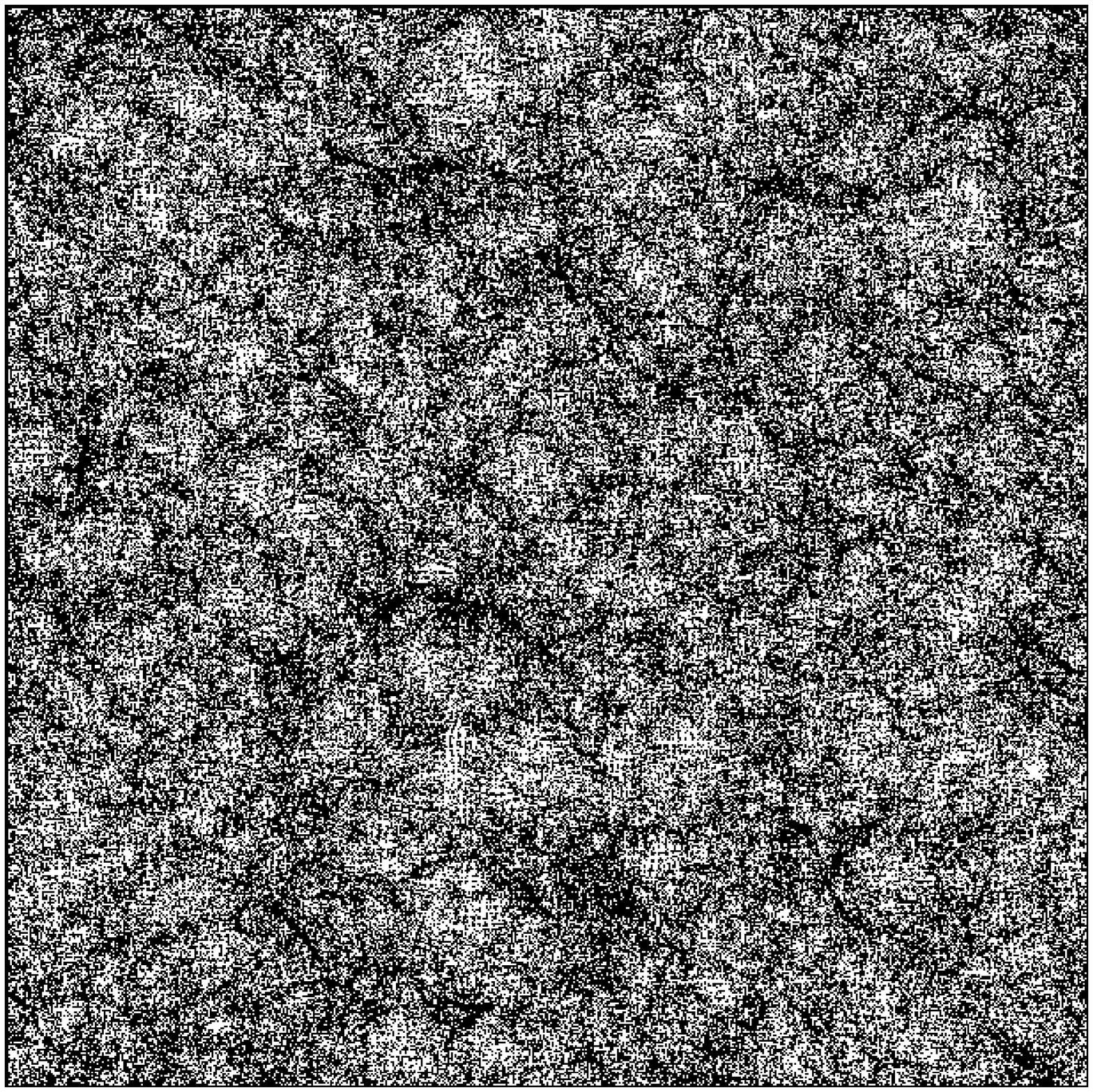,width=7.2cm,angle=0,clip=}}

\vspace{0.1cm}

\centering{
\epsfig{file=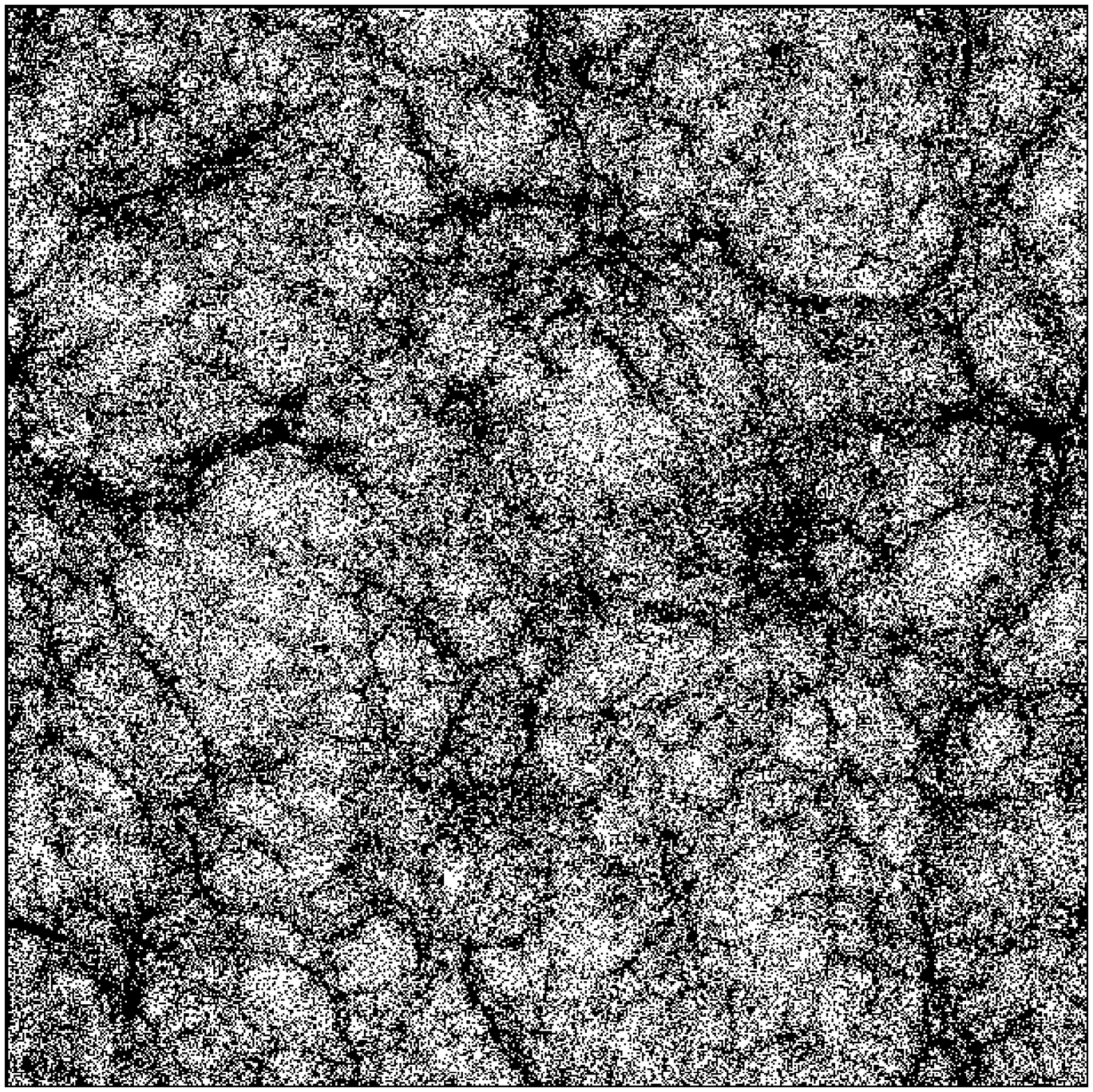,width=7.2cm,angle=0,clip=}
\epsfig{file=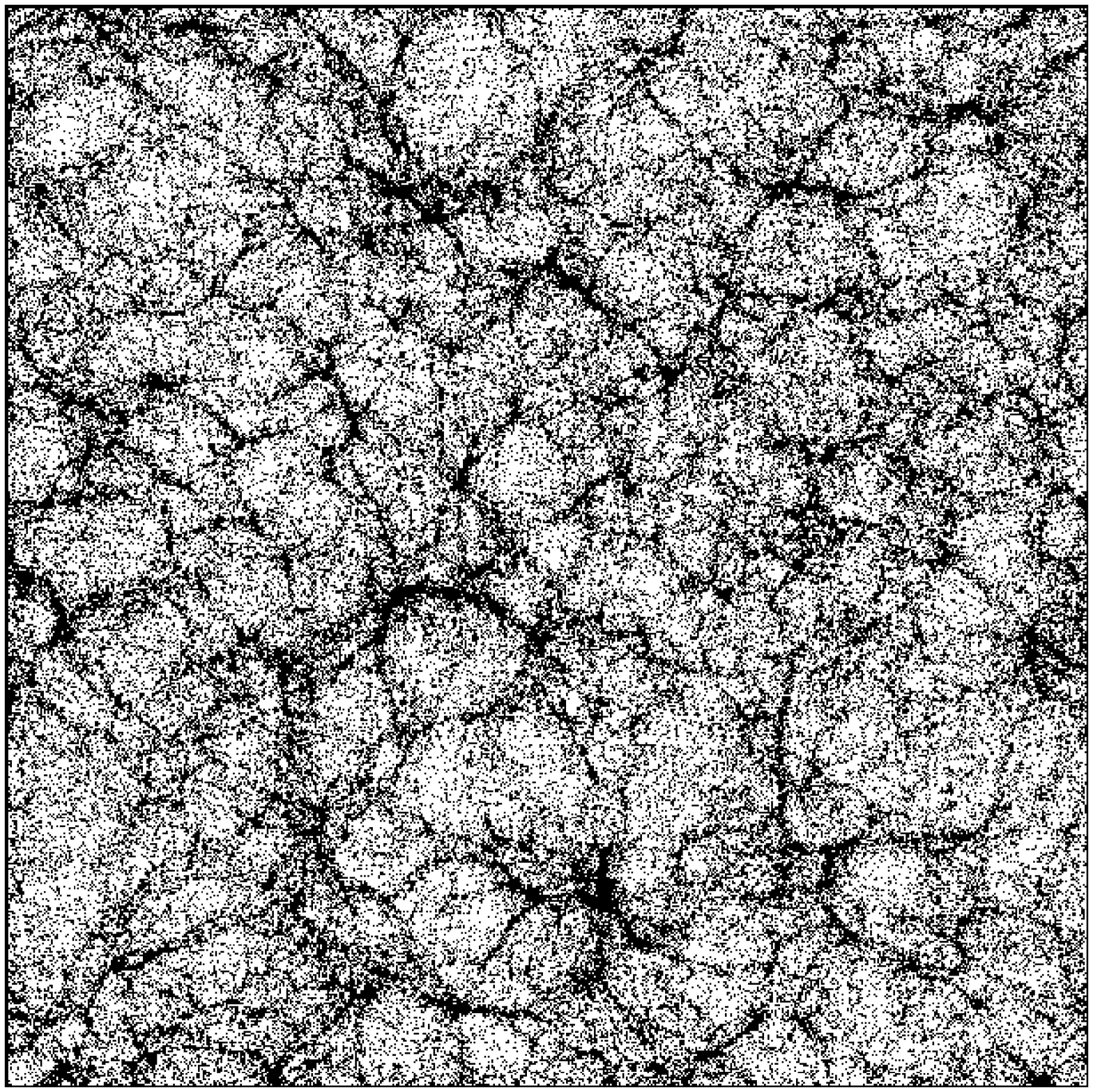,width=7.2cm,angle=0,clip=}}

\vspace{0.1cm}

\centering{
\epsfig{file=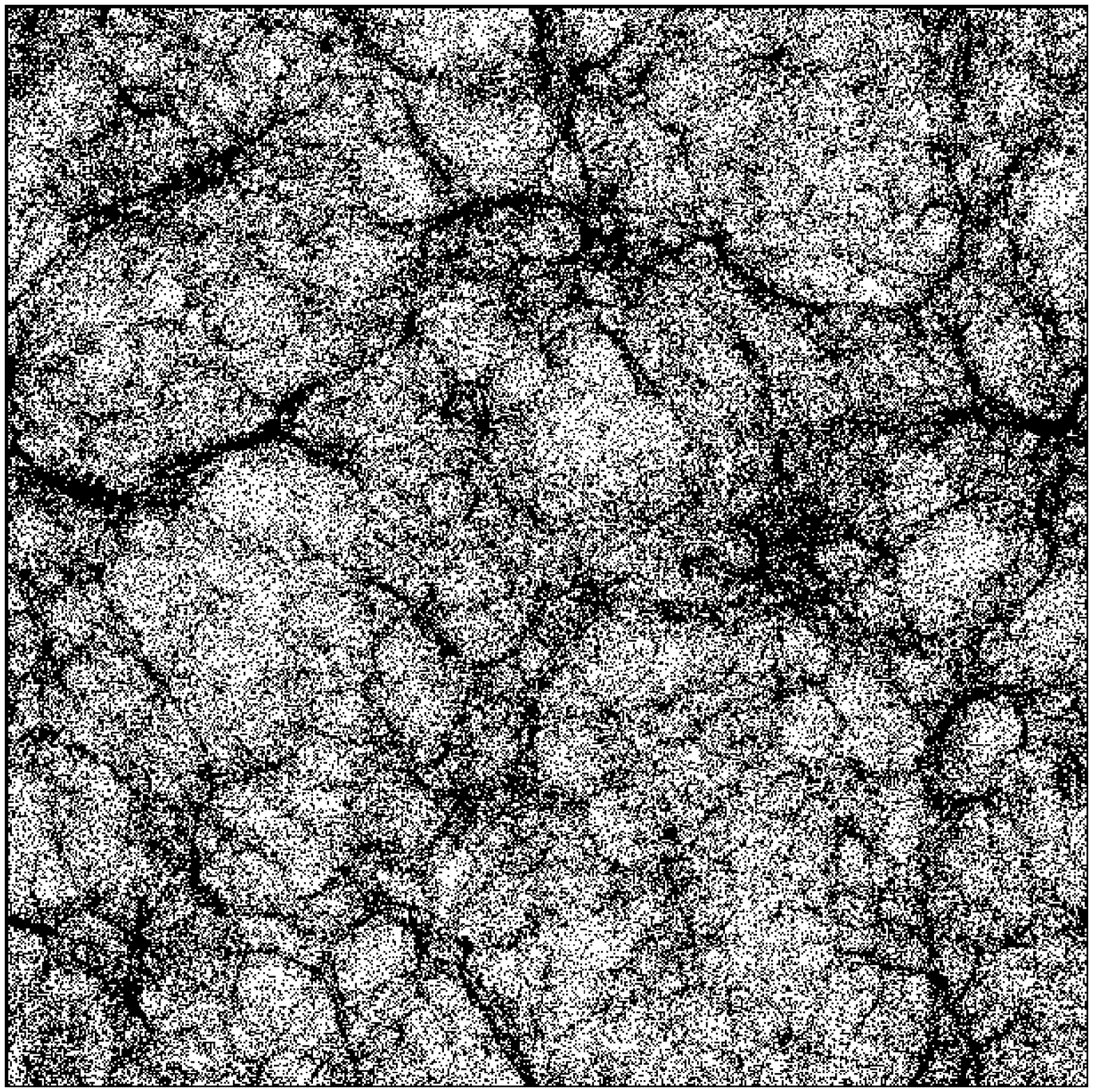,width=7.2cm,angle=0,clip=}
\epsfig{file=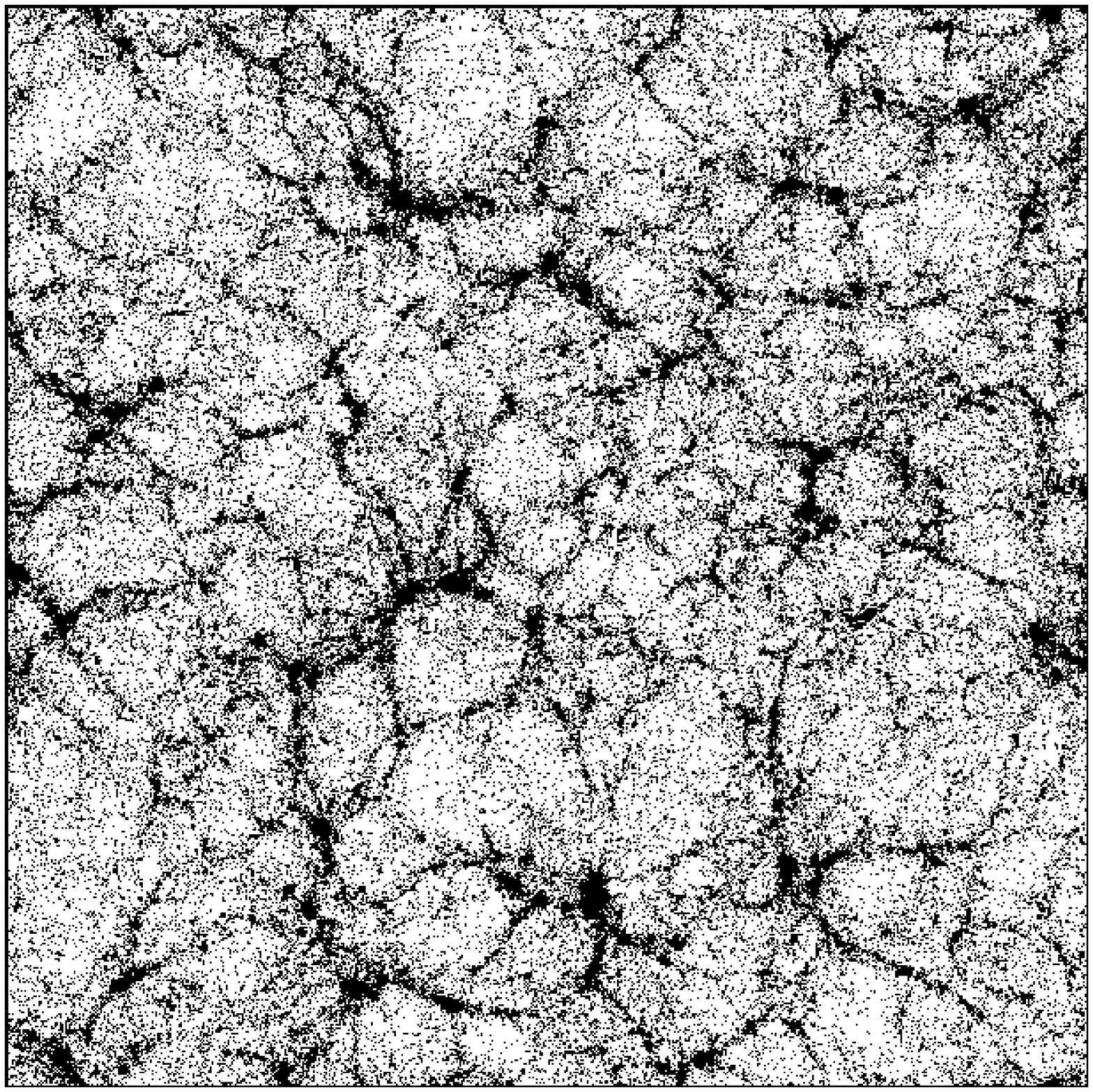,width=7.2cm,angle=0,clip=}}

\caption{\small{Slices showing the growth of structure in the glass
$n=-2$ simulation (left column) and `grid-start' $n=-1.5$ simulation
(right column).  All of the slices are of thickness $L/10$. From the
$n=-2$ simulation we show expansion factors $a=0.2, 0.45$ and $0.55$,
and from the $n=-1.5$ simulation we show epochs $a=0.25,0.63$ and
$1.0$. The normalization of the final states in the $n=-2$ and $n=-1.5$
runs were $\Delta^2(2\pi/L,a=1.0)= 0.133$ and $0.046$, respectively.}
\label{slice1}}
\end{figure*}


\begin{figure*}

\centering{
\epsfig{file=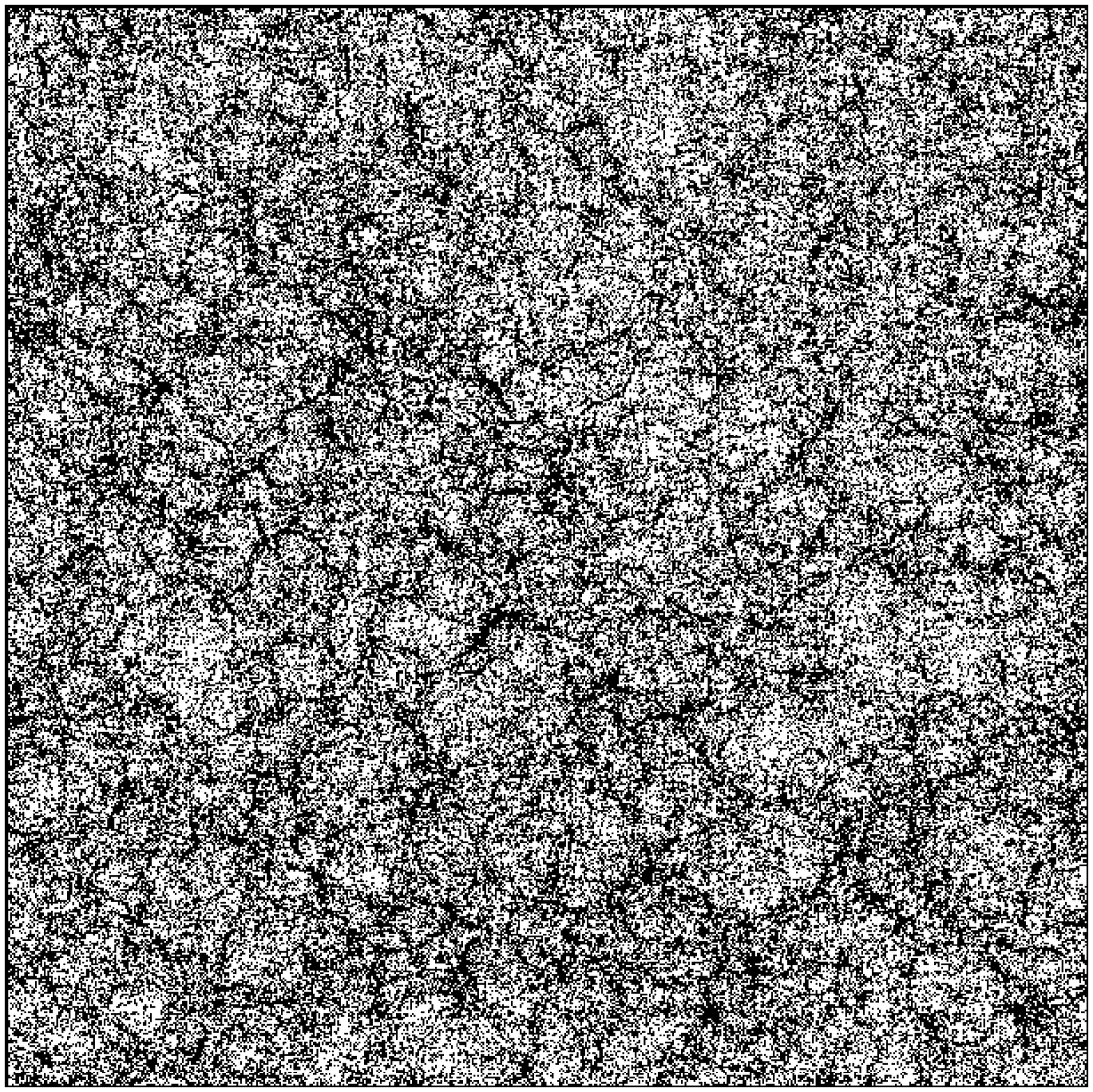,width=7.2cm,angle=0,clip=}
\epsfig{file=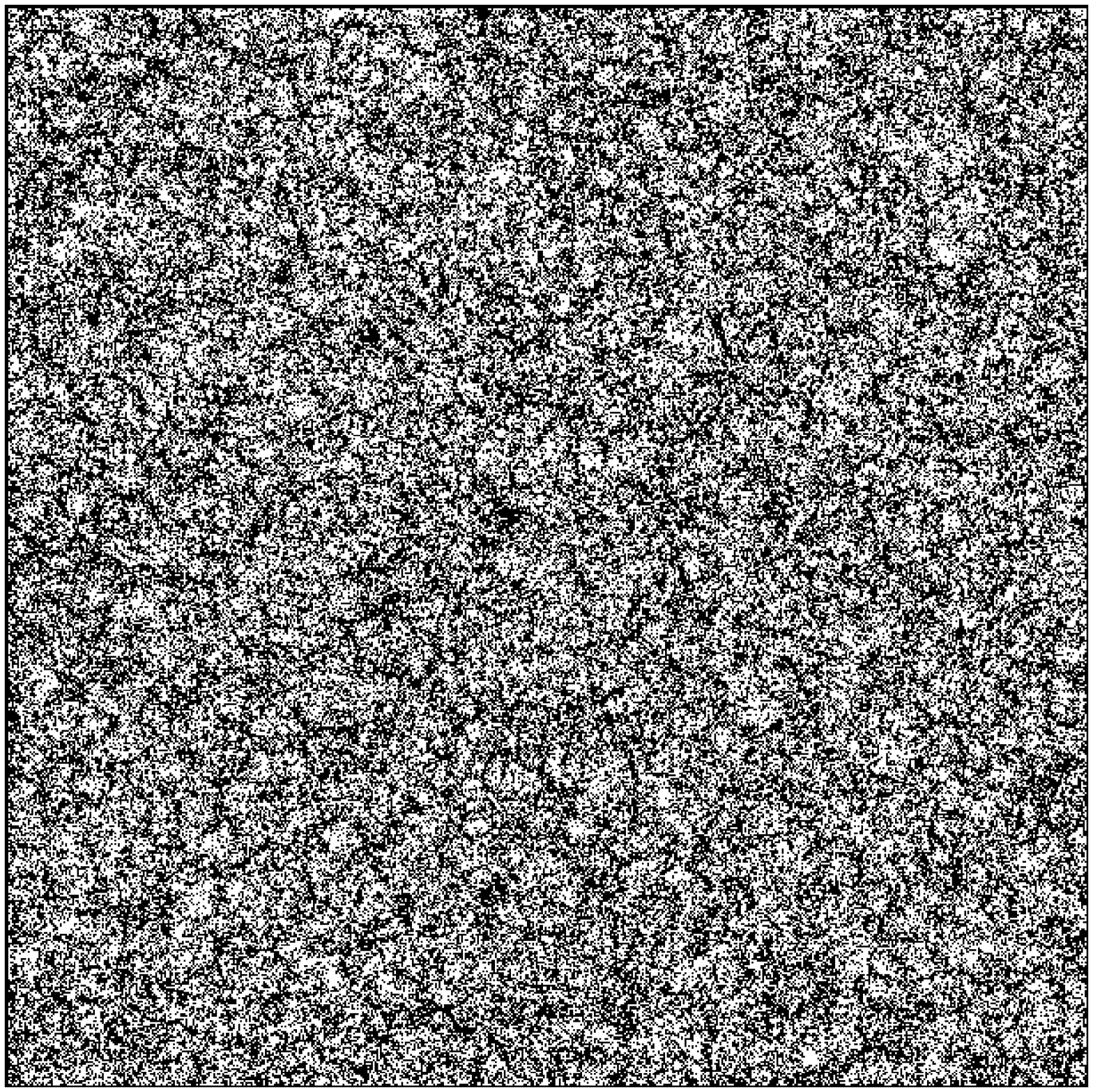,width=7.2cm,angle=0,clip=}}

\vspace{0.1cm}

\centering{
\epsfig{file=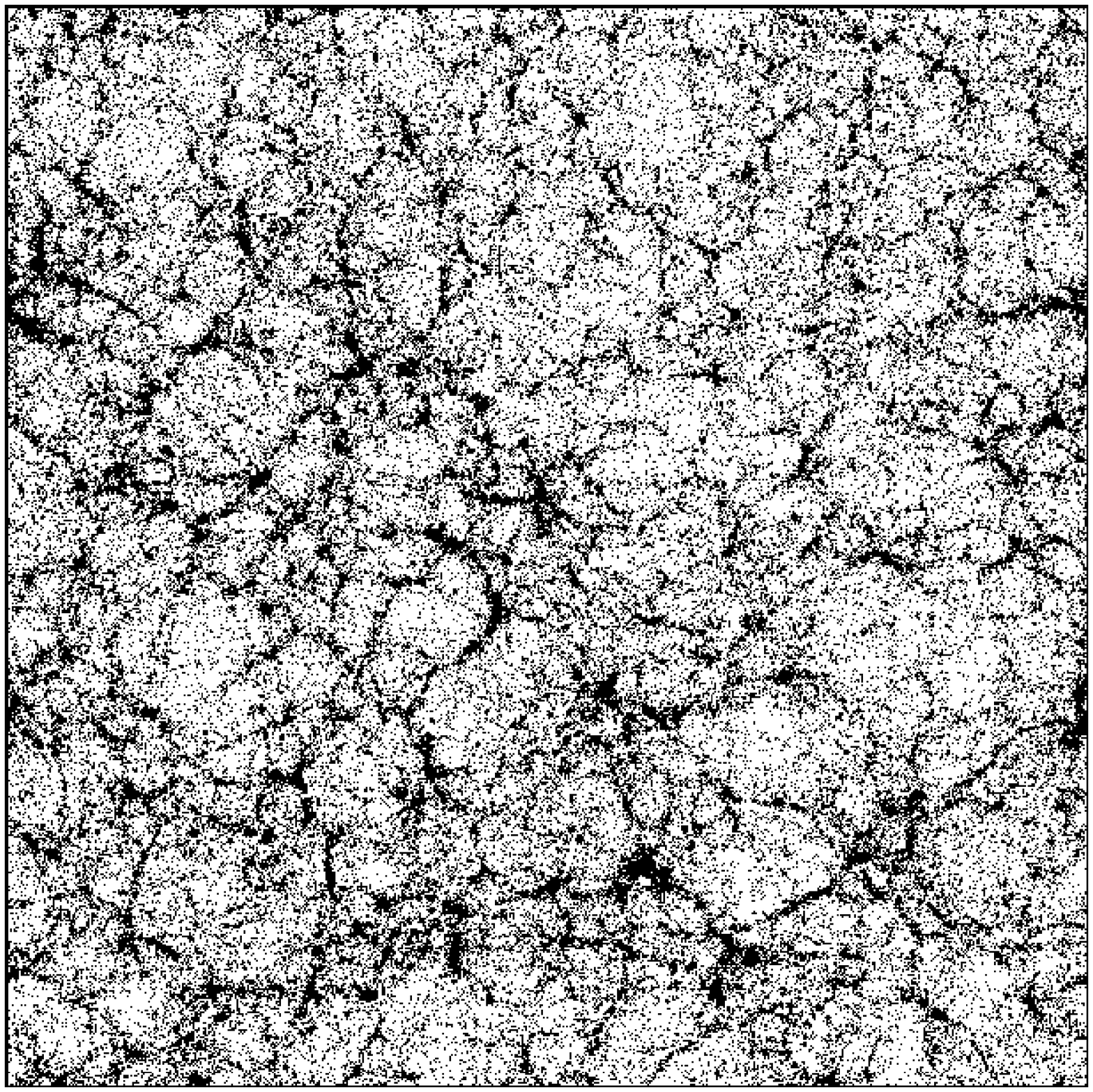,width=7.2cm,angle=0,clip=}
\epsfig{file=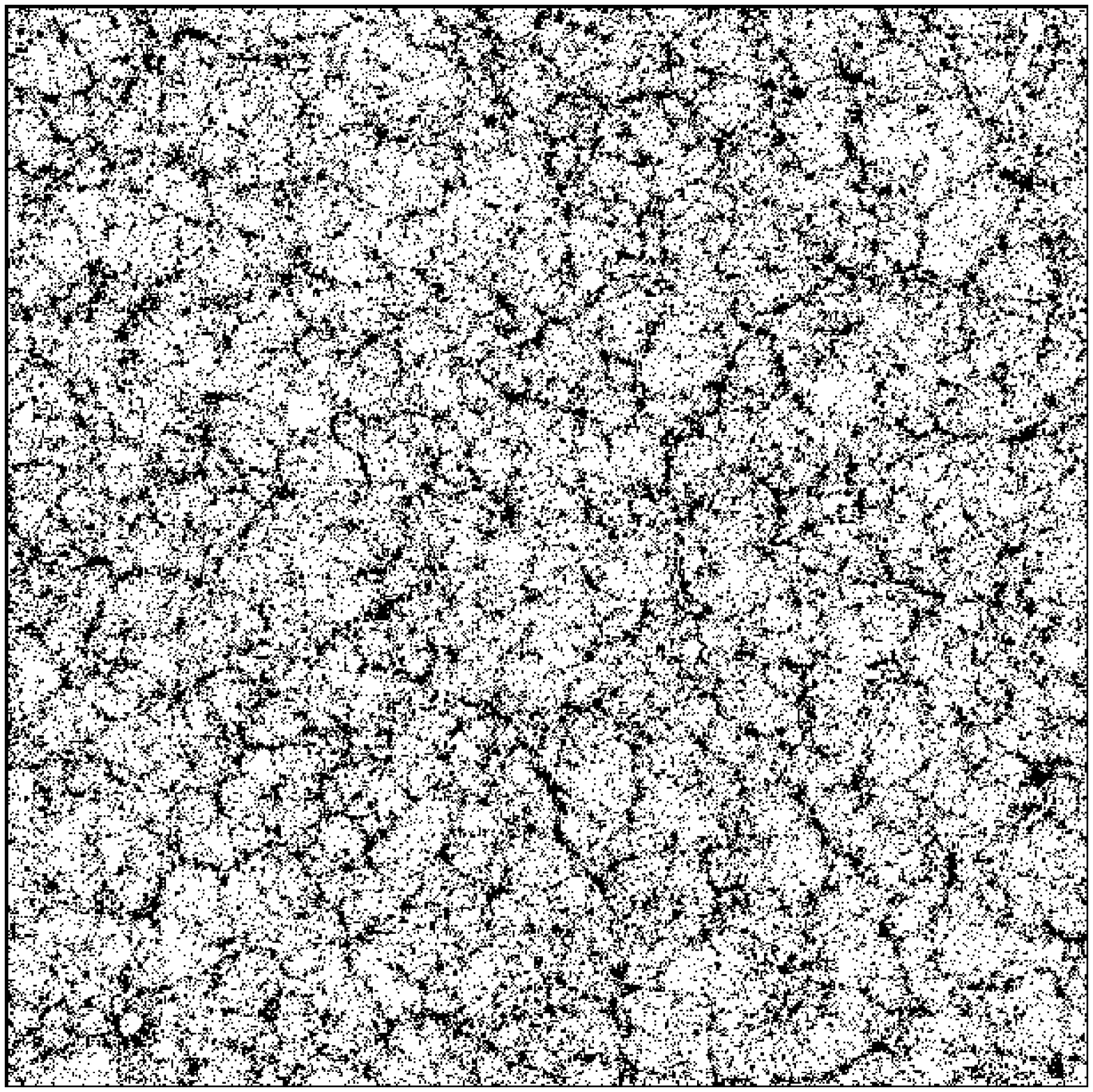,width=7.2cm,angle=0,clip=}}

\vspace{0.1cm}

\centering{
\epsfig{file=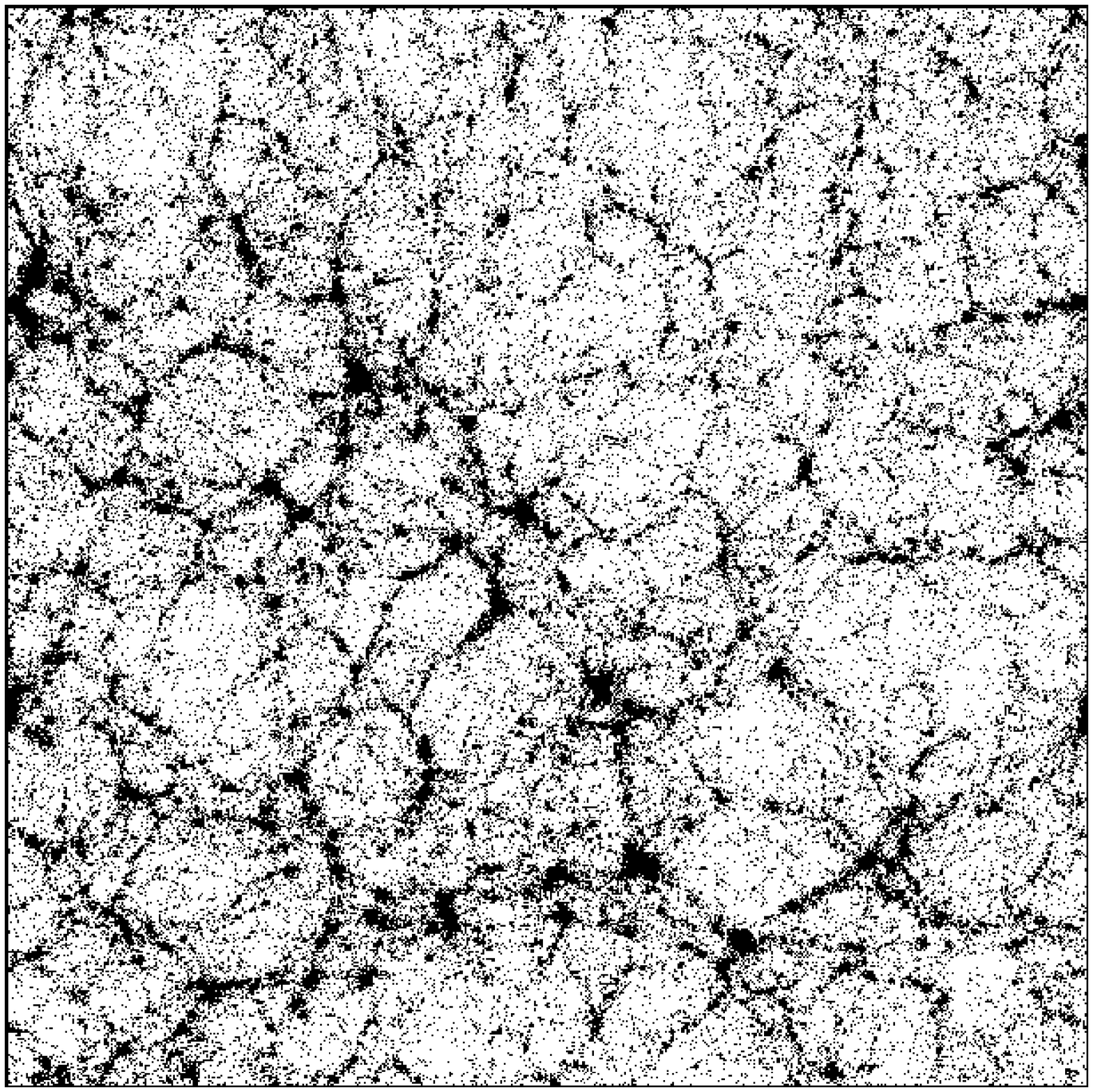,width=7.2cm,angle=0,clip=}
\epsfig{file=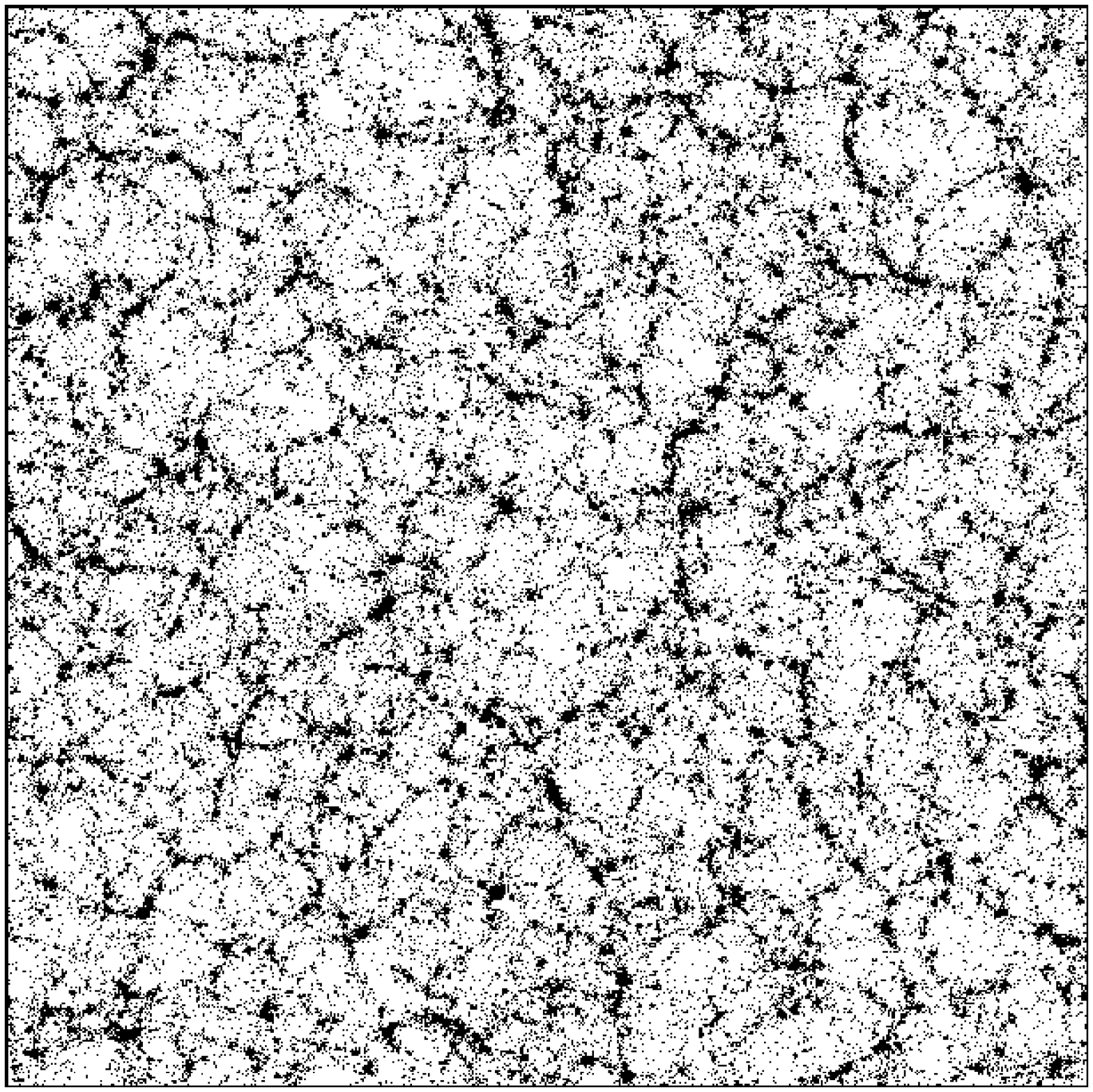,width=7.2cm,angle=0,clip=}}

\caption{\small{Same as Fig. \ref{slice1}, but this time showing the
comoving projection of particles in the glass $n=-1$ simulation
(left column) and glass $n=0$ simulation (right column). From the
$n=-1.0$ simulation we show epochs $a=0.25, 0.63$ and $0.83$, and from
the $n=0$ simulation we show expansion factors $a=0.1, 0.3$ and
$0.5$. The normalization of the final states in the $n=-1$ and $n=0$
runs were $\Delta^2(2\pi/L,a=1.0)= 0.017$ and $0.003$,
respectively.}
\label{slice2}}
\end{figure*}

Figs \ref{slice1} and \ref{slice2} provide a visual account of the
growth of structure in the four models. We show three epochs from the
four different models: the initial conditions; an intermediate epoch;
the final output epoch. The $n=-2$ simulations display a number of
large-scale fluctuations which collapse to form large filaments and
groups, whereas the $n=0$ simulations are characterized by a large
number of tightly bound objects and a paucity of large-scale
filamentary features, in accordance with the results of Efstathiou et
al. \shortcite{Efstathiouetal1988}.  Fig. \ref{slice1} also compares
the glass start to the grid starts. In the glass start no features
other than the prescribed fluctuations are observed, whereas the grid
start shows faint lattice patterns which are still
observable in the voids at the final epoch.


\subsection{Power-law open and flat simulations}

At late times the amplitude of the nonlinear power spectrum is very
sensitive to the density of the universe, and strongly modulates the
amplitude of the nonlinear clustering signal. This effect is
important to quantify if one wishes to construct general models for
evolving nonlinear power spectra.  We investigated this density
dependence by performing a further series of high resolution, $256^3$
particle, simulations for open universes where $\Omega=0.2$ at the
final epoch and for flat universes where $\Omega=0.26$ and
$\Lambda=0.74$ at the final epoch. The values for the density
parameter were selected so that each full integration would span a
large dynamic range in $\Omega$.  The amplitude of
the final box-scale mode was set slightly lower than in the $\Omega=1$
simulations, because of the greater small-scale nonlinearities that
are generated in low-density models. For all of these simulations we
have used the glass initial particle load. Table 2 displays all
of the relevant simulation parameters.


\subsection{The challenge of $n\rightarrow-3$}\label{finitevolume}

On small scales, the slope of the CDM power spectrum approaches
$n\simeq-3$, so it is important to understand how such spectra evolve
in the nonlinear regime. However, highly negative spectral indices
have proven difficult to simulate (Efstathiou et al. 1988; Jain, Mo \&
White 1995; PD96; Jain \& Bertschinger 1998), and this can be
attributed to two main effects.

First, the number of particles must be high enough to simulate
virialized clusters convincingly. Second, the finite size of the
simulation volume means that the longest wavelength fluctuations that
are present are $\lambda_{\b}=L\ ;\ k_{\b}=2\pi/L$. The absence of
modes beyond the box scale induces an error in the nonlinear
spectrum, since nonlinearity couples Fourier modes together and power
leaks from large to small scales; the importance of this effect
increases for increasingly negative spectral indices and dominates as
$n$ approaches $-3$.

The error in the power spectrum due to these missing modes can be
estimated from the linear power spectrum.  We can quantify the missing
variance as follows: 
\be \sigma^2_{\rm miss}=\int^{\infty}_{0}\Delta_{\L}^2(k)\; \frac{dk}{k} - 
\sum \frac{1}{4\pi}{\Delta_{\L}^2(k^\prime) \over(\ell^2+m^2+n^2)^{3/2}}\ , \ee
where the sum is over all integer triples $(\ell,m,n)$ except $(0,0,0)$
and the wavenumber $k^\prime=k_{\b}(\ell^2+m^2+n^2)^{1/2}$.  Strictly
speaking, both terms on the rhs
are divergent for power-law spectra with $n\ge-3$.  Nonetheless, if
one imposes a sufficiently smooth cut-off at $k_{\rm cut}$
in the power spectrum, then the difference is well defined in the
limit of $k_{\rm cut}\rightarrow\infty$.

We have estimated $\sigma^2_{\rm miss}$ numerically 
in this way for scale-free
power spectra as a function of $n$. To about 1\% accuracy
the result is given by:
\be 
\sigma^2_{\rm miss}=\frac{\Delta_{\L}^2(k_{\b})}{3+n}F(3+n),
\ee
where $F(y)=1-0.31y+0.015y^2+0.00133y^3$ and this expression is valid
for $-3\le n\le 1$. One can check the numerical result, not only by
confirming it is insensitive to the precise value of $k_{\rm cut}$,
but also for the special case $n=0$ where it is easy to see from
geometric considerations that the value of $F(3)$ is $3/4\pi$.  In the
limit of $n\rightarrow-3$ the missing variance is well approximated by
the quantity $\sigma^2_{\err}$ defined as:
\be 
\sigma^2_{\err}=\frac{\Delta_{\L}^2(k_{\b})}{3+n}\ .\label{missingmodes}
\ee
 So as to ensure that the missing variance does not become
significant for our simulations, we have
chosen to adopt the criterion
\be
 \sigma^2_{\err}\leq0.04 \ , \label{criterion}
\ee
for which the large-scale missing modes are safely linear. 

It is for these reasons that the relatively low resolution (compared
to modern standards) $32^3$ particle, $n=-2$ simulation of Efstathiou
et al. \shortcite{Efstathiouetal1988} could only reproduce the exact
similarity solution for the power spectrum over a narrow range of
expansion.  Also, for the more recent high-resolution $256^3$
particle, $n=-2$ simulation of Jain \& Bertschinger
\shortcite{JainBertschinger1998}, the box-scale power for their last
three outputs violates the condition (\ref{criterion}), rising to
$\sigma^2_{\err}\simeq 0.4$ for the last epoch.


\subsection{Simulation error and Layzer--Irvine energy}

A test for the global accuracy of the integration of the equations of
motion is to measure how well the Layzer--Irvine energy equation
\cite[equation 24.7]{Peebles1980} is obeyed \cite{Efstathiouetal1985}.
One way to characterize this is through the change in the
Layzer--Irvine integral, $I$, divided by the total potential energy $W$
(Couchman et al 1995):
\be I=K+W+\int \left[2K+W\right]\frac{da}{a}\ ,\ee
where $K$ is the total kinetic energy. In Tables \ref{table-scf} and
\ref{table-pow} we present the percentage error in each of the
simulations.  The accuracy of the integration decreases as the
spectral index steepens and as $\Omega$ decreases, the least accurate
integration being that of the open $n=0$ simulation, for which the
global error at the final epoch was of the order 1.8\%.


\section{Measuring the power spectrum}\label{scfmeasuring}

The Fourier modes of the particle distribution can be determined
exactly using the expression \cite{Peebles1980}
\be \delk=\frac{1}{N}\sum_{i=1}^{N}e^{i\kdotx_i}\ .\label{modes}\ee
Owing to the periodic
boundary conditions, wavenumbers are restricted to be integer
multiples of the fundamental mode, with an upper limit imposed by the
finite sampling of the mesh: the Nyquist frequency,
\be k_{\Nq}=\frac{\pi}{\Delta x} \ ,\ee
where $\Delta x=L/N_{\rm m}$ is the mesh spacing and $N_{\rm m}$ is
the dimension of the mesh. The power spectrum can then be estimated
through averaging over all of the modes in a thin shell in $k$ space:
\be
\hat P(k)\equiv \left<\left|\delta_k\right|^2\right>=\frac{1}{m}\sum_{i=1}^{m}
\left|\delta_{k_{i}}\right|^2. \label{discpower}\ee
where $m$ is the number of modes to be averaged. This method is
computationally inefficient, with the required cpu time scaling as
$MN$ for $M$ modes. A faster
method is to distribute particles onto a cubical mesh and perform a
Fast Fourier Transform (FFT). However, 
the assignment of mass to grid cells introduces some
systematic effects which must be corrected;
these issues will be discussed in detail in Section
\ref{scf-discrete}.

In the case of a 3D particle distribution, the task soon becomes
memory limited, a $512^3$ FFT requiring roughly 0.5~Gbyte
for a `real-real' transform and 1~Gbyte for a `complex-complex'
transform.  A solution to this problem was proposed by J98;
we now detail this method, since it is critical for the present
paper.


\subsection{Chaining the power}\label{chainpow}

Consider a 1D discrete density field $\delta(x)$, which is periodic
over a length scale $L$ and which has a discrete Fourier transform
given by equation (\ref{modes}). If we partition the density field
using a coarse mesh with $M$ grid cells, then the density at the
point $x$ can be described by the relation
\be \delta(x)=\delta\left(x^{\prime}+jL/M\right)\ ,\ee
where $x^{\prime}$ is the position of the particle in its grid cell
and $j$ labels the cell. If we now map all of the grid cells into one
cell, then the reduced density field, which is now periodic on the
scale $L/M$, is
\be \delta^{r}(x^{\prime})= \sum_{j=0}^{M-1}\delta(x^{\prime}+jL/M)\ .\ee
The discrete Fourier transform of this reduced density field is then,
\be \delta_k^r=\frac{1}{N}\sum_{i=1}^{N}\exp(ikx_i^{\prime})
= \frac{1}{N}\sum_{i=1}^{N}\exp[ik(x_i-jL/M)]\ .\ee
Provided that the $k$-modes are integer multiples of the new fundamental
mode, $k=\ell 2\pi/(L/M)$, then the last term in the exponential is a multiple of $2\pi$,
so the modes of the reduced field are equivalent to the modes of the
true field. There is, however, 
a reduction in the number of available modes,
since the smaller volume of the coarse mesh gives a lower density of states.


\subsection{Numerical effects on the power}\label{scf-discrete}

There are three important numerical effects which can modify the
`observed' power spectrum from the true nonlinear signal:
discreteness effects, charge assignment and force softening.


\subsubsection{Discreteness effects}

For a random distribution of particles with no
imposed clustering, the power does not
vanish. This result can be deduced by splitting 3D space into a large
number of cubical cells, so that the occupation number of each cell is
either $n_i=0$ or 1 (Peebles 1980). On computing the expectation of the power
spectrum, we obtain the shot-noise spectrum
\be \left<\left|\delta_k\right|^2\right>=\frac{1}{N}\ \ee
which in dimensionless form is written 
\be \Delta^2_{\shot}=\frac{4\pi}{N}\left[\frac{k}{k_{\b}}\right]^3
.\label{shotspec} \ee
This leads us to write the true power spectrum, in the
limit of large $N$ \cite{PeacockNicholson1991}
\be \Delta^2_{\rm true}(k)=\Delta^2_{\rm obs}(k)-\Delta^2_{\rm shot} ,
\label{power-corr1}\ee
where $\Delta^2_{\rm obs}$ is the observed power from equation (\ref{discpower}).

\begin{figure}
\centerline{\epsfig{file=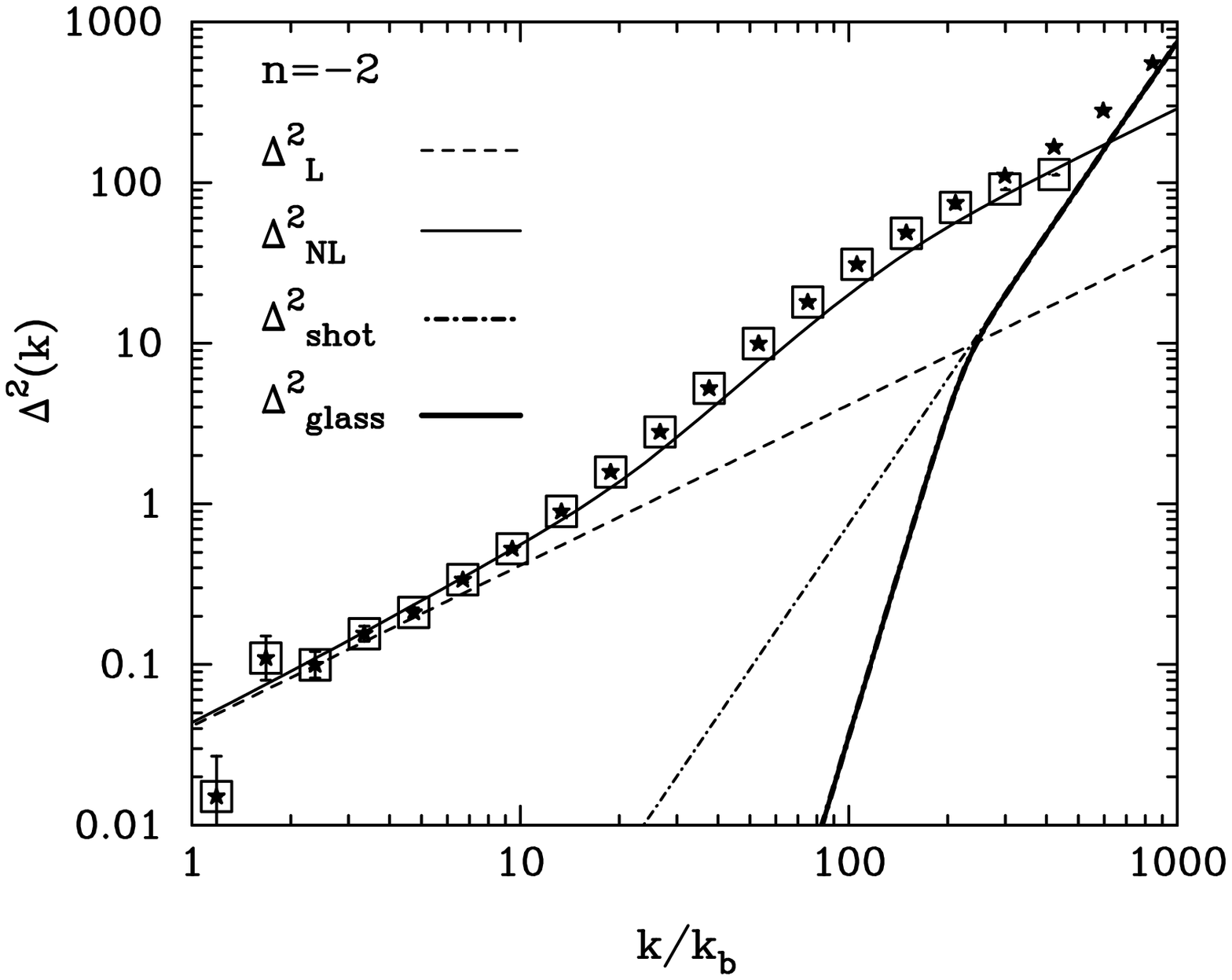,width=8.5cm,angle=0,clip=}}
\centerline{\epsfig{file=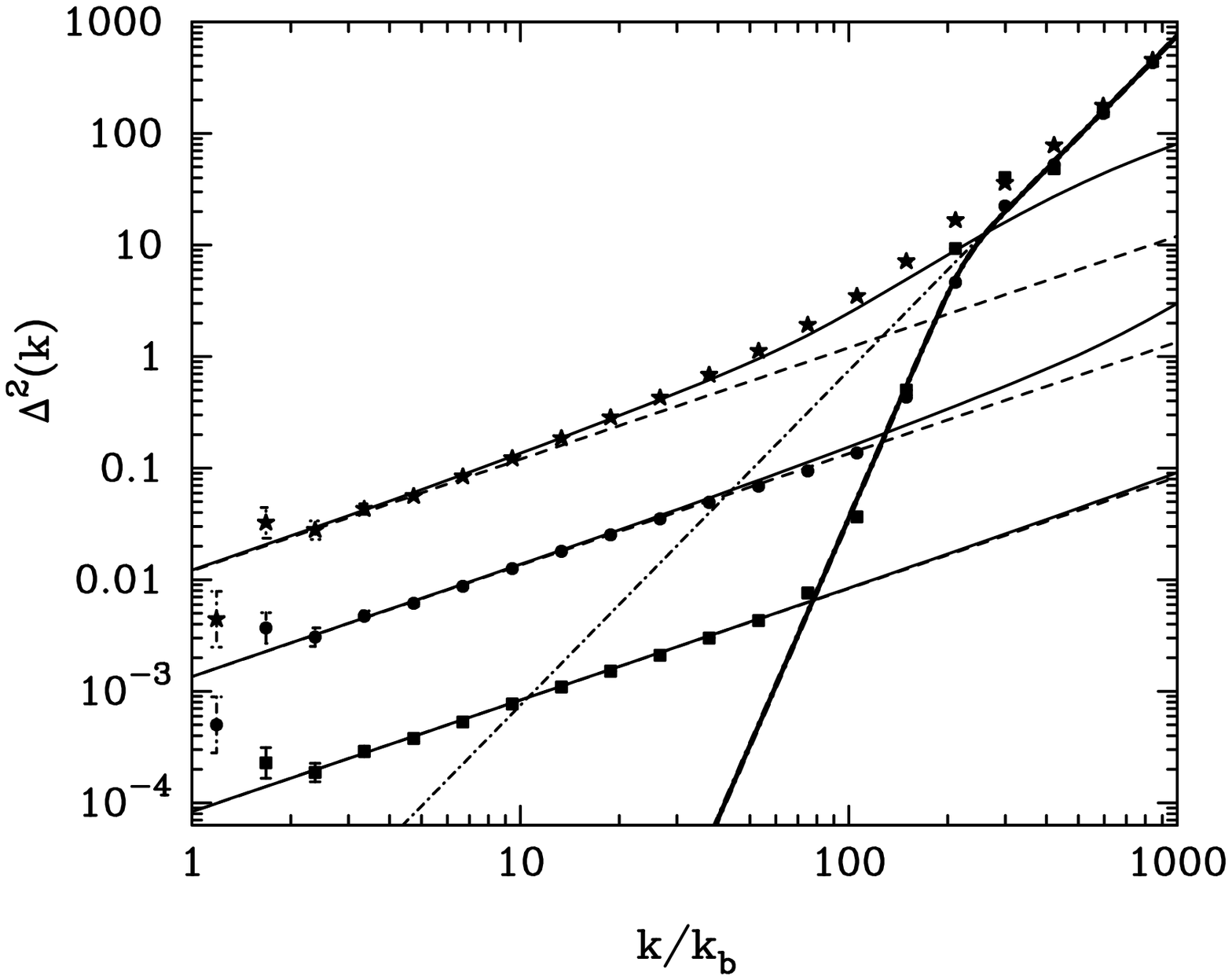,width=8.5cm,angle=0,clip=}}
\caption{\small{(Top) The glass-discreteness corrected (squares) and
uncorrected power spectrum (stars) of the glass $n=-2$ simulation at
an epoch $a=0.55$. We show the linear fluctuation spectrum (dashed
line), which demonstrates that the box scale mode is still linear, the
nonlinear spectrum according to the scaling formula of PD96 (thin
solid line), a shot noise spectrum (dot-dashed line) and our two
power-law discreteness model outlined in the text (thick solid
line). (Bottom) Three epochs from the early stages of the same $n=-2$
simulation. From bottom to top epochs are $a=0.025$ (squares), 0.1
(circles) and 0.3 (stars). This demonstrates that the discreteness
spectrum does not evolve and also that the linear spectrum has been
correctly established early on. Again, the lines are as in the top
panel, with the thick solid line representing our fit to
the discreteness spectrum.}
\label{glass}}
\end{figure}

However, this correction is invalid for the glass and grid starts
discussed in section \ref{scfsimulations}. To determine the
appropriate correction for these schemes we directly computed the
power spectrum of the initial conditions and then used these empirical
spectra to construct a simple correction model.  In Fig. \ref{glass}
(bottom) we show the raw power spectrum of the glass particle
load for the initial conditions and two subsequent epochs from the
$n=-2$ simulation. The glass power spectrum is characterized by a
two-power-law spectrum: on intermediate scales the spectrum is steep,
roughly the $n = 4$ `minimal slope' (see section 28 of Peebles 1980)
and at smaller scales this breaks to a shot noise
spectrum. Furthermore, the bottom panel of
Fig. \ref{glass} shows that the discreteness spectrum does not 
appear to evolve; we can therefore use the initial conditions to
determine a discreteness correction that can be applied to
correct the observed power at all subsequent epochs.  
This correction can be modelled as a transition between
shot noise on small scales and the almost minimal spectrum on intermediate scales:
\be \Delta^2_{\glass}=\left[\left(\Delta^2_{\shot}\right)^{-1/\alpha}
+\left(\Delta^2_{\min}\right)^{-1/\alpha}\right]^{-\alpha}\
,\label{glasscorr}\ee
where $\alpha=0.3$ and $\Delta_{\min}^2= (A\,k/k_{\b})^{\beta}$,
with best fitting values $A=0.0062$ and $\beta=6.8$.

For the grid, or `quiet' start, the issue of a discreteness correction
is fairly subtle, since there is initially no power added to the
distribution by particle placement except on the scales of the Nyquist
frequency of the mesh. However, as the simulation evolves under
gravity, the sparseness of particles on small scales forms a power
spectrum similar to a shot noise term on those scales. At late times
this can be remedied by subtracting the Poisson spectrum from the raw
power, since the large- and intermediate-scale modes in the evolved
distribution are of higher amplitude than the shot-noise spectrum. At
early times, when the true power is of relatively low amplitude, this
approach is incorrect.  We avoid the problem by excluding points whose
amplitudes are below the Poisson spectrum on the equivalent scale.

\begin{figure}
\centerline{\epsfig{file=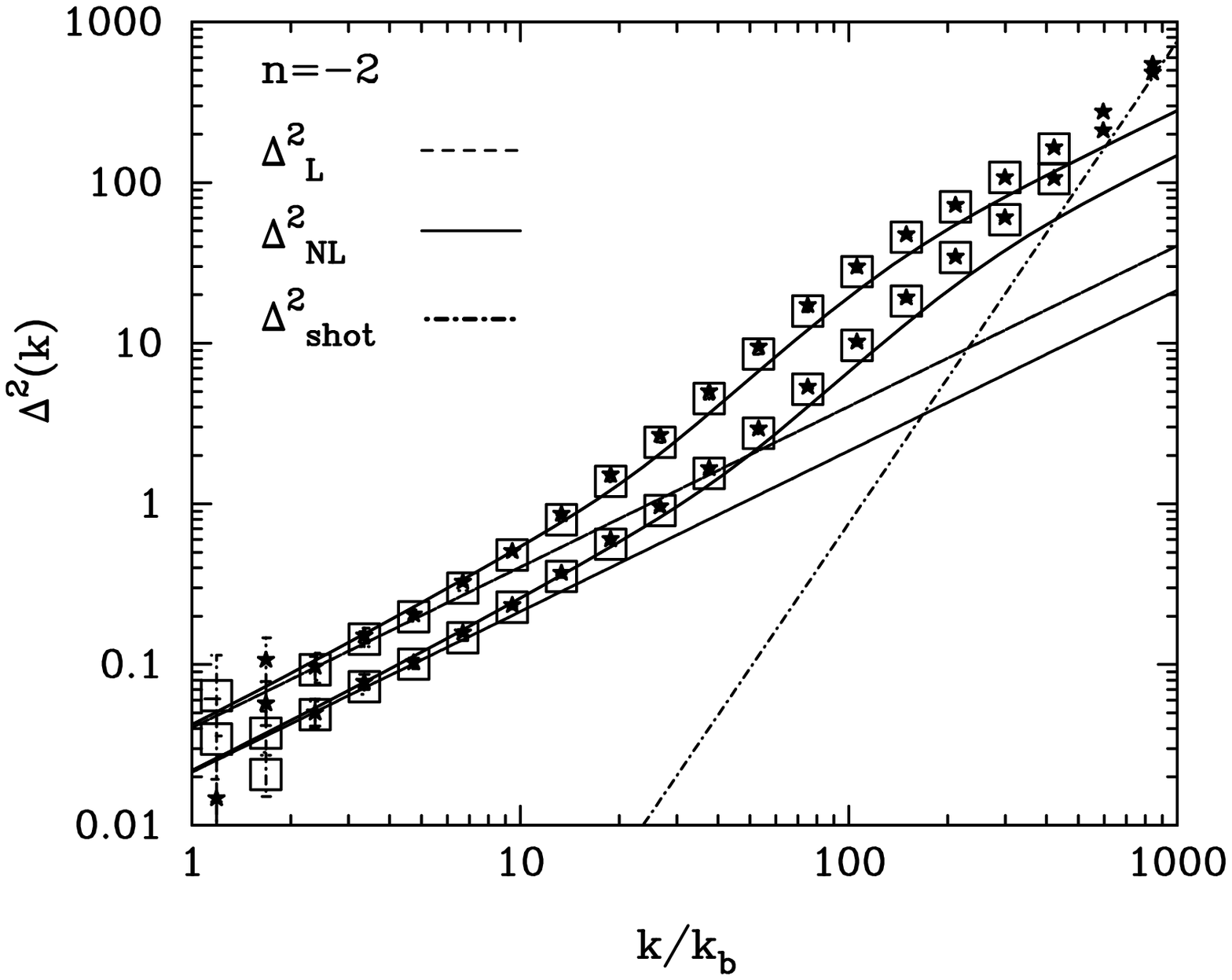,width=8.5cm,angle=0,clip=}}
\centerline{\epsfig{file=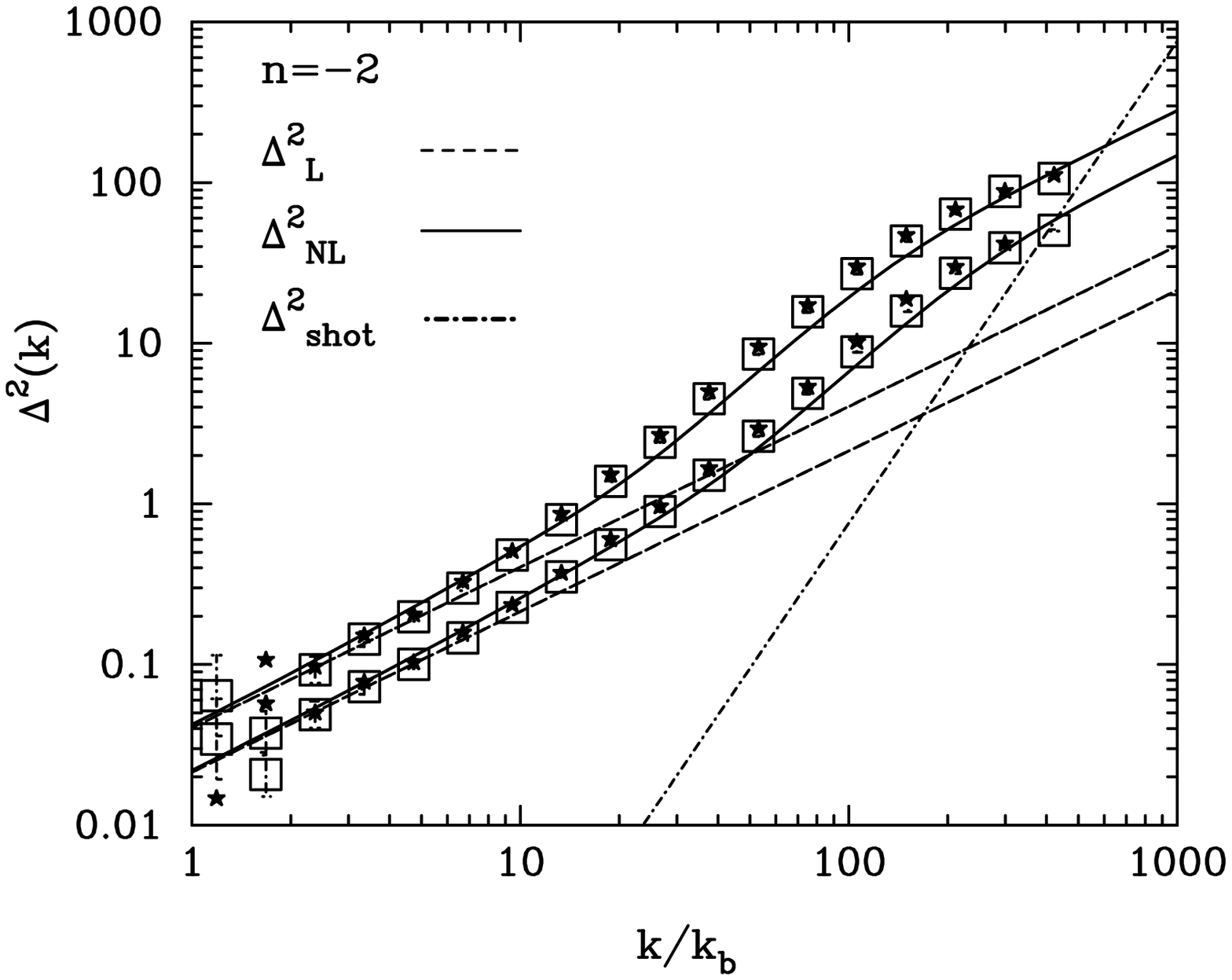,width=8.5cm,angle=0,clip=}}
\caption{\small{(Top) Comparison of discreteness uncorrected power
spectra measured from the quiet start (squares) and glass start
(stars) $n=-2$ simulations at epochs $a=0.4$ and 0.55. Line styles are
as in Fig. \ref{glass}. (Bottom) Comparison of discreteness
corrected power spectra for the same outputs from the two
simulations.}
\label{glassgrid}}
\end{figure}

Fig. \ref{glassgrid} (top) compares the uncorrected power spectra
for the glass and grid starts measured from the $n=-2$
simulations at two epochs $a=0.4$ and $a=0.55$. We observe that the
nonlinear loci defined by the data for these two simulations are
consistent and show no memory of the initial particle load. 
The only noticeable discrepancy between the two simulations is the
difference in large-scale power; this arises because
the simulations are independent realizations.
Fig. \ref{glassgrid} (bottom) contrasts the discreteness-corrected 
spectra; this shows that consistent final results are obtained
through simulating with grid or glass initial conditions.

\subsubsection{Mass assignment}

The assignment of mass onto the FFT mesh produces a finite sampling
error of the true density field. This problem was investigated for
power spectra by  Baugh \& Efstathiou \shortcite{BaughEfstathiou1994}, who proposed that
equation (\ref{power-corr1}) for the true field should be modified to
\be \Delta^2_{\true}(k)=\frac{\Delta^2_{\obs}(k)-\Delta^2_{\disc}(k)}
{w(y)}\ ; \ \ \ \ y=k/k_{\rm m}\ ,\ee
where $w(y)$ is the Fourier transform of the mass assignment window
function, $\Delta^2_{\disc}(k)$ is the appropriate discreteness
correction and $k_{\rm m}=2\pi/\Delta x$ is the wave number associated
with the inter-mesh spacing $\Delta x$. However, we believe that
there is a small flaw in their method. Any discreteness correction
should be made subsequent to the correction due to mass
assignment, since the discreteness correction accounts for the
representation of a continuous field with a point-like distribution.
We therefore implement the correction as
\be \Delta^2_{\true}(k)=\frac{\Delta^2_{\obs}(k)}
{w(y)}-\Delta^2_{\disc}(k)\ ; \ \ \ \ y=k/k_{\rm m}, \ .\ee

Several schemes exist for transferring mass onto the Fourier
mesh. The simplest scheme is nearest grid point (NGP), which assigns
all of the mass to the closest mesh point. 
%
%
More sophisticated methods such as cloud-in-cell (CIC) and
triangular-shaped-cloud (TSC) attempt to smear the mass across a
number of mesh points. We have adopted the TSC scheme to assign
particles to the mesh. 
However, the detailed correction is unimportant when using the
chained-power method of J98. Results at high $k$ can be obtained
either by making substantial binning corrections to the main FFT mesh,
or by moving to a sub-mesh of higher resolution. In practice,
we make this transition before the corrections from binning 
become significant.
Finally, Baugh \& Efstathiou showed that, even after correcting for the window function, 
the power is affected by aliasing close to the Nyquist frequency.
Again, when following the method described in Section \ref{chainpow},
aliasing errors can be avoided by only using modes that are a safe
distance from the Nyquist frequency of a given (sub)mesh
(a factor of 2, in practice).


\subsubsection{Force softening}\label{scfsoft}

The softening of the Newtonian force in the PP part of the $N$-body
calculation (described in Section \ref{scfsimulations}) induces an
error in the integration of particle trajectories for close pairs. By
considering the fractional error in the softened force from the true
Newtonian force, we can impose some constraints on the small-scale
cutoff, below which numerical effects dominate the clustering in our
simulations. For our spline-kernel force softening, we expect
numerical effects to suppress the true power on scales of a few times
the softening length. This corresponds to $k/k_{\rm b} \sim 1700$.

The simplest way to discriminate between the true nonlinear solution
and numerical artefacts is to use the self-similar evolution of the
scale-free simulations. Since the numerical features are of fixed
comoving length, the true density field will scale under the
transformations that were described in Section
(\ref{scfnonlinearscale}), whereas the numerical effects do not. We
provide evidence for this in section \ref{scfresults}.


\section{Numerical results}\label{scfresults}

\subsection{Similarity solution}

\begin{figure}
\centerline{\epsfig{file=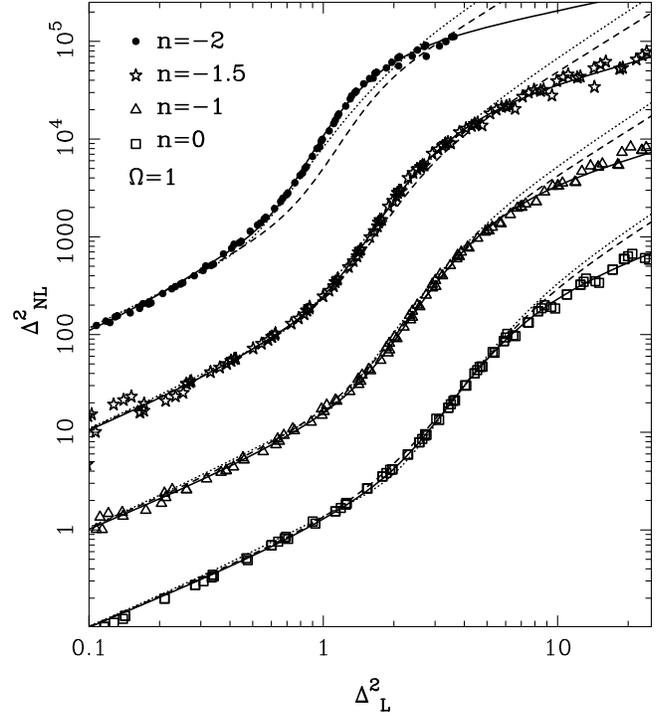,width=8.5cm,angle=0,clip=}}
\caption{\small{Nonlinear power plotted against linear power (points)
for the four scale-free simulations. For clarity, the data have been
separated from each other by one order of magnitude in the
$y$-direction, with the $n=0$ data untranslated. To determine the
linear power given a nonlinear data point, the appropriate linear
scale is required. In the HKLM method, this is found using the
transformation $k_{\L}=[1+\Delta^2_{\NL}(k_{\NL})]^{-1/3}k_{\NL}$. The
solid line represents the fitting formula for the Einstein--de Sitter
models presented in Appendix \ref{appHKLM}; the dashed line represents
the PD96 fitting formula; the dotted lines are the fits using the
formula of JMW95.}
\label{HKLMplot1}}
\end{figure}

\begin{figure}
\centerline{\epsfig{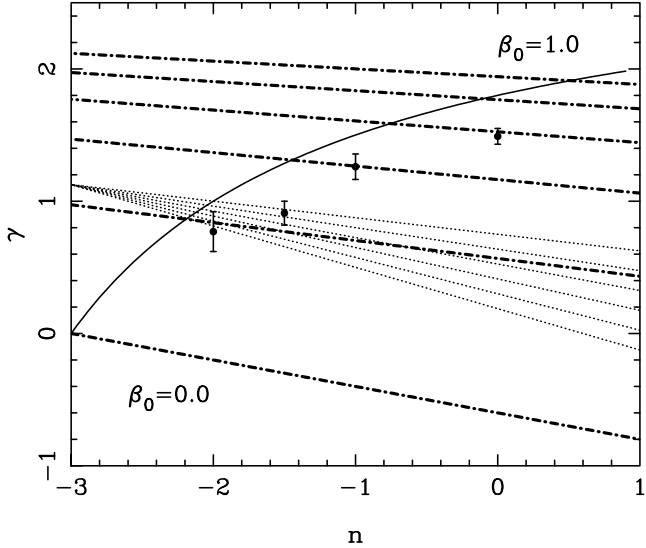}}
\caption{\small{Nonlinear slope of the power spectrum versus spectral
index. Points are measured from the scale-free simulation
power spectra. The solid line represents the stable-clustering
prediction. The dot-dash lines correspond to the predictions of the
halo model (Ma \& Fry 2000b), with an assumed $\epsilon=0.4$ and
$0.0\le\beta_0\le1.0$. The dotted lines correspond to the halo model
prediction with $\beta_0=0.25$ and $0.4\le\epsilon\le1.0$.
\label{gamma}}}
\end{figure}

Fig. \ref{HKLMplot1} shows the data for the four scale-free models in
the HKLM form: nonlinear power on the nonlinear scale plotted as a
function of the linear power on the linear scale. For clarity, the
data have been separated from each other by one order of magnitude in
the $y$-direction, with the $n=0$ data untranslated. In order to
determine the linear scale and power that correspond to a given
nonlinear data point, we use the nonlinear scaling relation
(\ref{NSR}).  Explicitly, given a nonlinear data point
$k_{i\NL},\Delta^2_{i\NL}$, its linear counterpart is
\be k_{i\L}=(1+\Delta^2_{i\NL})^{-1/3}k_{i\NL} \ ;\hspace{1cm} 
\Delta^2_{i\L}=\left(\frac{k_{i\L}}{k_0}\right)^{3+n} ,\ee 
where $k_0$ is a time-dependent normalization wavenumber defined by
$\Delta^2(k_0)\equiv 1$ and we have assumed an initial power-law power
spectrum for this example.

When plotted in this form the scaling nature of these models is
apparent. The power spectra measured from multiple epochs of the
simulations precisely overlay to define a single locus for each of the
spectral models considered. We confirm the observation of JMW95 and
PD96 that different spectral models produce different amounts of
nonlinear growth and that the more negative the spectral index the
steeper the locus in this plane. 
Fig. \ref{HKLMplot1} also shows that the $n=-2$ simulations
have produced a single, tightly defined locus. This was not observed
in previous studies (see Fig. 1 of Jain 1997, Fig. 1 of PD96  and
Fig. 7 of Jain \& Bertschinger 1998). 
This failure of scaling in earlier $n=-2$ results was probably
attributable to saturation of the box-scale mode.

The evolution in the data can be roughly broken down into three
regimes, the linear, the quasi-linear and the nonlinear. General
observations made about these regimes are:
\begin{enumerate}
\item Linear: $\Delta_{\NL}^2<1$: the `nonlinear' power for all of the models
converges to the linear power. 
\item Quasi-linear: $2<\Delta_{\Q}^2<30$: the slope of the $f_{\NL}$
curves are steep. Modelling the data in this regime with a single
power-law of the form, $\Delta_{\Q}^2\propto
\left[\Delta_{\L}^2\right]^{\alpha}$, we find for $n=-2$, $-1.5$, $-1$
and 0, that the spectral slopes are $\alpha=3.62 \pm 0.03$, $3.38 \pm
0.05$, $3.12 \pm 0.06$ and $2.96 \pm 0.1$.  This is reasonably close
to the suggestion of Padmanabhan \shortcite{Padmanabhan1996} that
$\smash{\Delta^2_{\Q} \propto [\Delta^2_{\L}]^3}$, although there is a
clear trend with $n$ that is not expected in Padmanabhan's
argument. This departure from a simple scaling relation is also
supported by the results from loop-correction perturbation theory (see
Fig. 19 of Scoccimarro \& Frieman 1996). However, it may be argued that
extended perturbation theory will fail at such large nonlinearities.
One caveat is that it has been suggested that the nonlinear scaling
relation may only truly be valid for $\bar{\xi}$
\cite{KanekarPadmanabhan2001} and not $\Delta^2$. The small scatter
observed in Fig. \ref{HKLMplot1} leads us to believe that this might
not be the case.
\item Nonlinear: $\Delta^2_{\NL}>30$: the $f_{\NL}$ curves break away
from the steep evolution which characterized the quasi-linear growth
to form loci that are much shallower. Again, we have performed a
simple power-law fit to the data of each locus. We find that for
$\Delta^2_{\NL}>50$ the $n=-2$ data have a nonlinear slope
$\alpha=1.05\pm0.09$, and for $\Delta^2_{\NL}>100$ the $n=-1.5$, $-1$
and 0 data have nonlinear asymptotes of $\alpha=0.87\pm0.04$,
$\alpha=1.08\pm0.04$ and $\alpha=0.99\pm0.04$. This result is
interesting for two reasons. Firstly, within the scatter in the
simulations there appears to be little dependence on the initial
spectrum for the nonlinear slope. Secondly, it is in clear
contradiction to stable clustering, which predicts that $\alpha=3/2$.
We note that this result agrees with the findings of Bagla, Engineer
\& Padmanabhan \shortcite{Baglaetal1998} for clustering in
2D. However, Fukushige \& Suto \shortcite{FukushigeSuto2001} found
that the stability on small scales, as measured from peculiar
velocities, was not preserved locally but did apply globally. Our results 
do not agree with this.
\end{enumerate}

The shallow slope at high $k$ may be interpreted in terms of the halo
model. Ma \& Fry (2000b) derived the following asymptotic limit for
the power spectrum:
\be \Delta^2(k)\propto k^{\gamma};\hspace{0.5cm}
\gamma=\frac{18\beta-\epsilon(n+3)}{2(3\beta+1)}\label{halo-gam} ,\ee
where $\beta\simeq0.8\beta_0$ is the power-law that governs the mass
dependence of halo concentrations: $c=r_v/r_s=(M/M_*)^{\beta_0}$;
$r_v$ and $r_s$ being the virial and characteristic radius; and
$\epsilon$ is the power-law index that governs the low-mass tail of the
mass function: $dn/dM\propto\nu^{\epsilon}; \nu\propto1/\sigma(M)$.
Realistic values for $\epsilon$ and $\beta_0$ are $0.4\le\alpha\le1.0$
and $0.0\le\beta_0\le0.5$.  This is illustrated in Fig. \ref{gamma},
which shows the nonlinear power spectral index $\gamma$ as a function
of the initial spectral index $n$. The values of $\gamma$ were
obtained from the above nonlinear scaling relations,
$\Delta^2_{\NL}\propto [\Delta^2_{\L}]^{\alpha}$, using the
relationship (PD96)
\be \Delta^2_{\NL}\propto k_{\NL}^{\gamma}; \hspace{0.5cm} 
\gamma=\frac{3\alpha(n+3)}{3+\alpha(n+3)}\ .\ee
We find $\gamma=0.77,\; 0.91,\; 1.26,\;1.49$ for spectral indices
$n=-2,\;-1.5,\;-1.0,\;0.0$.  Comparing these measured values against
the two predictions from equation (\ref{stab-gam}) and
(\ref{halo-gam}), we see that $\gamma$ increases with the steepness of
the spectrum, but that the data fall below the stable clustering
prediction. In terms of the halo model, if one assumes $\epsilon=0.4$ in
accord with Sheth \& Tormen (1999), then a strong dependence of
$\beta_0$ on $n$ is required in order to match the measured data. On
the other hand, if one adopts a value $\beta_0=0.25$ in the middle of
the current measured values, then it is impossible to match the
measured data with any value of $\epsilon$ in the plausible range 0.4 to
1.0. In summary, equation (\ref{halo-gam}) seems unable to predict the
observed trend of $\gamma(n)$ in a natural manner.  This is puzzling,
since we will show below that the general ideas of the halo model work
very well in describing our data. One possibility is that equation
(\ref{halo-gam}) is valid only on scales smaller than those probed by
current simulations.

Also, in Fig.  \ref{HKLMplot1} we contrast our data with the fitting
formula of JMW95 and PD96 (see Appendix \ref{appJMW95} and
\ref{appPD96} for these formulae).  Both models work reasonably well
in the quasi-linear regime, but with significant discrepancies. The
$n=-2$ results are poorly fit by both models, with the power being in
general underestimated; PD96 gives the poorer fit, and underestimates
the power by up to a factor 2.  The $n=-1.5$ locus is fairly well
characterized by the JMW function, but underestimated by PD96. The
$n=-1$ results are fairly well fit by both models, except around the
break between linear and quasi-linear slopes, where the functions
overestimate the power. Finally, the $n=0$ locus is slightly
overestimated at the linear to quasi-linear break by PD96 and
underestimated by JMW95.  We have produced a new HKLM fitting formula
that accurately fits the individual Einstein--de Sitter models, the
results of which are shown in Fig. \ref{HKLMplot1} as the thin solid
line. The formula is described in Appendix \ref{appHKLM}.

\begin{figure}
\centerline{\epsfig{file=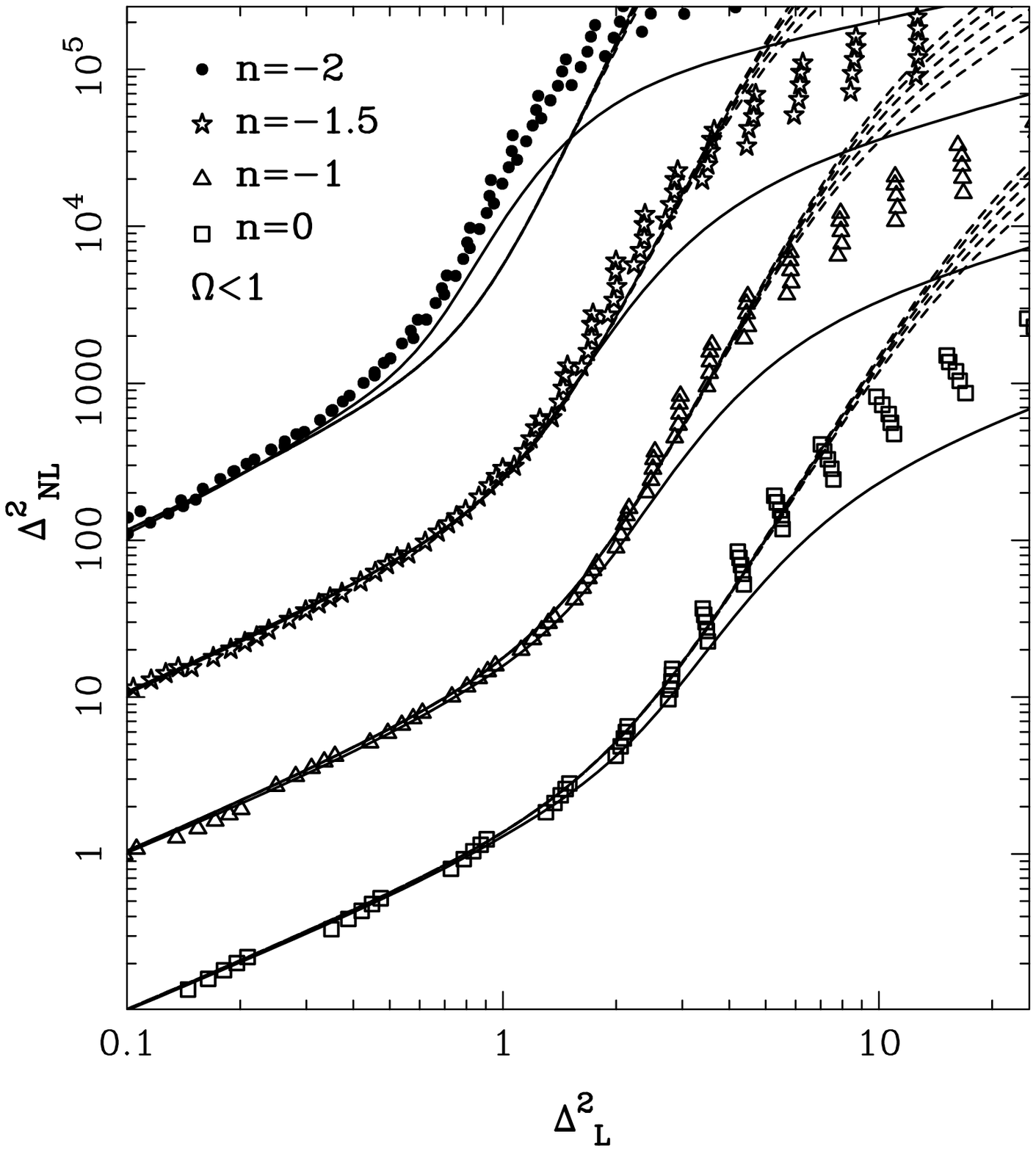,width=8.5cm,angle=0,clip=}}
\centerline{\epsfig{file=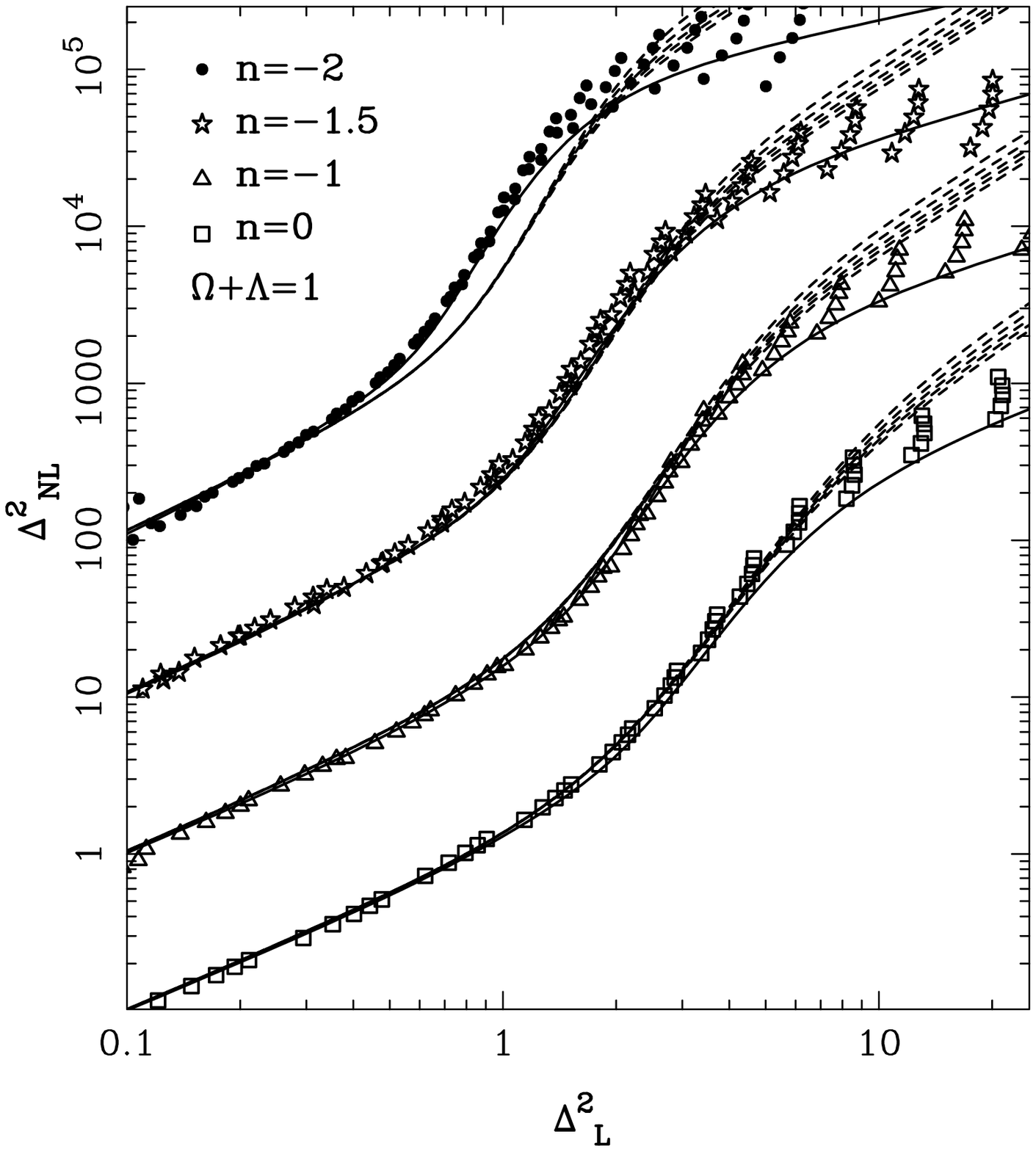,width=8.5cm,angle=0,clip=}}
\caption{\small{(Top) HKLM plot for the open models. (Bottom) HKLM
plot for the flat low-density models with a cosmological
constant. Again, for clarity, the data have been separated from each
other by one order of magnitude in the $y$-direction, with the $n=0$
data untranslated. For each model five epochs are shown, these are:
$a=0.6, 0.7, 0.8, 0.9, 1.0$, with the lowest locus for each model
corresponding to the $a=0.6$ epoch. In terms of the mass density
parameter of the universe, these epochs correspond to: $\Omega =0.294,
0.263, 0.238, 0.217, 0.200$ for the open models and $\Omega =0.619,
0.505, 0.407, 0.325, 0.260$ for the $\Lambda$ models. As in
Fig. \ref{HKLMplot1} the solid lines represent the fitting formula for
the Einstein--de Sitter models presented in Appendix \ref{appHKLM}; the
dashed lines represent fits from the PD96 function.}}
\label{HKLMplot2}
\end{figure}

\subsection{Low-density power-law models}

In Fig. \ref{HKLMplot2} we show how the nonlinear behaviour of the
power-law models deviates from the scale-free solutions (solid lines)
as the background density is lowered. Again, for clarity, the data
have been separated from each other by one order of magnitude in the
$y$-direction, with the $n=0$ data untranslated.  In the linear
regime, we again find that the nonlinear data follow the linear
power. In the quasi-linear regime, $2<\Delta_{\NL}^2<80$, as $\Omega$
decreases, the locus defined by the data increases in amplitude
relative to the scale-free models and the power-law slope
steepens. This density-dependent evolution of $f_{\NL}$ in the
quasi-linear regime was not apparent in previous studies (see
PD96). 
The quasi-linear slope steepens as both $n$ and $\Omega$ decrease. In the
nonlinear regime, $\Delta^2_{\NL}>80$, we again observe that the
slope of $f_{\NL}$ is lower than the $3/2$
value that is required by stable clustering.

In Fig. \ref{HKLMplot2}, we also compare the data with the density
dependent fitting formula of PD96. Again, the formula underestimates
the shallow spectra and slightly overestimates the steeper
spectra. However, the more striking discrepancy is that the formula
suppresses the onset of density dependent growth until evolution is
far into the nonlinear regime, 
and then tends to overestimate the highly nonlinear power.
These discrepancies can in fact be seen in the comparison
with the simulation data used by PD96. However, this library of
small ($N=80^3$) simulations was in most cases unable to probe
beyond $\Delta^2_{\NL} \simeq 200$, and so the deviations never
became substantial. 

The failure of the JMW95 and PD96 functions to accurately model the
Einstein--de Sitter data and account for the density dependence of
nonlinear growth has clearly been shown. On attempting to fit this
data set using the standard HKLM-PD96 procedure we were able to
produce an improved formula with an rms precision of 12\%.  However,
on attempting to integrate the CDM models into the formulation, we
could not find a satisfactory way to assign an effective spectral
index to the models.  We therefore decided to pursue an alternative
approach to the problem of general nonlinear fitting functions, 
which proved to be more accurate. 


\begin{figure}
\centerline{\epsfig{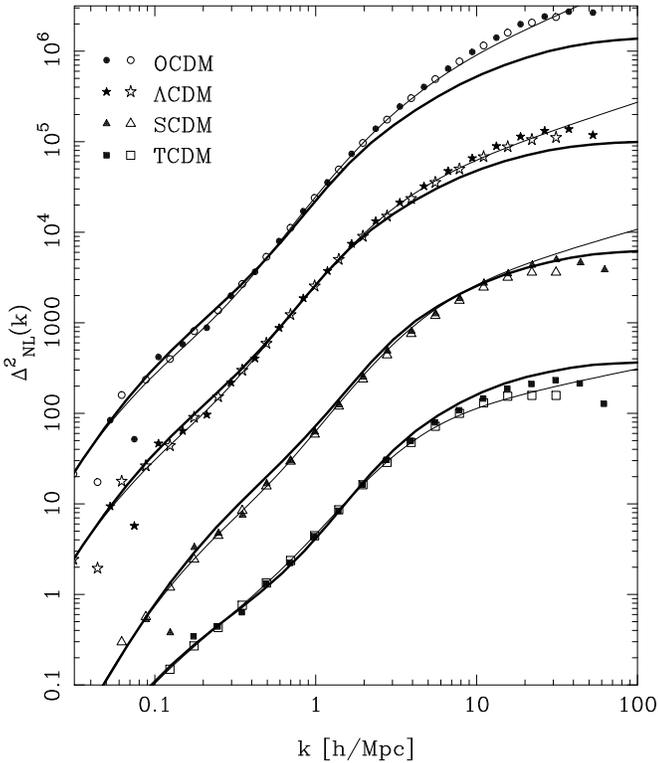}}
\caption{\small{Comparison of the full halo-model calculation as
described in the text (thick solid lines) with the CDM data
(points). Also shown is the halo-model fitting formula from this work
(thin solid lines). For clarity the four CDM models have been
separated from each other by one order of magnitude in the
y-direction, with the $\tau$CDM data untranslated.
\label{halomodel_comp_cdm}}}
\end{figure}

\section{The halo model fitting function}\label{scfhalo}

In this Section, we attempt to describe the above nonlinear results
by means of concepts abstracted from the `halo model'
\cite{PeacockSmith2000,Seljak2000,MaFry2000a}. The basic approach
suggested by the halo model is to decompose the density field into a
distribution of isolated haloes. Correlations in the field then arise
on large scales through the clustering of haloes with respect to each
other and on small scales through the clustering of dark matter
particles within the same halo. This then leads to a total nonlinear
power spectrum
\be P_{\NL}(k)=P_{\Q}(k)+P_{\H}(k), \label{eq:halomodel}\ee
where $P_{\Q}(k)$ is the quasi-linear term that represents the power
generated by the large-scale placement of haloes and where $P_{\H}(k)$
describes the power that results from the self-correlation of haloes.

It is remarkable that such a simple decomposition appears to work well
in describing the main characteristics of the two-point correlations of
the cosmological mass density. It is maybe yet still more impressive
when one considers that the present formulation knows nothing of the
large-scale filamentary structure of the density field (which is
governed by the correlation function of halo centres). Indeed this
deficiency was recently pointed out and addressed by Scoccimarro \&
Sheth \shortcite{ScoccimarroSheth2002}

In Fig. \ref{halomodel_comp_cdm} we directly compare the halo-model
calculations (thick solid lines) with the CDM simulations of J98 (data
points -- see section \ref{AppCDM} for full description).  Also shown
is the halo-model fitting function that we present later (thin solid
lines -- see Appendix \ref{halofit}). The halo model calculations are
exactly those of Peacock \& Smith (2000). From the figure, it can
clearly be seen that the calculations qualitatively reproduce the data
for all of the models, but that in detail only match SCDM and
$\tau$CDM closely.  Furthermore when one attempts to model the
power-law spectra the results are worse, with the $n=0$ case being an
extreme example (see the later discussion). Thus our aim in what
follows will therefore be to produce a simple fitting formula that
draws on the broad elements of the halo model, such as the above
decomposition of the power spectrum into two linearly summed terms,
but which is of very high accuracy.


\subsection{The quasi-linear term}

Consider the quasi-linear part first. Seljak \shortcite{Seljak2000}, Ma
\& Fry \shortcite{MaFry2000a} and Scoccimarro et al. 
\shortcite{Scoccimarroetal2001} assumed that one should use linear
theory filtered by the effective window corresponding to the
distribution of halo masses, convolved with their density profiles and
a prescription for their bias with respect to the underlying mass
field:
\be P_{\Q}(k)=P_{\L}(k) \left[ \frac{1}{\rhob}\int dM \; b(M)\, n(M)\;
 \tilde{\rho}(k,M) \right]^2, \label{quasi-halo}\ee
where $n\left(M\right)dM$ is the mass function,
$\tilde{\rho}\left(k,M\right)$ is the Fourier transform of the density
profile and $b(M)$ is the bias field of dark matter halo seeds.
Peacock \& Smith \shortcite{PeacockSmith2000} made the simpler
assumption that the quasilinear term corresponded to pure
linear theory:
\be P_{\Q}(k)=P_{\L}(k).\ee
This is equivalent to equation (\ref{quasi-halo}) on large
scales, since in this limit the filtering effect of haloes
is negligible, and we must have
\be \frac{1}{\rhob}\int dM \; b(M)\, M \,n(M) =1 \ .\ee

Neither of these approaches is really satisfactory, 
since $P_{\H}$ comes to dominate
only at scales where linear theory must break down to some extent
($\Delta^2_{\rm L}\sim 1$).  Quasi-linear effects must modify the
relative correlations of haloes away from linear theory, 
irrespective of whatever allowance may be made for
the finite sizes of haloes. One way of seeing this is
via the scaling part of the HKLM procedure: see equations (\ref{NSR}).
This shift of scales from gravitational collapse causes a
significant change in power at wavenumbers where $\Delta^2_{\L}$
is of order unity -- which is just the point where the filtering
effects of the largest haloes will also start to be important.
An alternative point of view is provided by perturbation theory,
which suggests that quasilinear effects should tend to suppress power
for $n>-1.4$, but enhance power for more negative indices
(e.g. section 4.2.2 of Bernardeau et al. 2001).
Again, such effects cannot be cleanly separated from the
convolving effects of halo profiles. We therefore take an empirical
approach,  allowing the quasilinear effects to depend on $n$.
Since the philosophy of the halo model is that $\Delta^2_{\Q}$ should
be negligible on small scales, we also build in a truncation
at high $k$:
\be \Delta^2_{\Q}(k)= \Delta^2_{\L}(k){ \left[1+\Delta^2_{\L}(k)\right]^{\beta_n}
\over 1 + \alpha_n\Delta^2_{\L}(k)} \exp{-f(y)};\ \ \ y\equiv k/k_{\sigma}
\label{QL3}.\ee
where $k_{\sigma}$ is a nonlinear wavenumber,
defined below in Section \ref{halononlscale}
$\alpha_n$ and $\beta_n$ are spectral dependent coefficients and
$f(y)$ is the polynomial $y/4+y^2/8$ that governs the decay rate.  We
adopt this expression for all spectra.


\subsection{The halo term}

In the halo model the self-halo term is
\cite{Seljak2000,PeacockSmith2000,MaFry2000a,Scoccimarroetal2001}
\be P_{\H}(k)=\frac{1}{\rhob^2\left(2\pi\right)^3}\int dM\; n(M)
\left| \tilde{\rho}\left(k,M\right)\right|^2 .\ee 
In order to model this we want something that looks like a shot-noise
spectrum on large scales, but is progressively reduced on small scales
by the filtering effects of halo profiles and the mass function. In
terms of the dimensionless power spectrum, a candidate form for this
is
\be \Delta^{2\ \prime}_{\H}(k) = { a_n\; y^3 \over 1 + b_n\; y
+ c_n\;y^{3-\gamma_n} };\quad\quad y\equiv k/k_{\sigma} \ ,\label{H1}\ee
where $(a_n,b_n,c_n,\gamma_n)$ are dimensionless numbers that depend
on the spectrum. However, with $P_{\H}$ defined in this way, the
formalism defined by equation (\ref{eq:halomodel}) breaks down for
steep spectra.  The self-halo power clearly dominates at small $k$ for
any spectrum that is asymptotically $n>0$ (e.g. all CDM models). This
has been independently noted by Sheth \& Cooray (2001).  The halo model
thus fails to respect low-order perturbation theory in such cases, and
this is a clear defect of the model.

In order to solve this problem, the self-halo power 
must become steeper than Poisson on the largest scales.
This makes sense if we think of the halo model as a two-stage
process: (i) fragment a uniform mass distribution into a
set of haloes; (ii) move these haloes according to a superimposed
large-scale displacement field. Since the first stage conserves
mass, the large-scale power spectrum must approach a `minimal'
form with $n=2$ (e.g. section 28 of Peebles 1980).
If one conserves momentum also, the minimal spectrum becomes
even steeper: $n=4$. 
It is a moot point which of these is the appropriate asymptote for
this problem, since the two-stage view of the halo model is
only a heuristic argument. Since we will never wish to consider
spectra that are asymptotically much steeper than $n=1$, it
will suffice to force the $n=0$ self-halo term to approach $n=2$
on sufficiently large scales.
This can be achieved if equation (\ref{H1}) is modified as follows
\be \Delta^2_{\H}(k) = {\Delta^{2\ \prime}_{\H}(k) \over 
1+\mu_ny^{-1}+\nu_ny^{-2}};\ \ \ y\equiv k/k_{\sigma}\ee
where we have introduced a term in $k^4$ in order to soften the
transition to the $k^5$ slope. Again, the parameters $\mu_n$ and $\nu_n$ are
spectral dependent coefficients. 

\subsection{The nonlinear scale}\label{halononlscale}

In order to implement these arguments, we need an appropriate general
definition of the nonlinear scale (see Section
\ref{scfnonlinearscale}), which should be related to the
characteristic mass in the halo mass function. As studies over many
years have shown with increasing accuracy
\cite{PressSchechter1974,ShethTormen1999,Jenkinsetal2001}, the halo
mass function appears to depend only on the dimensionless fluctuation
amplitude
\be \nu\equiv \delta_c / \sigma(R,t),\ee
where $\delta_c$ is a constant of order unity, usually identified with
the linear over-density for collapse in the spherical model and $R$ is
the effective filter radius. The multiplicity function for haloes,
which is defined as the fraction of mass carried by haloes with mass
in a logarithmic interval, peaks for systems where $\sigma(R,t)$ is
of order unity, and we can therefore choose to define the nonlinear
scale in this way:
\be \sigma(k_{\sigma}^{-1},t) \equiv 1\ .
\label{nonk}
\ee
This definition of scale depends on the functional form chosen to
filter the spectrum, but the main effects of changes in this choice
can be absorbed into the fitting coefficients.  We therefore take the
convenient choice of a Gaussian filter:
\be \sigma^2(R_{\rm G},t) \equiv \int \Delta^2_{\L}(k,t) \, \exp(-k^2 R_{\rm G}^2)\;
d\ln k.\ee
With this choice of filter, scale-free spectra have
\be 
\eqalign{
\Delta_{\rm L}^2(k,t) &=\left({k\over k_0(t)}\right)^{3+n} \; \Rightarrow \cr
{k_{\sigma}\over k_0(t)} &= \left( { [(1+n)/2]! \over 2} \right)^{-1/(3+n)}. 
\cr
}
\ee
%


\subsection{Application to CDM}\label{AppCDM}

\begin{figure}
\centerline{\epsfig{file=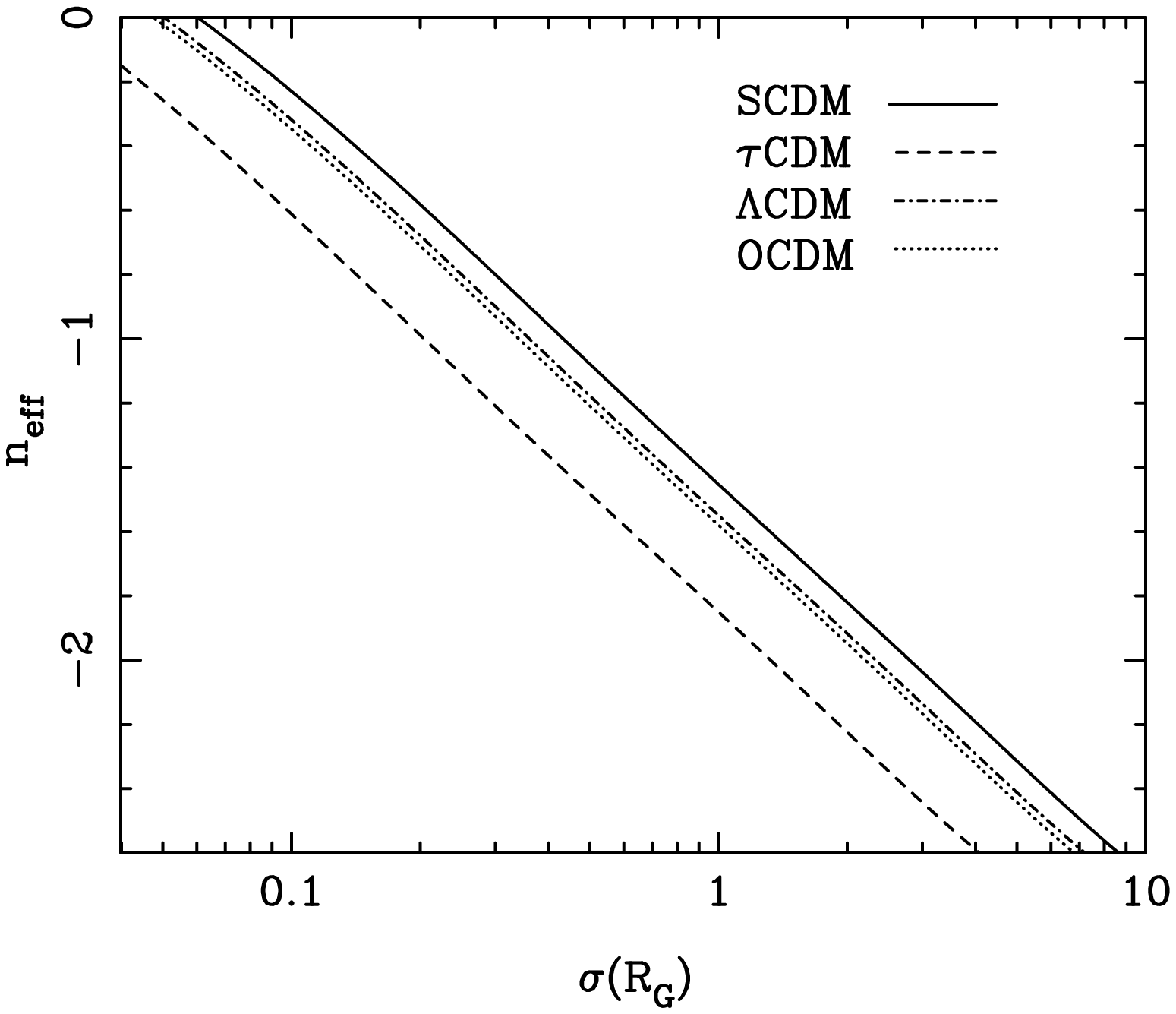,width=8.5cm,angle=0,clip=}}
\centerline{\epsfig{file=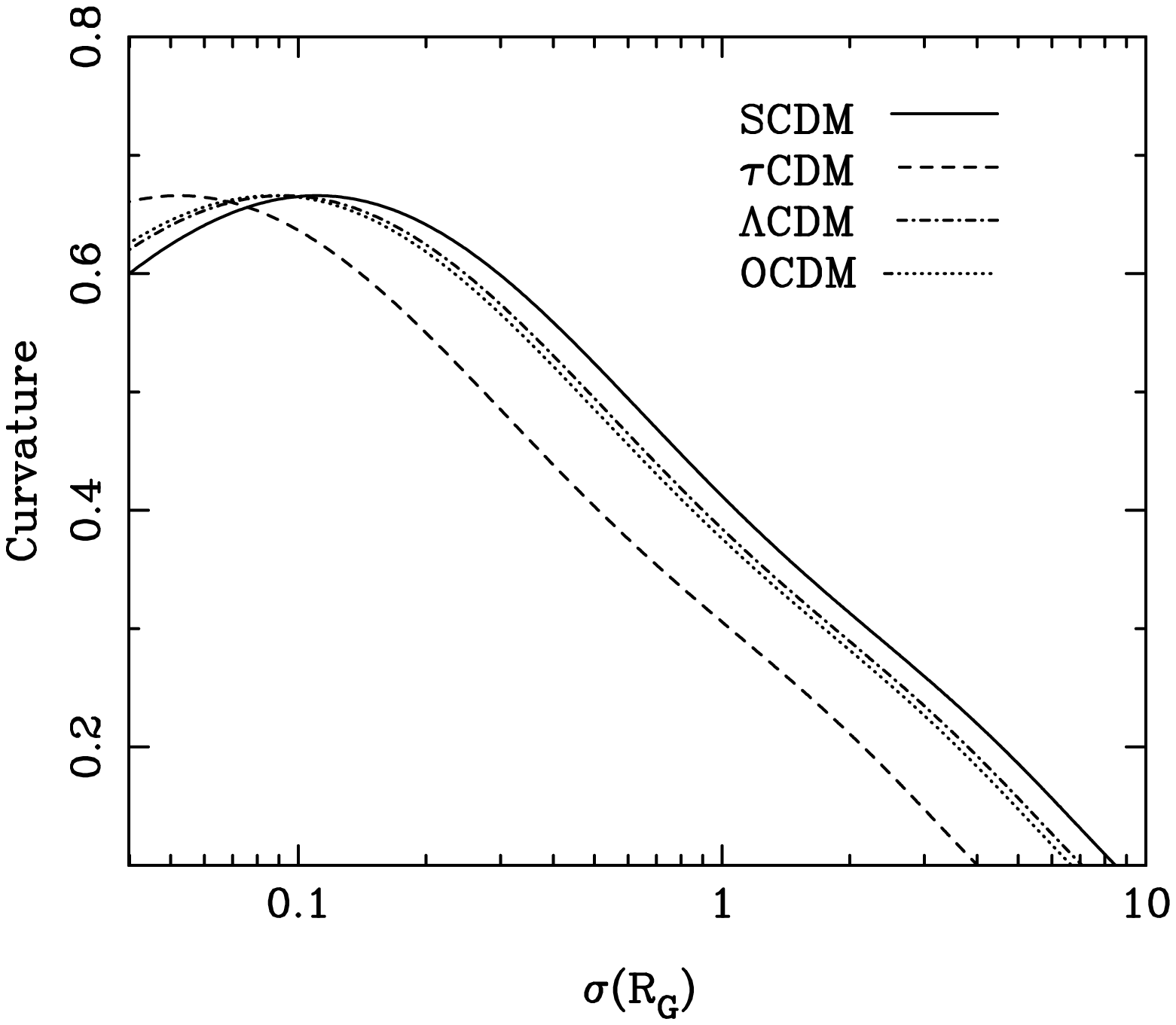,width=8.5cm,angle=0,clip=}}
\caption{\small{Variation of effective spectral index (top panel) and
curvature (bottom panel) as a function of the rms fluctuation in
Gaussian spheres of radius $R_{\rm G}$, for the four cosmological
models considered. Note the lower $\sigma_8$ values that corresponded
to the big-box simulations has been assumed for the SCDM and $\tau$CDM
models.}\label{SP01effective1}}
\end{figure}

\begin{figure}
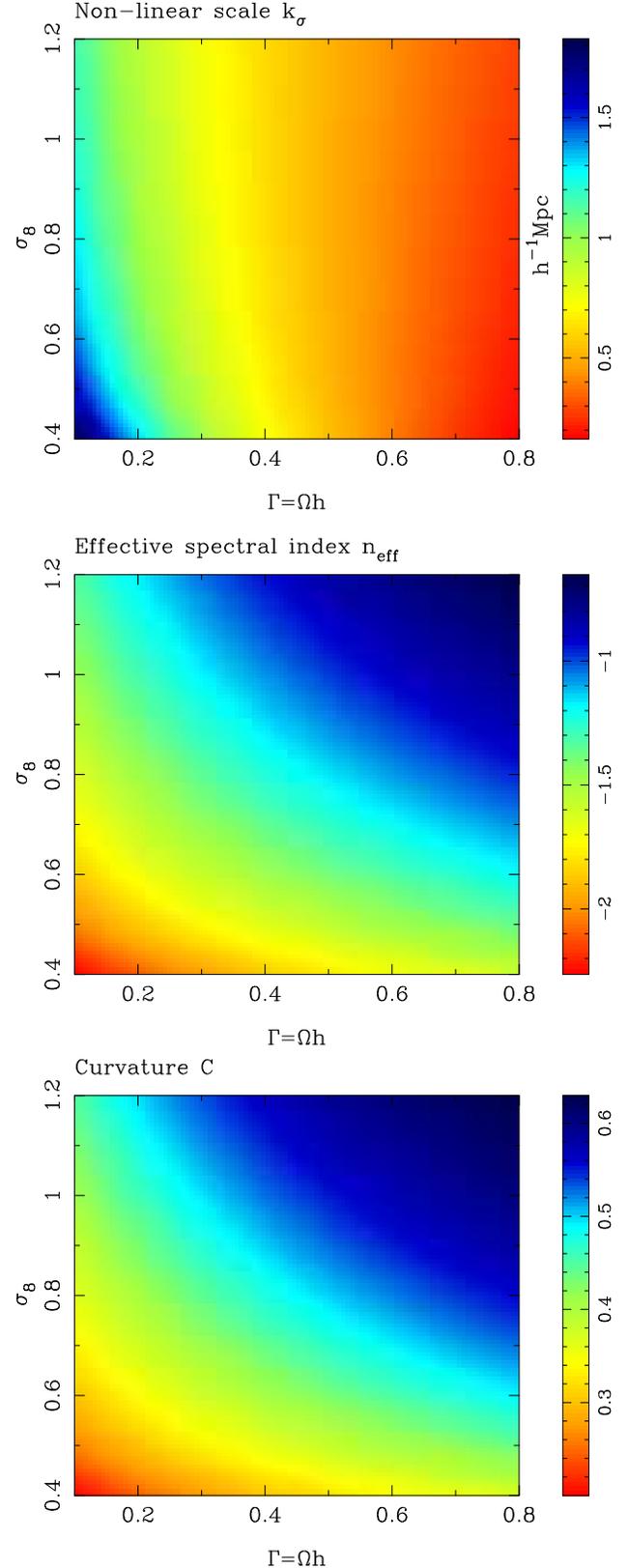

\centerline{\epsfig{file=fig10.1.eps,width=8.5cm,angle=0,clip=}}\vspace{0.2cm}
\centerline{\epsfig{file=fig10.2.eps,width=8.5cm,angle=0,clip=}}
\centerline{\epsfig{file=fig10.3.eps,width=8.5cm,angle=0,clip=}}
\caption{\small{Dependence of the nonlinear wave-number $k_\sigma$
(top panel), effective spectral index $n_{\rm eff}$ (middle panel) and
curvature parameter $C$ (bottom panel) on the shape parameter $\Gamma$
and normalization $\sigma_8$ of the linear power spectrum. The
parameters $n_{\rm eff}$ and $C$ are degenerate under $\Gamma$ and
$\sigma_8$. This degeneracy is, however, broken by the nonlinear
wavenumber.}
\label{cdm-eff}}
\end{figure}

\begin{table}
\caption{\small{The cosmological parameters of the $N=256^3$
CDM simulations from J98. For these CDM models $\Gamma\equiv
\Omega h$ is the shape parameter of the spectrum, $\sigma_8$ is the
rms fluctuation in spheres of $8 \mpcoh$ and $h$ is the
Hubble parameter }}\label{CDMpar} \centering{
\begin{tabular}{ccccccc}
\hline
\hline 
\noalign{\vglue 0.2em}
Model & $\Gamma$ & $\sigma_8$ & $\Omega$ & $\Lambda$ & $h$ & $L / \mpcoh$ \\
\noalign{\vglue 0.2em}
\hline 
\noalign{\vglue 0.2em}
SCDM  & 0.50 & 0.51 & 1.0 & 0.0 & 0.5 & 239.5\\
SCDM  & 0.50 & 0.6 & 1.0 & 0.0 & 0.5 & 84.55\\
$\tau$CDM  & 0.21 & 0.51& 1.0 & 0.0 & 0.5 & 239.5\\
$\tau$CDM  & 0.21 & 0.6& 1.0 & 0.0 & 0.5 & 84.55\\
$\Lambda$CDM  & 0.21 & 0.90 & 0.3 & 0.7 & 0.7 & 239.5\\
$\Lambda$CDM  & 0.21 & 0.90 & 0.3 & 0.7 & 0.7 & 141.3\\
OCDM  & 0.21 & 0.85 & 0.3 & 0.0 & 0.7 & 239.5\\ 
OCDM  & 0.21 & 0.85 & 0.3 & 0.0 & 0.7 & 141.3\\
\noalign{\vglue 0.2em}
\hline 
\hline
\end{tabular}}
\end{table}
 
We have generalized our formula to fit the Virgo and GIF CDM
simulations from J98, which comprise four models: SCDM; $\tau$CDM;
$\Lambda$CDM; OCDM. Table \ref{CDMpar} lists the cosmological
parameters for these models. The data are publicly available from {\tt
http://www.mpa-garching.mpg.de/Virgo/}.  We have re-measured the power
spectrum for the epochs $z=0.0,\;0.5\;1.0\;2.0$ and $3.0$ for both the
Virgo and GIF data, the results are presented in Figs. \ref{cdm1} and
\ref{cdm2}. The transfer function for these simulations was that of
Efstathiou, Bond \& White \shortcite{EfstathiouBondWhite1992}:
\be
\Delta^2(k)=\frac{Ak^4}{\left\{1+[aq+(bq)^{3/2}+(cq)^2]^{\nu}\right\}^{2/\nu}}\ee
where $q=k/\Gamma$, $a=6.4\; \mpcoh$, $b=3 \mpcoh$, $c=1.7
\mpcoh$. The normalization constant $A$ is chosen by fixing
$\sigma_8$.

In order to model these more general curved spectra, we define an
effective spectral index via
\be 3+n_{\rm eff} \equiv -\left.{d \ln \sigma^2(R,t) 
\over d \ln R}\right|_{\sigma=1} \label{neff} \ .\ee
Since the mass function should depend mainly on the Taylor expansion
of $\sigma$ about the nonlinear scale, we also allow dependence on the
spectral curvature:
\be C \equiv - \left.{d^2 \ln \sigma^2(R,t) \over d \ln R^2}
\right|_{\sigma=1}. \label{curve} \ee
For the case of a Gaussian filter these expressions have the explicit
forms,
\be 3+n_{\rm eff} = \frac{2}{\sigma^2}
\left.\int d \ln k \; \Delta^2_{\L}(k,t) \; y^2 \exp{\left(-y^2\right)}  
\right|_{\sigma=1} \ee
and
\ba C & = & \left(3+n_{\rm eff}\right)^2 
 + \frac{4}{\sigma(R)^2} \quad\times \nonumber\\ 
&& \left.\int d \ln k \; \Delta^2_{\L}(k,t) \; \left(y^2-y^4\right)
\exp{\left(-y^2\right)} \right|_{\sigma=1} ,\ea
where $y=kR_{\rm G}$ and where the explicit time dependence of the
power spectrum has been kept to indicate the redshift dependence of
the effective quantities. In Table \ref{CDMnonlin} we list the
nonlinear wavenumber, effective spectral index and curvature of the
spectrum on the nonlinear scale for the four Virgo (big-box) CDM
models, generated according to the above prescription.

\begin{table}
\caption{\small{The nonlinear wavenumber $k_{\sigma}$ in units of
$\hompc$, the effective spectral index $n_{\rm eff}$ and
curvature $C$ of the spectrum on the nonlinear scale, for the four
CDM models listed in the text.}}
\label{CDMnonlin}
\centering{
\begin{tabular}{cccc} 
\hline
\hline 
\noalign{\vglue 0.2em}
Model & $k_{\sigma}$ & $n_{\rm eff}$ & $C$ \\
\noalign{\vglue 0.2em}
\hline 
\noalign{\vglue 0.2em}
SCDM & 0.574 & $-1.455$ & 0.411\\ 
$\tau$CDM & 0.735 & $-1.850$ & 0.305\\ 
$\Lambda$CDM & 0.306 & $-1.550$ & 0.384\\ 
OCDM & 0.332 & $-1.581$ & 0.375\\
\noalign{\vglue 0.2em}
\hline
\hline 
\end{tabular}}
\end{table}

Fig. \ref{SP01effective1} shows the variation of the effective
spectral index (top panel) and curvature (bottom panel) for the four
Virgo CDM models with the rms fluctuation measured in Gaussian spheres
of effective radius $R_{\rm G}$. The effective spectral index is quite
sensitive to whether it is defined at $\sigma=1$ or at some other
value. However, including the curvature (which depends much more
weakly on $\sigma$), means that this uncertainty is automatically
allowed for.  With the nonlinear scale and effective spectral index
and curvature as defined through equations (\ref{nonk}-\ref{curve}),
we find that we can accurately model CDM spectra. As expected,
Fig. \ref{SP01effective1} shows that the OCDM and $\Lambda$CDM models
are almost indistinguishable: both possess nearly identical linear
power spectra, with only a slight difference in normalization. The
$\tau$CDM model has the shallowest effective spectral index, almost
approaching $n=-2$ and the SCDM model has the steepest, with
$n=-1.4$. The power-law models that we have simulated encompass this
range of $n_{\rm eff}$.  Thus, we are confident that the new fitting
function will be constrained by the appropriate range of spectral
models, with the notable exception of the $z>3$ $\tau$CDM data for
which $n_{\rm eff}<-2$. These models are the sole basis for the
fitting formulae in the $n<-2$ regime.

Fig. \ref{cdm-eff} shows the dependence of $k_{\sigma}$ (top panel),
$n_{\rm eff}$ (middle panel) and $C$ (bottom panel) on the shape
parameter and normalization of the linear power spectrum for
$(0.1\le\Gamma\le0.8)$ and $(0.4\le\sigma_8\le1.2)$.  In all of the
models dark contrast represents a higher value. The parameters $n_{\rm
eff}$ and $C$ are degenerate under $\Gamma$ and $\sigma_8$. This
degeneracy is, however, broken by including the nonlinear wavenumber.


\subsection{Parameter optimization}\label{scffit}

\begin{figure}
\centerline{\epsfig{file=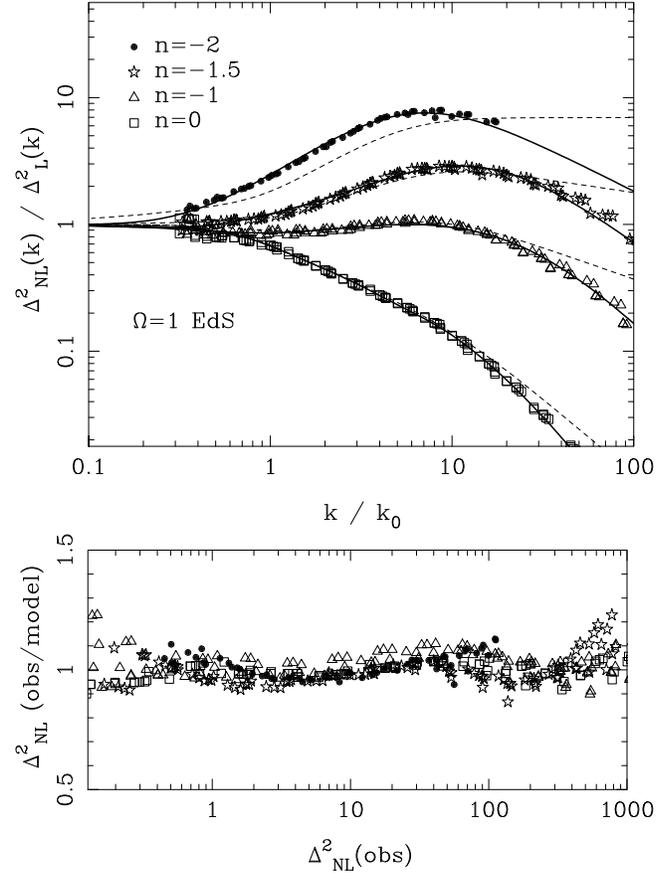,width=8.5cm,angle=0,clip=}}
\centerline{\epsfig{file=fig11.2.eps,width=8.5cm,angle=0,clip=}}
\caption{\small{(Top) nonlinear power ratioed to the linear power as a
function of wavenumber scaled in terms of the normalization wavenumber
$k_0$: where $\Delta^2(k_0)=1$. The data points are for the scale-free
simulations; the solid lines represent the fits from the new halo
based formula in Section \ref{scffit}; the dotted lines are PD96 fits.
(Bottom) The goodness of the new fit. The $y$ axis represents the
ratio of observed nonlinear power to nonlinear power predicted by the
halo based fitting function. The $x$ axis is observed nonlinear power.}
\label{SP01scf}}
\end{figure}

\begin{figure}
\centerline{\epsfig{file=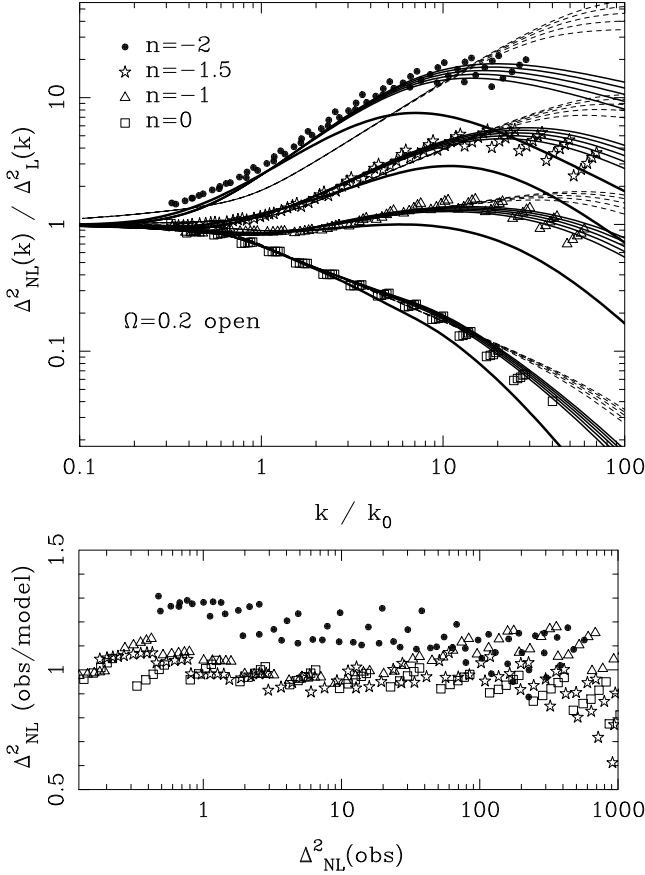,width=8.5cm,angle=0,clip=}}
\centerline{\epsfig{file=fig12.2.eps,width=8.5cm,angle=0,clip=}}
\caption{\small{Top and bottom panels are similar to
Fig. \ref{SP01scf}, but this time points represent open model
data. Five epochs are shown; these are $a=0.6, 0.7, 0.8, 0.9, 1.0$. In
terms of $\Omega$, these epochs correspond to: $\Omega =0.294, 0.263,
0.238, 0.217, 0.200$. The thick solid line represents the new
halo-model based fitting to the scale-free data.}
\label{SP01open}}
\end{figure}

\begin{figure}
\centerline{\epsfig{file=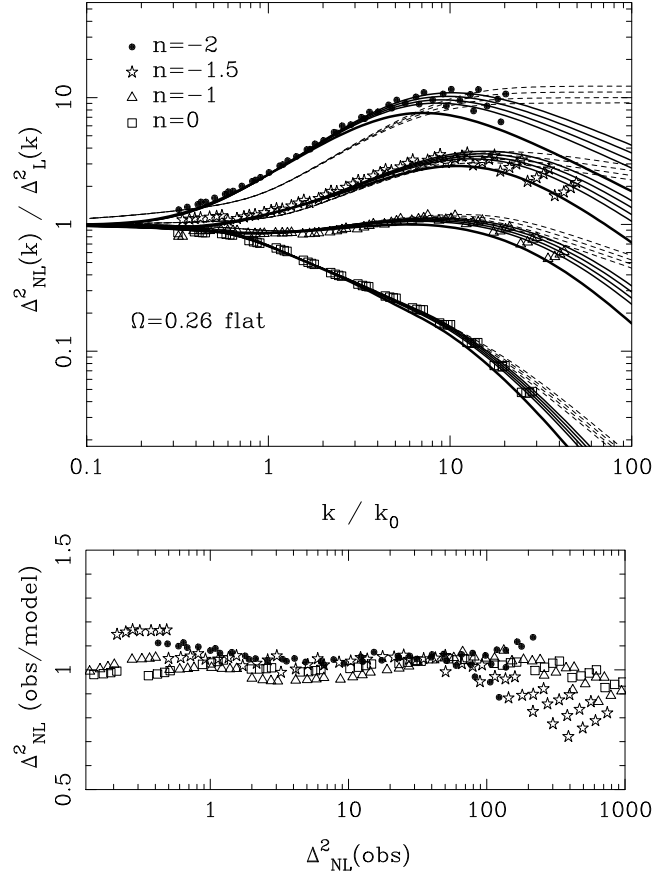,width=8.5cm,angle=0,clip=}}
\centerline{\epsfig{file=fig13.2.eps,width=8.5cm,angle=0,clip=}}
\caption{\small{Top and bottom panels are similar to
Fig. \ref{SP01scf}, but this time points represent $\Lambda$ model
data. Four epochs are shown; these are $a=0.7, 0.8, 0.9, 1.0$. In
terms of $\Omega$, these epochs correspond to: $\Omega = 0.505, 0.407,
0.325, 0.260$. Again, the thick solid line represents the fit to the
scale-free data.}
\label{SP01lambda}}
\end{figure}

We now give the best-fitting coefficients, including dependence on
cosmology. These coefficients were obtained by optimizing the formula
to fit the scale-free and $\Omega<1$ power-law simulations described
here; the CDM simulations of J98; and on large scales ($k<0.15
\hompc$), the results of 2nd-order perturbation theory (calculated
using the formulae of {\LL}okas et al. 1996).  Owing to the fact that
numerical simulations are susceptible to sample variance on large
scales, analytic perturbation theory results were preferred.  In the
halo model the cosmology dependence arises in a subtle way.  To the
extent that the mass function depends only on $\nu$ (when expressed as
a function of $R$) and that $\delta_c$ has no strong cosmology
dependence, the mass function for a given spectrum is also independent
of cosmology. Therefore, the only effect on the halo power spectrum
should be through the sizes of haloes; these depend on cosmology
because haloes that collapse at high redshift are smaller. Collapse
redshift is a function of mass and cosmology (see e.g. Appendix C of
Peacock \& Smith 2000).  High-mass haloes always have $z_c\simeq 0$;
these thus filter the large-scale part of the spectrum in a
cosmology-independent way. Conversely, low-mass haloes are important
at high $k$, and these do depend on cosmology -- which alters the
effective scale at which filtering occurs. However, there appears to
be no simple way to implement such a complicated dependence into the
fitting procedure. We therefore insert empirical functions of $\Omega$
into the procedure.  Also, motivated by the findings of Section 5, we
allowed the power-law indices that govern the quasi-linear regime to
be density dependent.

The prescription that was found to work best is given in Appendix C.
Code to evaluate the fitting function can be downloaded from the web
address listed in the abstract. Note that the above coefficients were
obtained by fitting the data over a restricted range of scales. The
scale-free data were constrained to have $k/k_{\sigma}>0.3$. The open
and $\Lambda$ data were constrained to lie in the range:
$(4.0<\Delta^2_{\L}<15.0)$ for $n=-2$; $(0.3<\Delta^2_{\L}<15.0)$ for
$n=-1.5$; $(0.3<\Delta^2_{\L}<20.0)$ for $n=-1$;
$(0.3<\Delta^2_{\L}<25.0)$ for $n=0$. The CDM data were fit under the
constraints: $k>0.3$ for the big box data and to $ k>0.5 \hompc$ for
the higher resolution small box calculations; the nonlinear power must
be 10\% greater than the discreteness correction, equation
(\ref{glasscorr}). On larger scales $k<0.15 \hompc$, the formula was
calibrated to the results of 2nd-order perturbation theory.

In Fig. \ref{SP01scf} we compare our new halo-based fitting function
with the scale-free simulations. The new model clearly reproduces the
data to a high degree of accuracy. Also, it is important to note that
when the data are plotted in this way the scaling nature is again
apparent and the departure from stable clustering, which is indicated
by the deviation away from PD96 for $k/k_0>10$, is pronounced.

In Figs \ref{SP01open} and \ref{SP01lambda} we compare the new halo
based model with the power-law data
for $\Omega<1$ and $\Omega+\Lambda=1$.
For all of the models the inclusion of the functions $f_1$,
$f_2$ and $f_3$, seems to well reproduce the observed
density-dependent growth. The only significant discrepancy is for the
$n=-2$ open data, where the power is underpredicted in the
quasi-linear regime.

\begin{figure*}
\centerline{\epsfig{file=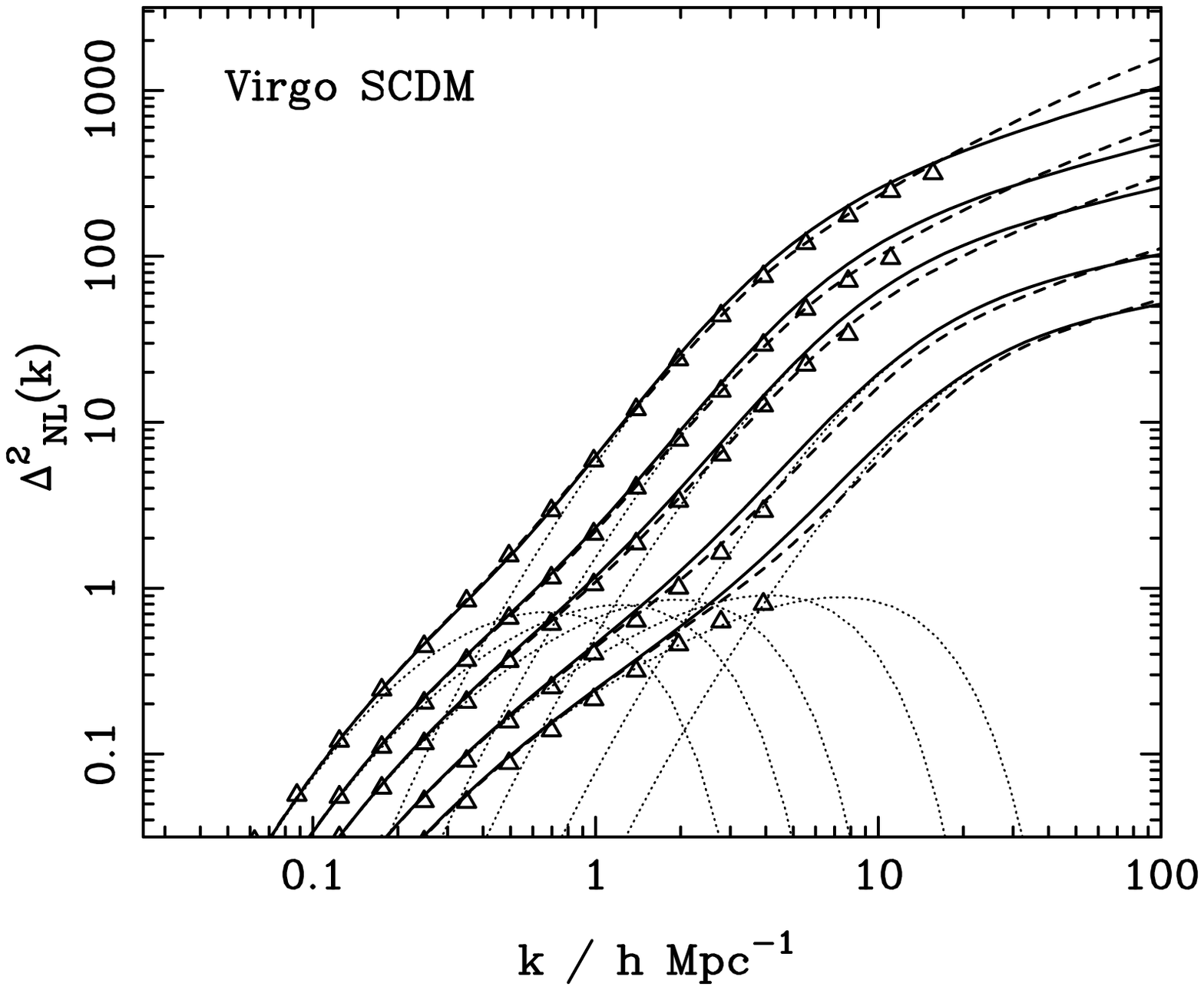,width=9.0cm,angle=0,clip=}
\epsfig{file=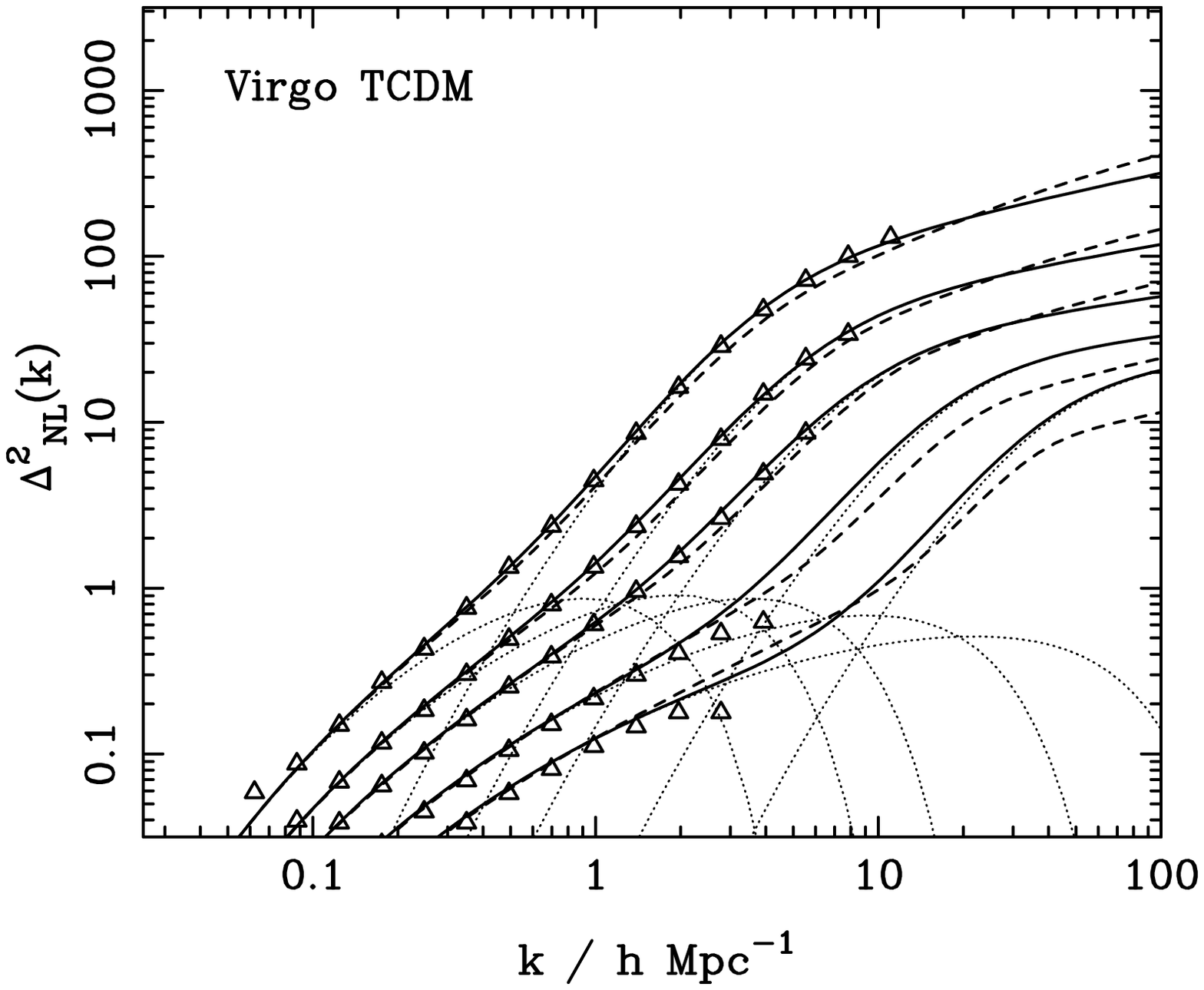,width=9.0cm,angle=0,clip=}}
\centerline{\epsfig{file=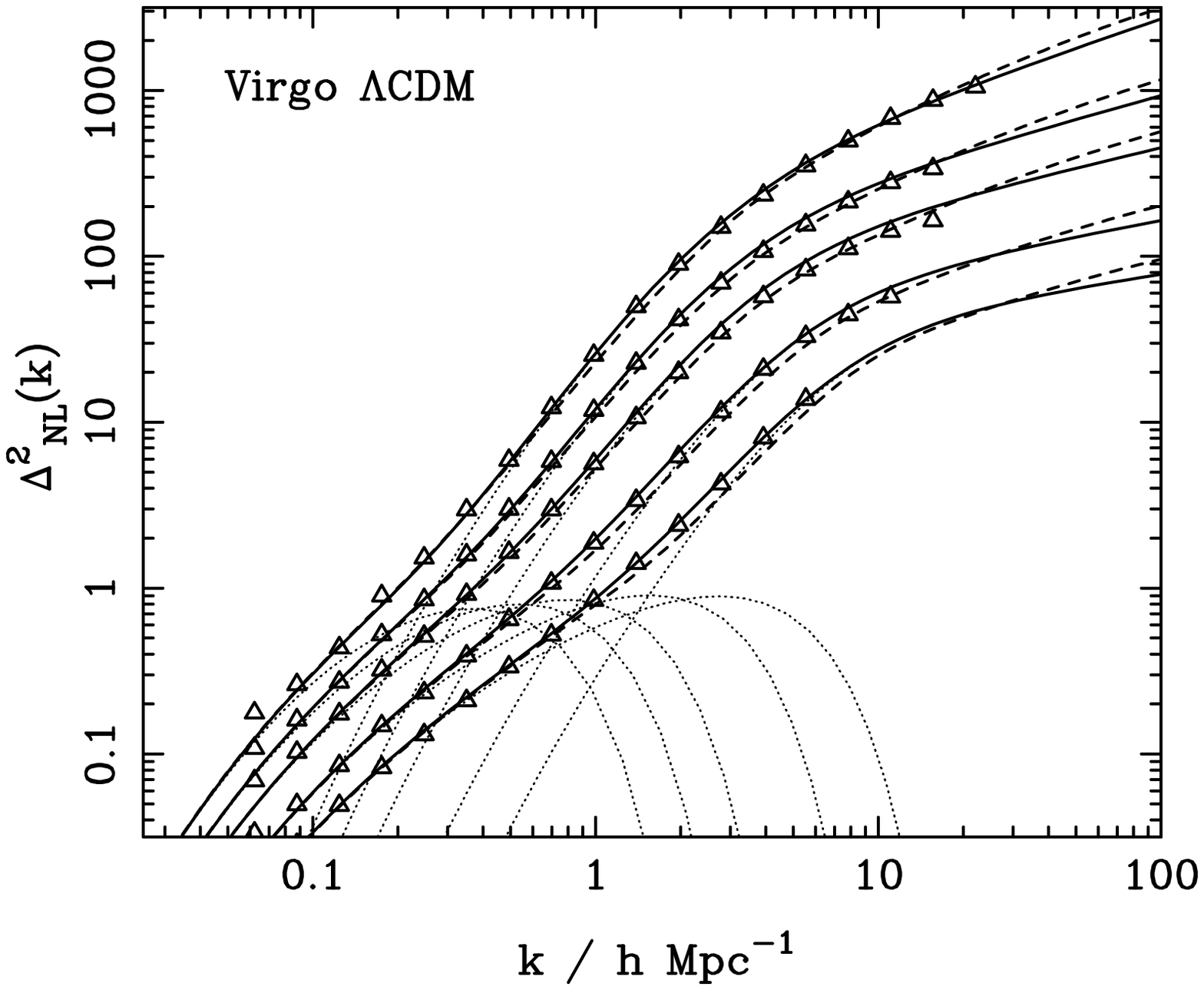,width=9.0cm,angle=0,clip=}
\epsfig{file=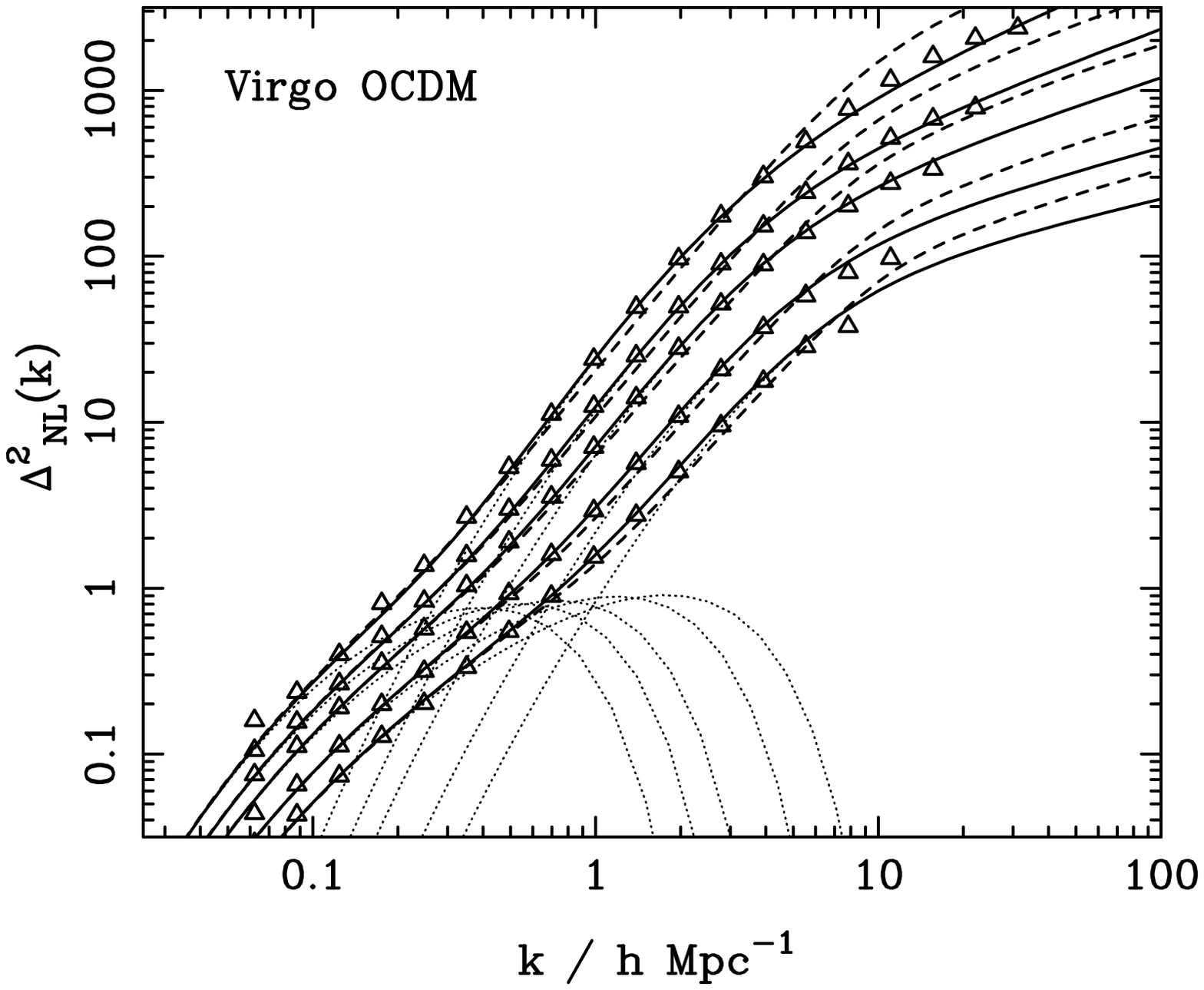,width=9.0cm,angle=0,clip=}}
\caption{\small{Power spectra for the four Virgo CDM simulations (J98)
in large cosmological volumes, $L=239.5 h^{-1}{\rm Mpc}$. Each panel
shows the evolution of structure with redshift. The data points
correspond, from low to high, to epochs $z=0,\; 0.5,\; 1.0,\; 2.0$ and
$3.0$. Note that only those points with a measured power above the
discreteness spectrum are plotted. The solid line represents the new
halo-model based fitting procedure, with dotted lines representing the
decomposition into the self-halo and halo-halo terms; the dashed line
is the PD96 fit.}
\label{cdm1}}
\end{figure*}

\begin{figure*}
\centerline{\epsfig{file=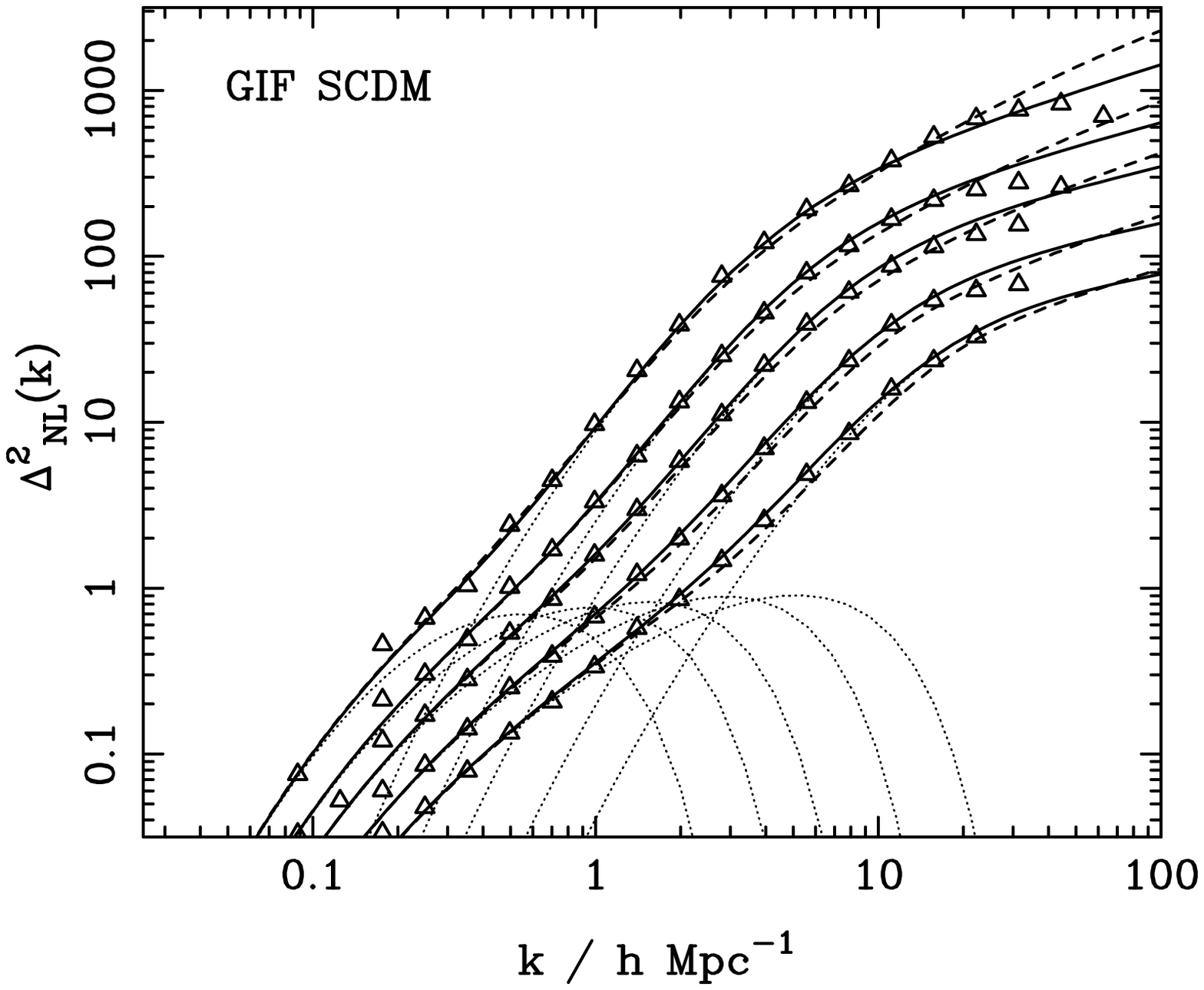,width=9.0cm,angle=0,clip=}
\epsfig{file=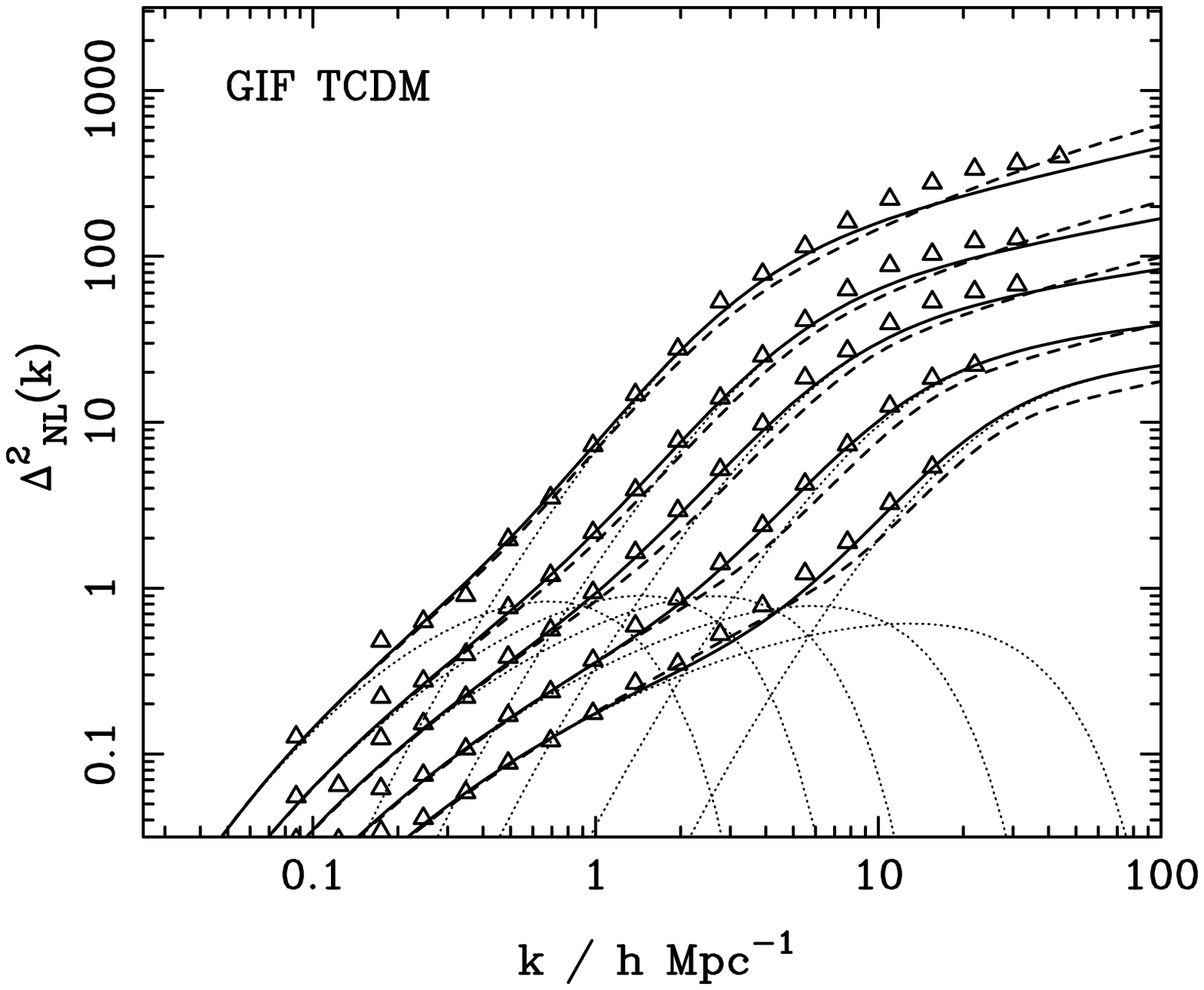,width=9.0cm,angle=0,clip=}}
\centerline{\epsfig{file=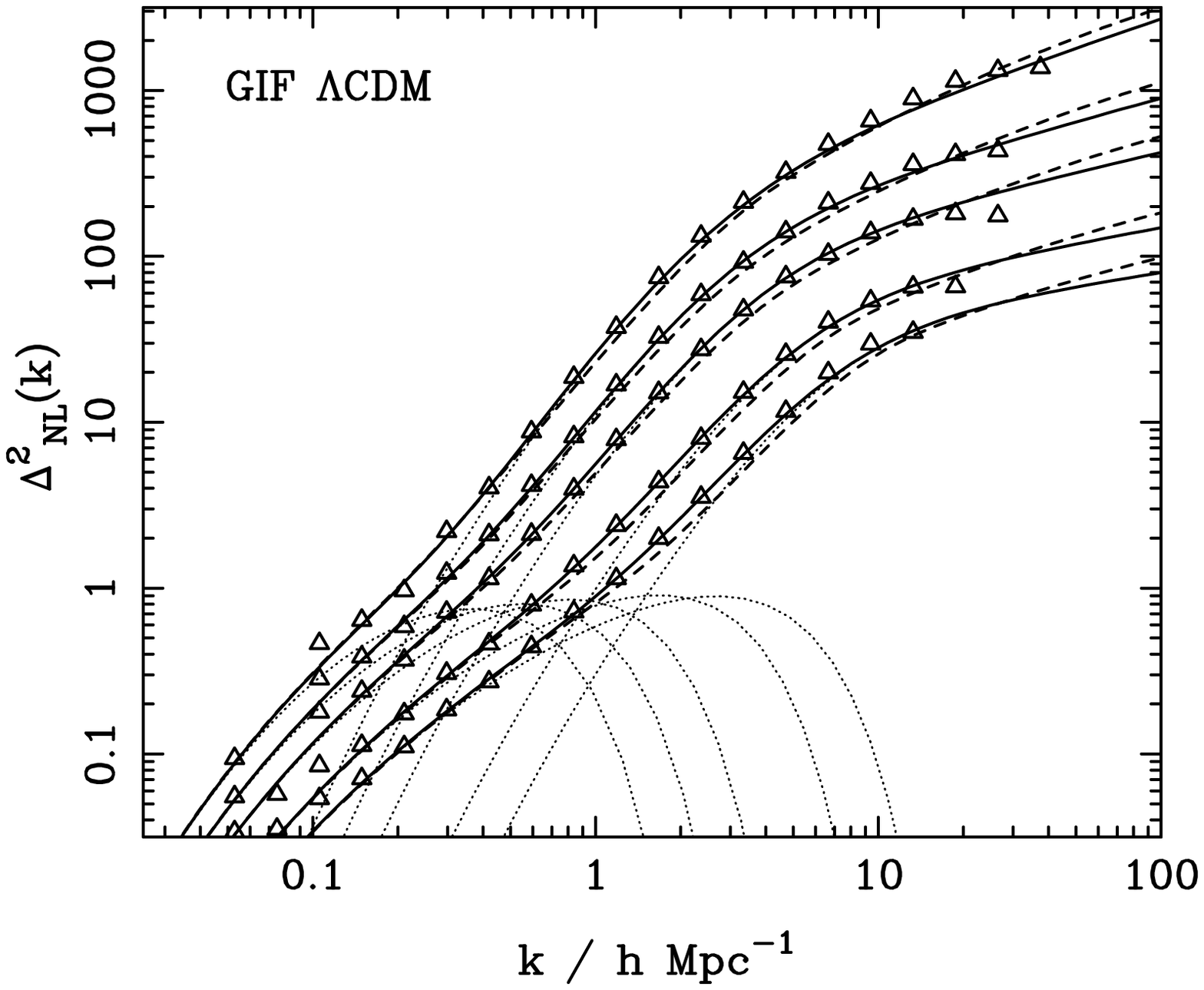,width=9.0cm,angle=0,clip=}
\epsfig{file=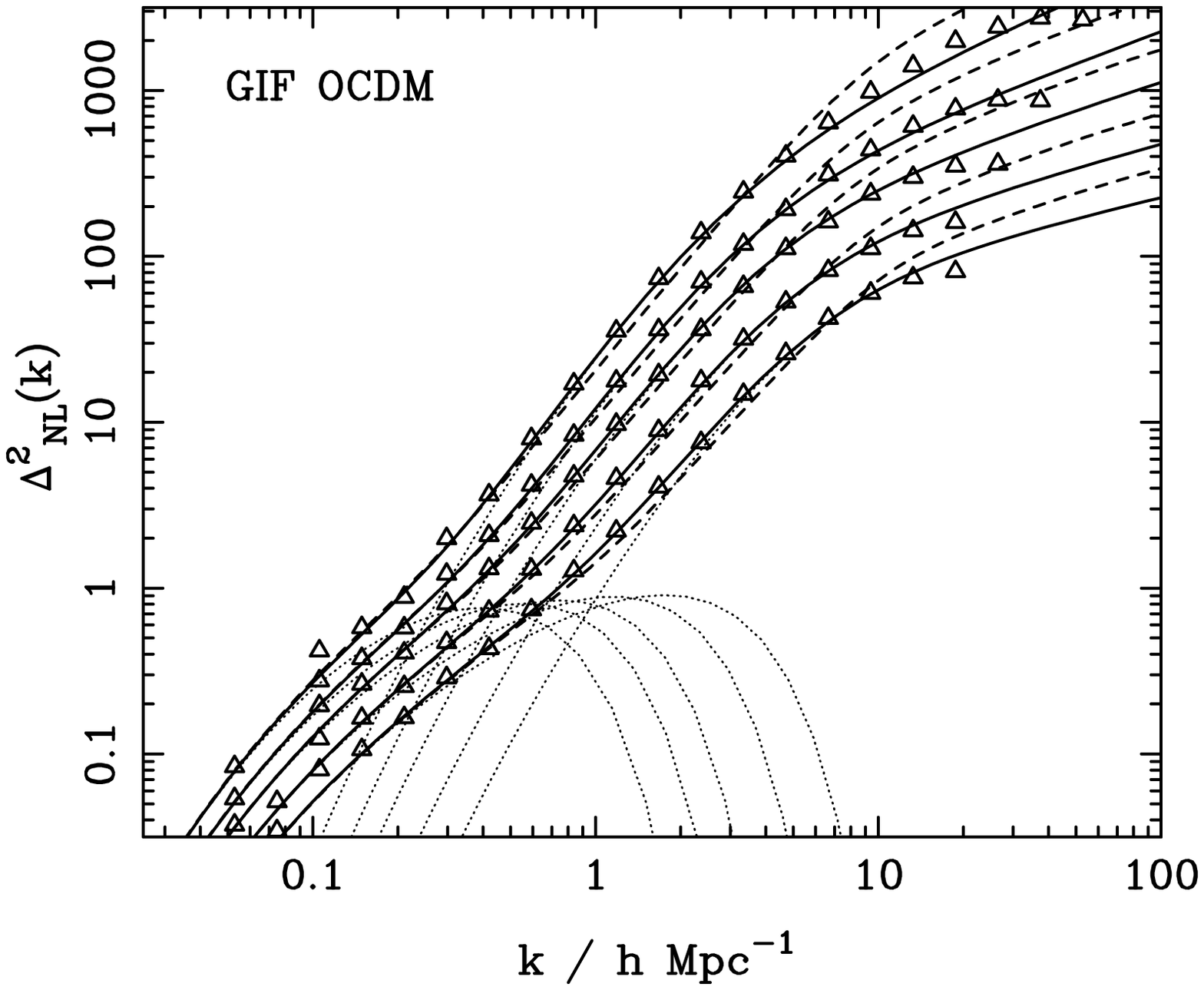,width=9.0cm,angle=0,clip=}}
\caption{\small{Same as for Fig. \ref{cdm1}, but this time for the 
smaller box GIF simulations.}
\label{cdm2}}
\end{figure*}

In Figs \ref{cdm1} and \ref{cdm2} we compare the model with the CDM data. Again,
the model does exceptionally well at reproducing all of the data over
the range of scales where we are confident that numerical effects are
unimportant. In particular, the OCDM and $\tau$CDM predictions are
very significantly improved using the new prescription.


\begin{figure}
\centerline{\epsfig{file=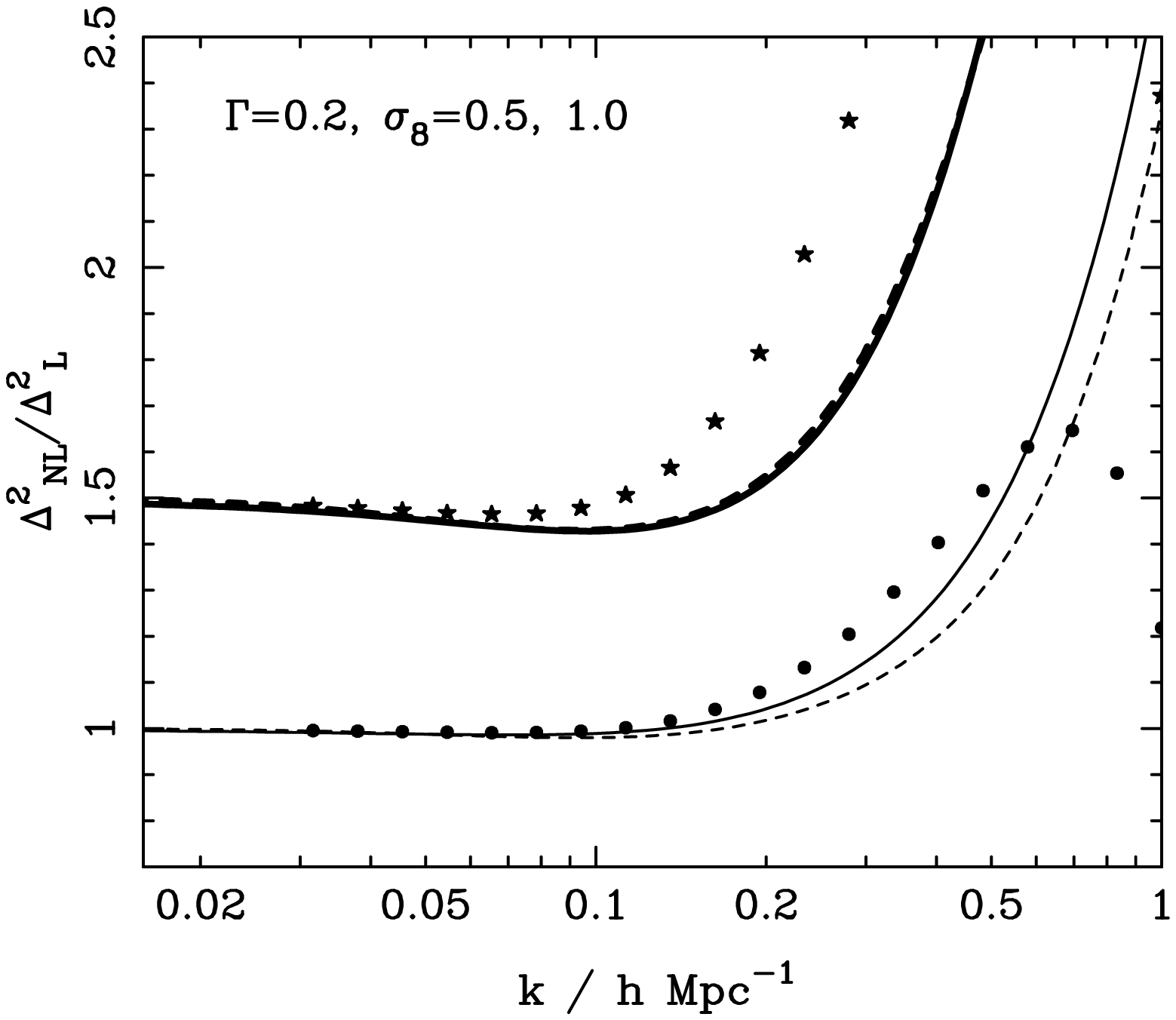,width=8.5cm,angle=0,clip=}}
\centerline{\epsfig{file=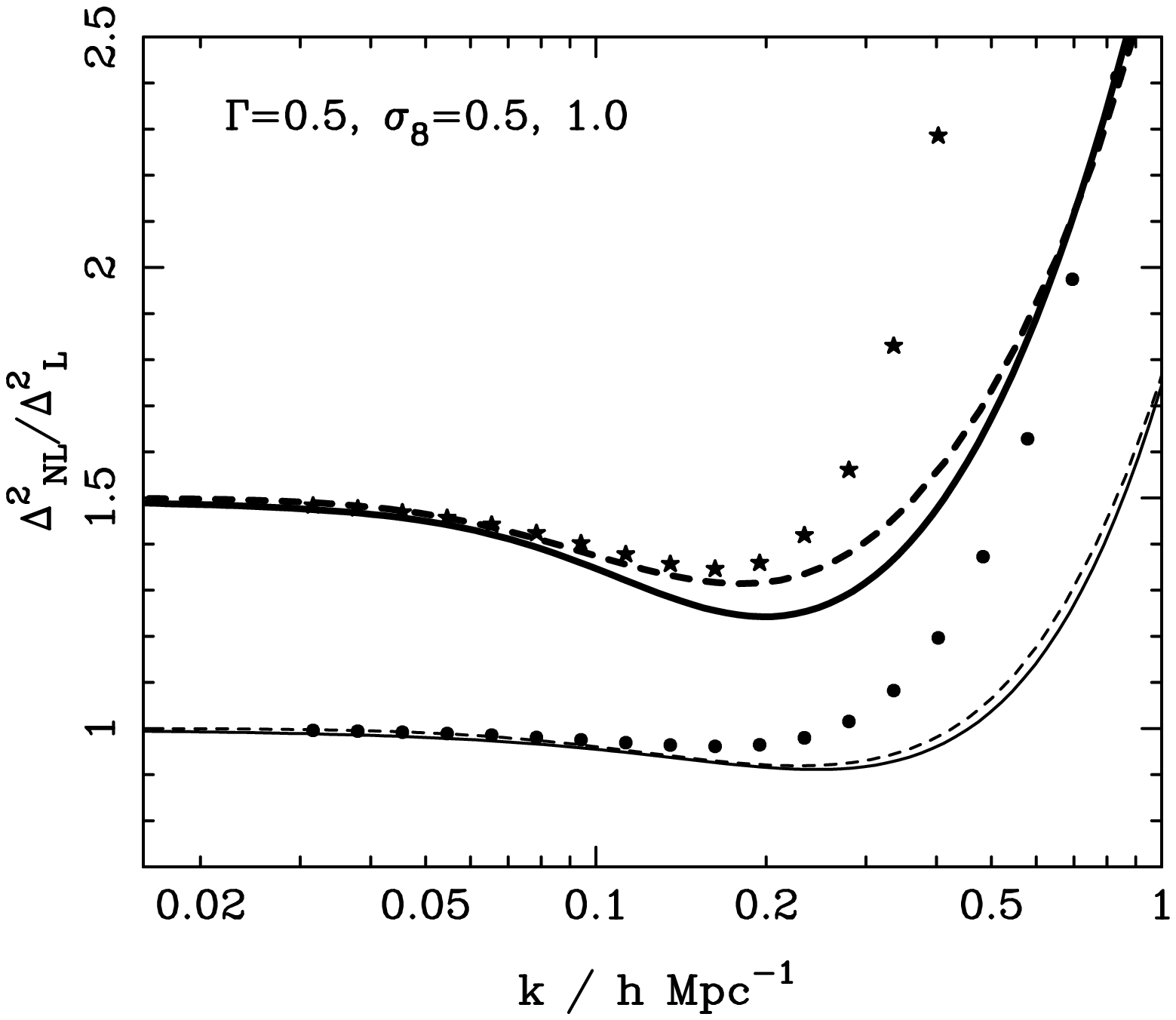,width=8.5cm,angle=0,clip=}}
\caption{\small{Comparison to perturbation theory. The top panel shows
the ratio of the evolved power to the linear power for two CDM models
with $\Gamma=0.2$, but with different normalizations. The model with
$\sigma_8=1.$ has been translated by a factor of 0.5 in the
$y$-direction.  The points represent perturbation theory; the thick
solid line is this work; the dashed line represents PD96. The bottom
panel is the same as the top, but with $\Gamma=0.5$.}
\label{SP02pt}}
\end{figure}


Having demonstrated the success of the halo fitting function on small
scales, we next consider the large scales.  We assess this using the
predictions derived from 2nd order perturbation theory (see Baugh \&
Efstathiou 1994). Fig. \ref{SP02pt} shows the ratio of nonlinear to
linear power for four CDM models. The current models match perturbation
theory for $k<0.1 \hompc$, but deviations exist at higher $k$.
These plausibly reflect a genuine breakdown of perturbation theory,
since the model was required to match perturbation theory as well
as possible for $k<0.15 \hompc$, and yet the fit is breaking down
slightly before this upper limit. 
Both the halo fitting function and PD96 agree well in this range.


\section{Conclusions and discussion}\label{scfdiscussion}

In this paper we have presented a set of high-resolution, $256^3$
particle, scale-free $N$-body simulations, designed to investigate
self-similar gravitational clustering and in particular the effects of
nonlinear evolution. We have also performed a further series of
numerical simulations, with the same resolution, to explore how the
evolution of clustering depends upon the background density of the
universe. Together, these simulations represent the best calculations
that exist to date for the set of models explored, with a factor 512
improvement in mass resolution over the ground-breaking work of 
Efstathiou et al. \shortcite{Efstathiouetal1988}.

We verified that the final output power spectra were robust by
considering grid and glass particle loads. However, at early times
the problem of discreteness correction is simpler to handle if a
glass start is applied; we have described a detailed method for
correcting the clustering signal in this case.  We have implemented
the power spectrum estimation technique of J98, which allowed us to
probe high spatial frequencies without aliasing effects or errors due
to mass assignment to the Fourier mesh.  The simulation results may be
summarized as follows:

\begin{enumerate}

\item Scale-free simulations with $0<n<-2$ show
self-similarity under the scaling $k_{0}(a)\propto a^{-2/(n+3)}$. This
conclusion is in agreement with the results of Efstathiou
\shortcite{Efstathiouetal1988} and Jain \& Bertschinger
\shortcite{JainBertschinger1998}.

\item In the quasi-linear regime, the power spectrum is characterized
by a steep power law. The exact slope depends upon the spectral index
$n$ of the input spectrum and the value of $\Omega$, the slope
steepening as $n$ becomes more negative and as $\Omega$ is reduced.

\item The observed nonlinear asymptote of the Einstein--de Sitter
simulations was found to be inconsistent with the
$\Delta_{\NL}^2\propto [\Delta_{\L}^2]^{3/2}$ prediction of stable
clustering.  A shallower slope with $\Delta_{\NL}^2\propto
[\Delta_{\L}^2]^{1}$ is preferred. This result makes sense in terms of
the halo model: calculations using the extended Press-Schechter
apparatus show that haloes will tend to merge with systems of similar
mass to their own \cite{LaceyCole1993}. Mergers of this kind will
disrupt the virial equilibrium of the system, violating the basic
assumption that underlies stable clustering. However, if this process
were rare then stable clustering could be upheld in a statistical
sense.  

\item The nonlinear fitting formulae of PD96 and JMW95 failed to
reproduce the $n=-2$ results and were only marginally successful at
reproducing the steeper spectra. The low-density power-law data were
poorly fit by PD96. 

\item For the $\Omega<1$ simulations, it is interesting to consider
how the nonlinear slope changes with density. In the nonlinear limit
equation (\ref{app-halo1}) (appendix C) becomes
\be \Delta^2(k)\propto k^{\;[3(f_1-1)+\gamma_n]}\ .\ee
For a given $n$, $f_1$ increases as $\Omega$ decreases, and so the
power-law slope steepens. This result supports the idea that small
scale clustering is more closely related to the emergence of the
internal density structure through the continual accretion and merger
of haloes. The reasoning is as follows: for a low-density universe
mergers are less frequent and so haloes have more time to virialize. 
This means that stable clustering may be considered to be a
better approximation for these systems. From the arguments in Section
1 and 5, this would then be manifest as a steepening of the nonlinear
slope.

\end{enumerate}

In the second part of this paper, we proposed an improved fitting
function for mass power spectra to replace the much-used PD96
formula. We have adopted a new approach to fitting power spectra,
based upon a fusion of the halo model and a HKLM scaling.  The method
was generalized to fit more realistic curved spectra, by introducing
two new parameters, $n_{\rm eff}$ the effective spectral index on the
nonlinear scale, and the spectral curvature, $C$. We found that the
halo model as previously envisaged in the literature fails to approach
linear theory on large scales for $n\ge 0$.  We have argued that this
should be cured by changing the self-halo power from $n=0$ to $n=2$ on
large enough scales, and we have shown empirically that this approach
allows an accurate description of a very wide range of power spectrum
data.  Our new fitting formula reproduced the scale-free
power spectrum data and also the CDM results of J98 with an rms error
better than 7\%. This is to be preferred to the widely-used PD96
prescription, and should be useful for a variety of cosmological
investigations.  In particular, our preliminary investigations show
that the present formalism should cope naturally with spectra
containing a realistic degree of baryonic features (e.g. Meiksin,
White \& Peacock 1999).

The halo model provides a novel way to view structure formation, and
has yielded useful insights into the origin of nonlinear aspects of
galaxy clustering.  This work has concentrated on the low-order
statistics of the density field, but it is also possible to consider
higher-order statistics such as the bispectrum.  This three-point
function in Fourier space probes the shapes of large-scale structures
that are generated by gravitational clustering. No shape information
is included in the current formalism, so it will be interesting to see
how well the model can account for higher-order statistics.  Initial
results in this direction (Scoccimarro et al. 2001) seem to be
promising.  In general, the important question is the extent to which
the halo model can encapsulate the phase information in the density
field, since fields with identical power spectra can possess
completely different real-space distributions (e.g. Chiang \& Coles
2000).  The halo model will inevitably fail to encompass these details
of the density field in full, although it may still offer useful
insights.  However, at the two-point level, we have shown that the
model is far more than an educational device, and it can be used as a
tool for a high-precision description of the evolution of the
dark-matter power spectrum.


\section*{Acknowledgements}
The authors would like to thank the anonymous referee for useful
comments, in particular for suggesting the inclusion of
Fig. \ref{halomodel_comp_cdm}. RES thanks Bhuvnesh Jain, Peter Watts,
Andy Taylor and Frank van den Bosch for helpful discussions during
this work. The simulations in this paper were carried out as part of
the programme of the Virgo Consortium for cosmological simulation
({\tt http://www.mpa-garching.mpg.de/Virgo}) using computers based at
the Computing Centre of the Max-Planck Society in Garching and at the
Edinburgh Parallel Computing Centre.  RES acknowledges a PPARC
research studentship and a PPARC postdoctoral research assistantship.


%


\appendix


\section{HKLM fitting functions}\label{app-HKLM}


\subsection{The JMW95 function}\label{appJMW95}

The JMW95 function was designed to model the
$n$-dependence of the nonlinear evolution of scale-free power
spectra. The formula was also used to model $\Omega=1$, CDM-like
models through the adoption of an effective spectral index; see
equation (\ref{JMW95neff}). JMW's formula described their numerical
data with an rms accuracy of 15-20\%, but for our higher
resolution scale-free data the fit is much worse, having an rms error
of 56\%. Their formula is
\be
\frac{\Delta_{\NL}^2(k_{\NL})}{B(n)}=f_{\JMW}
\left[\frac{\Delta_{\L}^2(k_{\L})}{B(n)}\right]\
,\ee
where $B(n)$ is a constant which depends upon the spectral index $n$
and where $f_{\JMW}(y)$ remains independent of $n$. The explicit forms
are:
\be B(n)=\left(\frac{3+n}{3}\right)^{1.3}\ee
and
\be f_{\JMW}(y)=y\left[\frac{1+0.6y+y^2-0.2y^3-1.5y^{3.5}+y^4}{1+0.0037y^3}
\right]^{0.5}\ee
where $y\equiv\Delta_{\L}^2(k_{\L})/B(n)$ .


\subsection{The PD96 function}\label{appPD96}

PD96 performed a similar study to JMW95, but extended the set of
cosmological models to include $\Omega<1$ open and flat
universes. They also improved on JMW95 by including CDM data in the
optimization procedure and by proposing that the effective spectral
index would vary continuously with scale: equation
(\ref{PD96neff}). They reported that their fitting formula described
their simulation data to an accuracy of about 14\%, but it describes
our complete data set with an rms error of 54\%. The PD96 fitting
formula is
\be f_{\PD}(y) = y \left[ \frac{1+B\beta y+[Ay]^{\alpha\beta}}
{1+([Ay]^{\alpha}g^3(\Omega,\Lambda)/[Vy^{1/2}])^\beta}\right]^{1/\beta}\ ,\ee
where $y\equiv\Delta_{\L}^2(k_{\L})$. $B$ describes a second order
deviation from linear growth; $A$ and $\alpha$ parameterize the power
law that dominates the function in the quasi-linear regime; $V$ is the
virialization parameter that gives the amplitude of the
$f_{\NL}\propto y^{3/2}$ asymptote; $\beta$ softens the transition
between these regions; $g(\Omega)$ is the density dependent growth
factor of \cite{Carrolletal1992}, which is the ratio of the linear
growth factor to the expansion factor. This has the functional form
\be 
\eqalign{
g(\Omega) &= \frac{D(a)}{a} \cr &= 
\frac{5}{2}\Omega\left[\Omega^{4/7}-\Lambda
 +(1+\Omega/2)(1+\Lambda/70)\right]^{-1}\ .
\cr
}
\ee
The best-fitting parameters were
\ba
A & = & 0.482\,(1+n/3)^{-0.947}\nonumber\\
B & = & 0.226\,(1+n/3)^{-1.778}\nonumber\\
\alpha & = & 3.310\,(1+n/3)^{-0.244}\nonumber\\
\beta & = & 0.862\,(1+n/3)^{-0.287}\nonumber\\
V & = & 11.55\,(1+n/3)^{-0.423}.
\ea


\section{New HKLM fits to the present data}\label{appHKLM}


We have performed a nonlinear least squares fitting to the individual
scale-free loci (see Fig. \ref{HKLMplot1}) using a single
formula. The individual fitting functions are accurate to
$\simeq9\%$. The formula is
\be f_{\rm EdS}(y)= y\left[
\frac{1+y/a+(y/b)^2+(y/c)^{\alpha-1}}
{1+(y/d)^{(\alpha-\beta)\gamma}}\right]^{1/\gamma}\ ,\ee
where $y\equiv\Delta_{\L}^2(k_{\L})$ 
and the relevant parameters for each $n$ are presented below
\[
\begin{array}{cccccccc} 
n & a & b & c & d & \alpha & \beta & \gamma \\
-2 & 3.138 & 0.358 & 0.527 & 0.940 & 8.247 & 0.508 & 0.330\\
-1.5 & 2.710 & 0.710 & 0.919 & 1.852 & 0.707 & 0.647 & 0.332\\
-1 & 10.37 & 1.115 & 1.403 & 2.873 & 6.655 & 0.697 & 0.366\\
0  & 29.26 & 1.394 & 1.941 & 3.753 & 6.547 & 0.847 & 0.351
\end{array}
\] 


\section{The halo model fitting function}\label{halofit}

The halo model decomposes the power into a sum of two contributions:
\be \Delta^2_{\NL}(k)=\Delta^2_{\Q}(k)+\Delta^2_{\H}(k).\ee
These are given separately by
\be \Delta^2_{\Q}(k)= \Delta^2_{\L}(k) \left[\frac{(1+\Delta_{\L}^2(k))^{\beta_n}}
{1+\alpha_n\Delta^2_{\L}(k)} \right]
\exp{[-f(y)]};\ee
where $y\equiv k/k_{\sigma}$ and $f(y)=y/4+y^2/8$; and 
\be \Delta^2_{\H}(k) = {\Delta^{2\ \prime}_{\H}(k) \over 1+\mu_ny^{-1}+\nu_ny^{-2}},\ee
where
\be \Delta^{2\ \prime}_{\H}(k) = \frac{a_n\;  y^{3f_1(\Omega)}
}{1+b_ny^{f_2(\Omega)}
+\left[ c_nf_3(\Omega)\;y\right]^{3-\gamma_n}}\label{app-halo1}\ee
and $y\equiv k/k_{\sigma}$. 

The parameters of the spectrum are defined via Gaussian filtering:
\be \sigma^2(R_{\rm G}) \equiv \int \Delta^2_{\L}(k) \, \exp(-k^2 R_{\rm G}^2)\;
d\ln k.\ee
In these terms,
\be
\sigma(k_{\sigma}^{-1}) \equiv 1.
\ee
The effective index is
\be 3+n_{\rm eff} \equiv -\left.{d \ln \sigma^2(R) 
\over d \ln R}\right|_{\sigma=1},\ee
and the spectral curvature is
\be C \equiv - \left.{d^2 \ln \sigma^2(R) \over d \ln R^2}
\right|_{\sigma=1}. \ee
Allowing $(a_n,b_n,c_n,\gamma_n,\alpha_n,\beta_n,\mu_n,\nu_n)$ to vary
as a function of spectral properties, the following coefficients fit
our simulation data and the CDM simulations of J98 to an rms precision
of 8.6\% (very much better than PD96).  In particular, the model
describes the $\Lambda$CDM data of J98 extremely well. For redshifts
$z<3$, the deviation in power between model and the average of the
large-box and small-box data from J98 is always less than 3\% for
$k<10\hompc$.  This represents a perfect fit with present knowledge,
since the two datasets themselves can differ by at least this much.
Note the use of terms up to $n^4$ in the fit for $a_n$; these are
required in order to describe the rapid rise in amplitude of the halo
term for $n<-2$. For less negative $n$, the higher-order terms are
unimportant. The coefficients are: 
\[ \log_{10} a_n = 1.4861 + 1.8369\;n + 1.6762\;n^2 + 0.7940\;n^3\] 
\be \hspace{1.5cm}+ 0.1670\;n^4 -0.6206\;C\;;\ee
\be \log_{10} b_n = 0.9463+0.9466\; n +0.3084\;n^2-0.9400\; C \;;\ee
\be \log_{10} c_n = -0.2807+0.6669\; n + 0.3214\; n^2\; -0.0793\;C \;;\ee
\be \gamma_n = 0.8649 + 0.2989\;n + 0.1631\;C \;;\ee
\be \alpha_n = 1.3884+0.3700\;n-0.1452\;n^2 \;;\ee
\be \beta_n = 0.8291+0.9854\;n +0.3401\;n^2\;;\ee
\be \log_{10}\mu_n = -3.5442+0.1908\;n \;;\ee
\be \log_{10}\nu_n = 0.9589+1.2857\;n \;;\ee
and the $\Omega$ dependent functions are:
\be 
\left. 
\begin{array}{lll}
f_{1a}(\Omega) & = & \Omega^{\;-0.0732} \\
f_{2a}(\Omega) & = & \Omega^{\;-0.1423}\\
f_{3a}(\Omega) & = & \Omega^{\;0.0725}\\
\end{array}
\right\} \ \ \Omega\le1
\ee
\be 
\left. 
\begin{array}{lll}
f_{1b}(\Omega) & = & \Omega^{\;-0.0307}\\
f_{2b}(\Omega) & = & \Omega^{\;-0.0585}\\
f_{3b}(\Omega) & = & \Omega^{\;0.0743}\\
\end{array}
\right\} \ \ \Omega+\Lambda=1
\ee
For models in which $\Lambda$ is neither zero nor
$1-\Omega$, we suggest interpolating the functions
$f_1$ etc. linearly in $\Lambda$ between the open and flat cases.

\end{document}